\documentclass{emulateapj}

\usepackage{graphicx}

\usepackage{subfigure}
\usepackage{amsmath}

\newcommand{\kms}{\text{km\,s$^{-1}$}}
\newcommand{\ale}{\ensuremath{\alpha}-enhancement}
\newcommand{\sbr}[1]{_{\rm #1}}

\newcommand{\expec}[1]{\ensuremath{\langle #1 \rangle}}

\slugcomment{The Astrophysical Journal, submitted}

\shorttitle{Star Formation Histories of Red Galaxies}
\shortauthors{Allanson et al.}

\begin{document}

\title{The Star Formation Histories of Red-Sequence Galaxies,
	Mass-to-Light Ratios and the Fundamental Plane}
 
\author{Steven P. Allanson and Michael J. Hudson}
\affil{Department of Physics and Astronomy, University of Waterloo, Waterloo, ON N2L 3G1, Canada}

\author{Russell J. Smith and John R. Lucey}
\affil{Department of Physics, University of Durham, Durham DH1 3LE, United Kingdom}
\begin{abstract}

This paper addresses the challenge of understanding the typical star
formation histories of red sequence galaxies, using linestrength
indices and mass-to-light ratios as complementary constraints on their
stellar age distribution.  We first construct simple parametric models
of the star formation history that bracket a range of scenarios, and
fit these models to the linestrength indices of low-redshift cluster
red-sequence galaxies. For giant galaxies, we confirm the downsizing trend, i.e.\ the stellar populations are younger, on average, for lower $\sigma$ galaxies. We find, however, that this trend flattens or reverses at $\sigma \lesssim 70 \kms$. We then compare predicted stellar
mass-to-light ratios with dynamical mass-to-light ratios derived from
the Fundamental Plane, or by the SAURON group.  For galaxies with
$\sigma\sim{}70$ \kms, models with a late ``frosting'' of young stars
and models with exponential star formation histories have stellar
mass-to-light ratios that are larger than observed dynamical
mass-to-light ratios by factors of $1.7$ and $1.4$, respectively, and
so are rejected.  The single stellar population (SSP) model is
consistent with the Fundamental Plane, and requires a modest amount of
dark matter (between 20\% to 30\%) to account for the difference
between stellar and dynamical mass-to-light ratios. A model in which
star formation was ``quenched'' at intermediate ages is also
consistent with the observations, although in this case less dark
matter is required for low mass galaxies.  We also find that the
contribution of stellar populations to the ``tilt'' of the Fundamental
Plane is highly dependent on the assumed star-formation history: for
the SSP model, the tilt of the FP is driven primarily by
stellar-population effects. For a quenched model, two-thirds of the
tilt is due to stellar populations and only one third is due to dark
matter or non-homology.

\end{abstract}


\section{Introduction}

Red-sequence galaxies (RSGs) dominate the stellar mass in clusters and comprise $\sim 70\%$ \citep{BelMcIKat03} of the stellar mass in the Universe.  However, other than the fact that the \emph{bulk} of their stellar population is older than $\sim 1$ Gyr, a detailed understanding of their star formation histories remains elusive.  RSGs follow a tight color-magnitude relation \citep[hereafter CMR,][]{SanVis78,BowLucEll92} and scaling relations such as the Fundamental Plane \citep[hereafter FP,][]{DreLynBur87,DjoDav87} with little scatter. At the same time, the dependence of the colors on magnitude suggests that there are indeed variations in their stellar properties as a function of luminosity or a mass-related parameter such as stellar mass or velocity dispersion. Variations in the stellar populations would also affect the ``tilt'' of the Fundamental Plane with respect to the virial scaling, although the degree to which the tilt is due to stellar populations compared with dark matter (DM) and non-homology remains hotly debated \citep{PruSim96,PahdeCDjo98, GerKroSag01,TruBurBel04,CapBacBur06,LaBBusMer08}.

Broadband optical colors are not good discriminants of stellar populations because of the age-metallicity degeneracy \citep{Wor94}. These degeneracies can be broken by using stellar absorption lines \citep{Wor94,ThoMarBen03}. Recent results suggest that metallicity, \ale\ and age vary along the mass or velocity dispersion sequence \citep{CalRosCon03,NelSmiHud05,ThoMarBen05}, and also vary as a function of environment \citep{ThoMarBen05, SmiHudLuc06,SmiLucHud09}.

A limitation of the above analyses is that, in general, the absorption line indices are sensitive only to weighted mean stellar age, with a  weighting that strongly favors the most recent star formation \citep{SerTra07}. The indices provide only weak information about the range of stellar ages within a galaxy. For convenience, then, stellar population ages are often quoted as the equivalent age of a Single Stellar Population (hereafter SSP). In practice, other scenarios have been studied. For example, \cite{TraFabWor00} considered ``frosting'' models in which a small (few percent) ``icing'' of young stars on top of a ``cake'' of older stars yields a young luminosity-weighted age; \cite{ThoMarBen02,ThoMarBen05} modeled star formation histories as a Gaussian in age; \cite{BelMcIKat03} assume an exponential star formation history; \cite{KauHecWhi03} model SDSS galaxies as a combination of an exponential with small bursts; and \cite{HarSchWei06} considered models in which star formation starts early and is later quenched. In terms of predicting the line indices, these scenarios are largely, but not completely, degenerate, as we will show in Section \ref{sec:fitsline}.

However, it may be possible to break these degeneracies using other, complementary, observations. One approach is via chemical abundances: because different elements are produced by different progenitors at various times, it is possible (in principle) to use the abundances as a clock \citep[e.g.][]{ThoMarBen05}. In practice, however, there are a large number of unknowns (masses of progenitors, yields, gas inflows and outflows) which render the quantitative conversion of element abundances into time-scales problematic. 
An alternative approach is to compare the predictions of various star-formation history scenarios against other observational relations besides line indices. At low redshifts, the observables include the slope of the CMR, the tilt of the FP and the scaling of mass-to-light ratios with mass or velocity dispersion. At higher redshift, one can compare the predictions for the evolution of the zero-point, slope and scatter of the CMR and FP, and of the RSG dwarf-to-giant ratio. 

In this paper we fit the observed linestrength indices adopting six generic models of the star formation history.  Previous studies \citep{TraFabWor00,BerSheNic05} find velocity dispersion to be the ``driving parameter'' of stellar populations. \cite{SmiLucHud09c}, using the same spectroscopic data as in this paper, not only confirm velocity dispersion to be the primary parameter of stellar populations, but find \textit{no} additional dependence of the stellar populations on stellar mass. Therefore, for each model we assume a mean scaling and intrinsic scatter in metallicity, \ale\ and ``age'', all assumed to be functions of velocity dispersion. Here, ``age'' is some parameter related to the timescale of star formation, the definition differing from model to model. Having constrained the scalings via the line strength data, we generate synthetic clusters using the stellar population parameters to determine mass-to-light ratios, colors, magnitudes etc. For these synthetic clusters, we construct the CMR and FP relations and compare these to observational data from low redshift rich clusters, under the assumption that these relations are universal for rich clusters \citep{LopBarYee04,McIZabRix05}. In a future paper, we will extend this approach to comparisons with high redshift clusters.

An outline of this paper is as follows: Section \ref{sec:data} describes the spectroscopic and photometric data sets; Section \ref{sec:csp} describes the models of star formation history. The results of the fits to these models are presented in Section \ref{sec:fit}.  Having fixed the parameters of the models with the spectroscopic data, we create synthetic clusters based on these models in Section \ref{sec:synth} and compare these to dynamical mass-to-light ratios, colors, and the Faber-Jackson and Fundamental Plane relations in Section \ref{sec:results}.  We discuss the effect of systematics, the IMF and of dark matter on our results in Section \ref{sec:mod} and discuss the impact of our results on derived stellar mass densities and the tilt of the FP in Section \ref{sec:discuss}. We summarize the results in Section \ref{sec:conc}.

Throughout this paper, we assume the following cosmological parameters: $(\Omega_M,\Omega_{\Lambda},h)=(0.3,0.7,0.7)$.\\

\section{Data}
\label{sec:data}

In this paper, we will make use of two datasets for the RSG population in rich clusters. We assume that RSG populations in rich clusters are universal, in the sense of having the same distribution of star-formation histories at a given RSG mass. Absorption linestrength data from deep observations of the RSGs in three rich clusters in the Shapley supercluster are used to derive the ages and metallicities of the RSG population.  The resulting predicted colors, magnitudes and Fundamental Plane parameters for the RSG population are compared to data in the Coma cluster. 

\subsection{Linestrength Data from Cluster Galaxies in Shapley Concentration}
\label{sec:shapley}

The Shapley cluster sample consists of $\sim{}180$ R $<$ 18 emission-free cluster galaxies \citep{SmiLucHud07} that cover a wide range in central velocity dispersion ($38\,\kms < \sigma < 313\,\kms$). There is no morphological selection in this dataset: the only selection is on H$\alpha$ emission, and this is strongly correlated with color: $\sim{}99$\% of the emission-free galaxies are red. Furthermore, red-selection and morphology are tightly correlated: for example, only $\sim 20$\% of NFPS galaxies with $\sigma > 70 $ \kms\ are types Sa or later (Hudson et al. 09, in preparation). Finally, we note that nearly all of the galaxies in the Shapley cluster sample lie within half the virial radius ($r_{200}/2$) of the center of one of the three rich clusters (A3556, A3558, A3562) in the supercluster.

The Lick linestrength indices were measured from 8-hour, high signal-to-noise ratio ($\sim 60$ \AA$^{-1}$) spectra, obtained with the AAOmega instrument at the 3.9m Anglo-Australian Telescope. The spectra were obtained through a 1'' radius fiber, corresponding to a physical radius of 0.95 kpc. In this paper, we use six Lick linestrength indices (three Balmer indices and three metallic indices) to measure age-related quantities and metallicities, as discussed in Section \ref{sec:fit} below.

\subsection{Coma Cluster Galaxy Data}
\label{sec:comadata}

\begin{figure*}[htbp]
	\centering
	\includegraphics[angle=270,width=\linewidth]{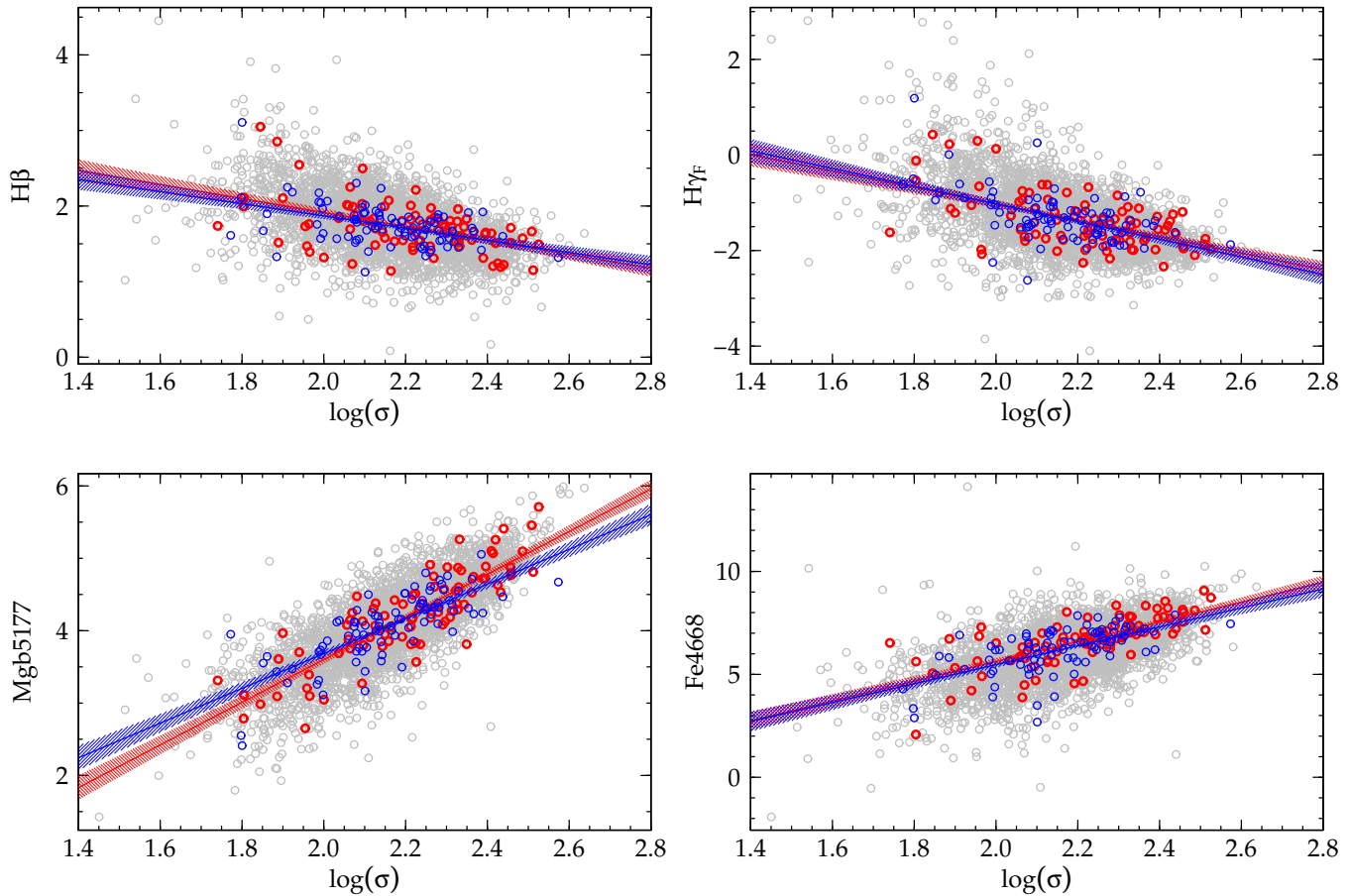}
	\caption[Linestrength-$\sigma$ relations]{Linestrength-$\sigma$ relations from the NFPS sample for Coma (blue) and the Shapley clusters (red).  NFPS data for other clusters are shown in grey. In all cases, galaxies within 0.5 Mpc are used. There is no significant difference between Coma and Shapley clusters.
	}
	\label{fig:ComaShap}
\end{figure*}

We will compare predictions from our models to photometric data for $218$ galaxies in the Coma Cluster from SDSS DR 6 \citep{AdeAguAll08} and 2MASS \citep{SkrCutSti06} with measured velocity dispersions from the literature. We compare with Coma rather than Shapley itself because no published photometry is available for Shapley, whereas a large range of wavebands is available in Coma ($u$ through $K$). We note that the Coma cluster is of similar richness to the Shapley clusters, and \cite{LopBarYee04} have shown that for Abell clusters, the cluster-to-cluster dispersion in CMR colour (at the characteristic luminosity $L_{\ast}$) is only $0.05$ in $B-R$. Moreover, the CMR of Coma is typical, deviating by no more than 0.05 mags from the average relation. Figure \ref{fig:ComaShap} compares the linestrength indices from the homogeneous NFPS survey \citep{NelSmiHud05}.
There is no significant difference between the index-$\sigma$ relations of the two clusters.

Derived photometric parameters in Coma include half-light radii, surface brightnesses within the half-light radius, and total magnitudes measured in the SDSS $r$-band. These parameters have been corrected for the effects of seeing by fitting a model using GALFIT \citep{PenHoImp02} and deriving corrections to observed values from the difference between the seeing-convolved model and the unconvolved  model. In order to compare colors with ages and metallicities predicted from the Shapley linestrengths, we measure the colors within a 2'' radius aperture corresponding to a physical aperture of 0.97 kpc, which matches the 1'' aperture of the AAOmega data in Shapley. (Shapley is approximately twice as distant as Coma.) These aperture magnitudes are corrected for seeing in the same way as described above. Data are k-corrected using formulae from \cite{FreGun94}, and corrected for Galactic extinction using \cite{SchFinDav98}. Finally, because there is evidence for gradients in the stellar populations as a function of cluster-centric radius \citep{SmiHudLuc06,SmiMarHor08}, we limit fits to Coma galaxies within $r_{200}/2$ (i.e. within 1.24 Mpc, 0.71\arcdeg) of the cluster center.

\begin{figure}[htbp]
	\centering
	\includegraphics[width=\linewidth]{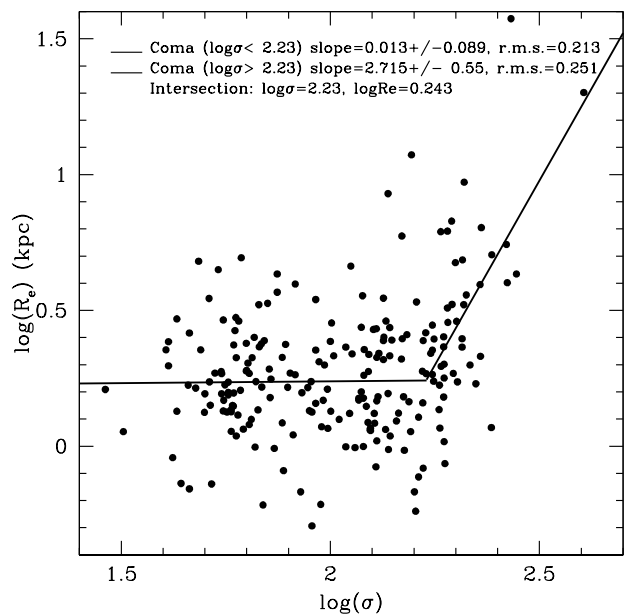}
	\caption[Coma $R_e$-$sigma$ relation]{Coma $\log(R_e)-\log(\sigma)$ relation. A broken power law (Equation \ref{eq:Resig}) was fit to Coma galaxies with cluster-centric radius less than $r_{200}/2$ (filled circles) to match the Shapley sample.
	}
	\label{fig:ComaRe}
\end{figure}

In order to model stellar masses, as well as to calculate aperture corrections, we will require the relationship between effective radius and velocity dispersion. This relationship in Coma is shown in Figure \ref{fig:ComaRe}. We find that a broken power-law is required to fit the data:
\begin{eqnarray}
\log R_e &\  =\ & 0.013 \log \sigma + 0.21; \qquad \log(\sigma)<2.23 \\
\log R_e &\  =\ & 2.715 \log \sigma -5.81; \qquad \log(\sigma)>2.23 \nonumber
\label{eq:Resig}
\end{eqnarray}
where $R_e$ is in units of kpc and $\sigma$ is in units of km/s. The break in $R_e$ is closely related to the well-known break in the surface brightness behavior \citep{Kor85}, and has been seen in the Faber-Jackson relation \citep{MatGuz05} and the $R_e - L$ relation \cite{HydBer08a}. The dispersion in $\log(R_e)$, is 0.213 for $\log(\sigma) < 2.23$, and is 0.251 for larger values. The scatter at the low-$\sigma$/faint end is comparable to that found at fixed luminosity for the SDSS for faint early-type galaxies; for comparison \citep{SheMoWhi03} find a dispersion $\gtrsim 0.22$ in $\log(R_e)$ for $M_r > -19$.

\section{Synthetic Linestrengths for Galaxies with Complex Star Formation Histories}
\label{sec:csp}

Our goal is to extend the SSP analyses of previous work to more complex star formation histories (CSFHs). A CSFH model can be generated by convolving the SSP response with the star formation rate. Here we combine the SSP SED's of Maraston \citep{Mar98,Mar05} with the $\alpha-$enhanced Lick-index absorption linestrengths \citep[together hereafter the TMBK models]{ThoMarBen03,ThoMarKor04} to construct linestrength indices for CSFHs. Specifically, the TMBK line indices are based on SSP models. To obtain predictions for complex star formation scenarios, we reconstruct the fluxes in the sidebands and in the central (absorption) band for each step of star formation. By integrating these fluxes over all time steps, appropriately weighted, we can produce a set of indices for any formation scenario. A very similar method applies to mass-to-light ratios.

The range of possible star formation histories is clearly large and highly uncertain. Here we consider six simple star formation scenarios that are intended to bracket more complicated, realistic scenarios. All but one of the models we consider have a single age-related quantity, whose definition depends on the specific model detailed below. The exception is the Old SSP model, which has a single fixed age. Note that ``ages" from Lick indices have poor age sensitivity at old ages, so it is impossible to discriminate, for example, an old instantaneous burst from an old burst of the same age, but with a short duration of 1 Gyr. The models and their abbreviations are as follows:

\begin{itemize}
\item Old SSP (hereafter OSP): a single burst of star formation at a fixed age of 13 Gyr.
\item Single Stellar Population (SSP): a single burst of star formation, with fitted age $t\sbr{SSP}$.
\item Exponential star formation rate (EXP): star formation begins 13 Gyr ago, ends 0.1 Gyr ago, with a fitted exponential decay parameter $\tau$.
\item ``Abruptly'' Quenched (AQ) star formation: a model with constant star formation beginning 13 Gyr ago, and an abrupt truncation of star formation at a fitted lookback time, $t\sbr{AQ}$.
\item A 2-component ``frosting'' model (FR) consisting of a dominant (98\% by mass) 13 Gyr SSP and a secondary (2\% by mass) burst with fitted age $t\sbr{\textrm FR}$. 
\item A ``Strangulation'' model (STR): a quenched (constant SFR) model beginning 13 Gyr ago up to fitted $t\sbr{STR}$, followed by an exponential SFR from $t\sbr{STR}$ to 0.1Gyr ago with fixed $\tau=1$ Gyr. The mass fractions in each component are a function of $t\sbr{STR}$, such that the SFR at $t\sbr{STR}$ is continuous.
\end{itemize}

In each model, we assume a single metallicity and a single \ale\ for all generations of stars, and fit these parameters to the data. Our models therefore do not attempt to include chemical enrichment in any physically motivated way and so the abundance parameters are intended to represent luminosity-weighted mean values. 

Note that in the OSP, AQ, EXP, FR and STR models, we have chosen to fix the age of the old population or the ``starting time'' for star formation at 13 Gyr. While, in principle, one would like to leave the starting time as a free parameter, in practice, the multiple age parameters become highly degenerate. We will discuss the implications of varying the age of the oldest stars in Sections \ref{sec:sysparam} and \ref{sec:sysfx}.

These models can be subdivided into two classes. In one class of models (OSP, EXP, FR), the stars are predominantly old. In the case of the EXP and FR models, intermediate ($\sim 5$ Gyr) luminosity-weighted ages can arise, but these are due to a small fraction (by mass) of very young ($\lesssim 1$ Gyr) stars in addition to the dominant (by mass) old population. In the other class of models (SSP, AQ, STR), there is little star formation at late times and the luminosity-weighted intermediate ages are due to stars which indeed formed at intermediate times.  We will show that these two classes have different properties, particularly in terms of their stellar mass-to-light ratios, even when constrained to match the same line index data. 

\section{Fitting Star Formation History Models to Linestrength Indices}
\label{sec:fit}

The goal of this paper is to create synthetic clusters of simulated galaxies with distributions of age and metallicity that are consistent with the observed line indices.  While the observational data are of high $S/N$, the fitted ages and metallicities of individual galaxies have substantial correlated errors, particularly for the faintest, low velocity dispersion galaxies.  Thus, rather than \emph{invert} the data, we instead \emph{model} the mean and scatter in the stellar population parameters, $P$ (``age'', metallicity and \ale) that are required to match the median linestrength and the scatter in the observed linestrengths.

By comparing the scatters in either linestrengths \citep{TraFabWor00}, colors \citep{BerSheNic05} or total $M/L$ \citep{CapBacBur06} as a function of velocity dispersion, with the same scatters as a function of luminosity, one finds that the correlations with velocity dispersion are always tighter than those with luminosity.  Indeed, we have confirmed this using our Coma cluster data described above. These results strongly suggest that the fundamental or ``driving" parameter of stellar populations is the velocity dispersion \citep{SmiLucHud09c}, an assumption we will make in this paper. Thus we divide our sample into five bins by $\log(\sigma)$ and measure median Lick indices and scatters for each bin, both of which are then fit to each star formation history model.

We will use the six Lick indices given in Table \ref{tab:corr} to break the degeneracies between ``age'' (traced primarily by the Balmer lines: H$\delta$F, H$\gamma$F, H$\beta$), Fe (Fe4383, Fe5015) and $\alpha$-element enhancement (Mgb5177).  We assign the galaxies to five bins in velocity dispersion, and for each bin we calculate the observed median linestrength index.   Our stellar population parameters include one age-related parameter ($t\sbr{SSP}$,$t\sbr{AQ}$,$t\sbr{STR}$,$\tau$,$t\sbr{FR}$), $[Z/H]$ and $\alpha/$Fe. 

In general, however, we expect that the linestrength data may have small zero-point offsets. To compare with previous results, and following \cite{SmiLucHud09c}, we force the median line indices in the highest-velocity dispersion bin to equal the SSP-parameters recovered by \cite{NelSmiHud05} in their highest velocity dispersion bin: (age, [Z/H], [$\alpha$/Fe]) = (10.8 Gyr, 0.24, 0.28). This corresponding offset given in Table \ref{tab:corr} in each line is then applied to all bins.

Our goal is not only to fit the median line indices at each velocity dispersion, but also fit the spread in ages, metallicities and \ale\ to the spread in each line index of each bin.  We first calculate the total observed scatter $S\sbr{tot}$ in a given index $I$ at a fixed $\sigma$  by measuring the semi-interquartile-range (SIQR) and convert this to its Gaussian equivalent: $S\sbr{tot}=\textrm{SIQR}/0.67$. This is more robust to outliers than the usual root-mean-square. We then estimate the \emph{intrinsic} dispersion in line index properties as follows:
\begin{equation}
S^2\sbr{int}=S^2\sbr{tot} - S^2\sbr{meas}
\end{equation}
where $S\sbr{meas}$ is taken to be the median measurement error in the bin for each specific absorption line.

\begin{deluxetable}{lc}
\tablecaption{Best-Fit Line Index Corrections}
\tablehead{
\colhead{Line Index} &
\colhead{Correction\tablenotemark{a} ($\AA$)}}
\tablewidth{0pt}
\startdata
 H$\delta$F & -0.063\\
 H$\gamma$F & +0.026\\
 Fe4383     & +0.257\\
 H$\beta$	& +0.061\\
 Fe5015     & +0.667\\
 Mgb5177    & +0.274
\enddata
\tablenotetext{a}{Applied to all observed bins}
\label{tab:corr}
\end{deluxetable}

\subsection{Results}
\label{sec:linestrengthresults}

\subsubsection{Fits to Median Linestrengths}
\label{sec:fitsline}
\begin{figure}[htbp]
	\centering
			\includegraphics[width=\linewidth]{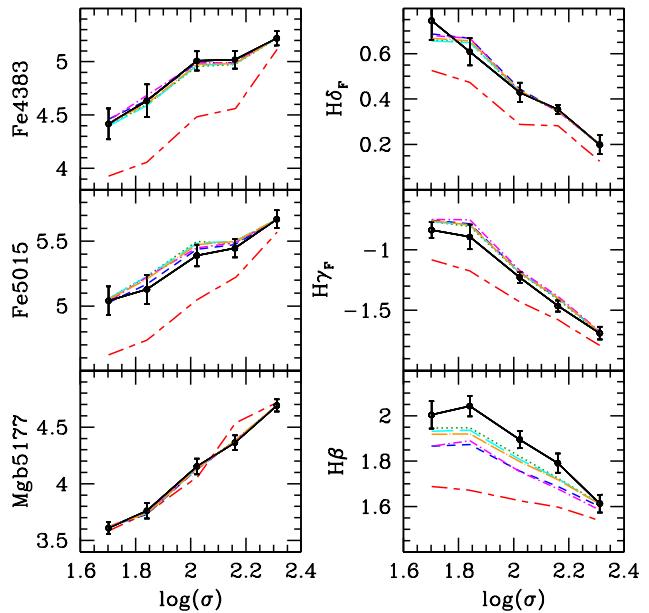}
			\caption[6 Fit Absorption Lines]{Linestrengths as a function of velocity dispersion.  The Shapley binned linestrength data are shown by the black lines and circles, with error bars representing the uncertainty in the median. The predicted line index values for the best fit model parameters for each model are also shown: OSP (red short - long dashed); SSP (green dotted); EXP (blue short dashed line); AQ (cyan long dashed); FR (magenta dot - short dashed), STR orange dot - long dashed). Note that the OSP model is offset even in the highest velocity dispersion bin because we have adopted an age of 13 Gyr for this model rather than the default of 10.8 Gyr. It is clear that, with the exception of OSP, the models are nearly degenerate in most line indices.}
	\label{fig:6fitlines}
\end{figure}

We employ a simple $\chi^2$ fitting scheme to the observed median linestrength indices. The results are shown in Figure \ref{fig:6fitlines} and are tabulated in Table \ref{tab:indices}, along with the Shapley binned values to which the models were fit.  Corresponding model parameters appear in Table \ref{tab:params}. Note that, for the Balmer lines H$\gamma$F and H$\beta$, the scaling is roughly linear at high velocity dispersions, but the lowest-$\sigma$ bin appears to deviate from this linear relationship. This will yield older ages for this bin. We will discuss this further in Section \ref{sec:lowvd}. 

From Fig. \ref{fig:6fitlines}, we see that all models, except the Old SSP Model which is clearly a poor fit,  yield similar predicted linestrengths. However, closer inspection of H$\beta$ shows the SSP, AQ and STR models are a better fit to this index. To quantify this, we calculate a $\chi^2$ for each model, for each line, as well as an aggregate $\chi^2$ over all six lines for the four lowest mass bins (since we have fit a free offset to match the high velocity dispersion bin), and these values are tabulated in  Table \ref{tab:indices}.  Formally, the EXP, FR and OSP are poor fits, the latter being rejected at $>99.99$ CL.  However, this is somewhat driven by our choice to correct the linestrengths to obtain a given age in the highest velocity dispersion bin, particularly for H$\beta$.  If instead, we attempt to fit H$\beta$ in all velocity dispersion bins simultaneously, then EXP and FR remain poorer fits than SSP but are no longer formally rejected. The OSP model, however, is still rejected at a high confidence level.

\begin{deluxetable*}{c|cccc|cccccc}
 \tablewidth{\linewidth}
\tablecaption{Best-Fit Model Linestrength Indices by Bin and Model}
\tablehead{
\colhead{Bin} &
\colhead{Line} &
\colhead{Median ($\AA$)} &
\colhead{$^{Error\ in}\sbr{Median}$} &
\colhead{S} &
\colhead{SSP} &
\colhead{AQ} &
\colhead{STR} &
\colhead{EXP} &
\colhead{FR} &
\colhead{OSP}
}
	\startdata
	\multicolumn{1}{c}{$\sigma$ in km/s} \vline &
	\multicolumn{4}{c}{Observed Data} \vline &
	\multicolumn{6}{c}{Model Best Fits}\\ \hline			
Range of $\sigma$	&	H$\delta$F	&	0.746	&	0.085	&	0.287	&	0.660	&	0.655	&	0.667	&	0.687	&	0.679	&	0.525	\\	
$=35\rightarrow59$;	&	H$\gamma$F	&	-0.836	&	0.065	&	0.205	&	-0.767	&	-0.766	&	-0.762	&	-0.756	&	-0.748	&	-1.084	\\	
$\langle\log(\sigma)\rangle$	&	Fe4383	&	4.417	&	0.144	&	0.510	&	4.402	&	4.397	&	4.419	&	4.460	&	4.447	&	3.925	\\	
$=1.706$	&	H$\beta$	&	2.004	&	0.060	&	0.222	&	1.946	&	1.932	&	1.917	&	1.867	&	1.866	&	1.687	\\	
	&	Fe5015	&	5.041	&	0.110	&	0.405	&	5.058	&	5.058	&	5.050	&	5.026	&	5.034	&	4.621	\\	
N$\sbr{gal}=33$	&	Mgb5177	&	3.610	&	0.052	&	0.168	&	3.621	&	3.620	&	3.620	&	3.623	&	3.623	&	3.582	\\	\hline
	\multicolumn{4}{c}{} &
	\multicolumn{1}{r}{$\chi^2$:} &
	3.14 & 3.84 & 4.31 & 7.44 & 7.88 & 75.6 \\ \hline
Range of $\sigma$	&	H$\delta$F	&	0.608	&	0.060	&	0.207	&	0.648	&	0.648	&	0.656	&	0.668	&	0.668	&	0.472	\\	
$=59\rightarrow84$	&	H$\gamma$F	&	-0.893	&	0.102	&	0.466	&	-0.806	&	-0.794	&	-0.795	&	-0.785	&	-0.752	&	-1.177	\\	
$\langle\log(\sigma)\rangle$	&	Fe4383	&	4.634	&	0.153	&	0.674	&	4.596	&	4.588	&	4.603	&	4.637	&	4.683	&	4.055	\\	
$=1.844$	&	H$\beta$	&	2.043	&	0.045	&	0.181	&	1.947	&	1.936	&	1.919	&	1.873	&	1.891	&	1.671	\\	
	&	Fe5015	&	5.128	&	0.111	&	0.489	&	5.228	&	5.228	&	5.205	&	5.168	&	5.223	&	4.733	\\	
N$\sbr{gal}=37$	&	Mgb5177	&	3.761	&	0.068	&	0.300	&	3.739	&	3.739	&	3.736	&	3.728	&	3.750	&	3.738	\\	\hline
	\multicolumn{4}{c}{} &
	\multicolumn{1}{r}{$\chi^2$:} &
	6.64 & 7.91 & 9.67 & 16.5 & 15.1 & 107 \\ \hline
Range of $\sigma$	&	H$\delta$F	&	0.429	&	0.042	&	0.153	&	0.431	&	0.434	&	0.436	&	0.444	&	0.435	&	0.287	\\	
$=84\rightarrow124$;	&	H$\gamma$F	&	-1.228	&	0.045	&	0.184	&	-1.195	&	-1.190	&	-1.190	&	-1.184	&	-1.169	&	-1.433	\\	
$\langle\log(\sigma)\rangle$	&	Fe4383	&	5.008	&	0.090	&	0.387	&	4.956	&	4.966	&	4.971	&	4.991	&	4.988	&	4.482	\\	
$=2.022$	&	H$\beta$	&	1.895	&	0.038	&	0.164	&	1.821	&	1.808	&	1.797	&	1.756	&	1.754	&	1.625	\\	
	&	Fe5015	&	5.391	&	0.083	&	0.362	&	5.498	&	5.485	&	5.472	&	5.437	&	5.448	&	5.045	\\	
N$\sbr{gal}=37$	&	Mgb5177	&	4.153	&	0.070	&	0.329	&	4.132	&	4.132	&	4.132	&	4.128	&	4.127	&	4.054	\\	\hline
	\multicolumn{4}{c}{} &
	\multicolumn{1}{r}{$\chi^2$:} &
	6.49 & 7.55 & 8.70 & 15.0 & 16.2 & 136 \\ \hline
Range of $\sigma$	&	H$\delta$F	&	0.354	&	0.019	&	0.166	&	0.346	&	0.347	&	0.347	&	0.348	&	0.346	&	0.281	\\	
$=124\rightarrow171$;	&	H$\gamma$F	&	-1.464	&	0.046	&	0.200	&	-1.409	&	-1.401	&	-1.401	&	-1.399	&	-1.384	&	-1.580	\\	
$\langle\log(\sigma)\rangle$	&	Fe4383	&	5.018	&	0.082	&	0.344	&	4.972	&	4.978	&	4.980	&	4.990	&	4.992	&	4.557	\\	
$=2.160$	&	H$\beta$	&	1.792	&	0.042	&	0.192	&	1.725	&	1.723	&	1.718	&	1.688	&	1.678	&	1.597	\\	
	&	Fe5015	&	5.446	&	0.071	&	0.307	&	5.480	&	5.501	&	5.496	&	5.473	&	5.488	&	5.216	\\	
N$\sbr{gal}=37$	&	Mgb5177	&	4.363	&	0.066	&	0.309	&	4.382	&	4.377	&	4.377	&	4.375	&	4.377	&	4.533	\\	\hline
	\multicolumn{4}{c}{} &
	\multicolumn{1}{r}{$\chi^2$:} &
	4.64 & 5.51 & 5.80 & 8.28 & 10.9 & 89.9 \\ \hline
Range of $\sigma$	&	H$\delta$F	&	0.198	&	0.042	&	0.181	&	0.198	&	0.199	&	0.200	&	0.201	&	0.200	&	0.124	\\	
$=171\rightarrow314$;	&	H$\gamma$F	&	-1.689	&	0.052	&	0.240	&	-1.689	&	-1.687	&	-1.686	&	-1.682	&	-1.673	&	-1.794	\\	
$\langle\log(\sigma)\rangle$	&	Fe4383	&	5.219	&	0.067	&	0.290	&	5.219	&	5.222	&	5.220	&	5.227	&	5.225	&	5.116	\\	
$=2.312$	&	H$\beta$	&	1.613	&	0.039	&	0.179	&	1.613	&	1.611	&	1.609	&	1.601	&	1.584	&	1.537	\\	
	&	Fe5015	&	5.671	&	0.068	&	0.305	&	5.671	&	5.669	&	5.667	&	5.666	&	5.668	&	5.569	\\	
N$\sbr{gal}=36$	&	Mgb5177	&	4.692	&	0.054	&	0.250	&	4.692	&	4.693	&	4.693	&	4.692	&	4.690	&	4.714	 \\ \hline
	\multicolumn{4}{c}{} &
	\multicolumn{1}{r}{$\chi^2$:} &
	0.00 & 0.00 & 0.02 & 0.14 & 0.63 & 15.8	\\ \hline
	\multicolumn{1}{c}{} &
	\multicolumn{4}{r}{Bins $1\rightarrow4$, Total $\chi^2$:} &
	20.91 & 24.8 & 28.5 & 47.2 & 50.1 & 409	
	
	\enddata
\label{tab:indices}
\end{deluxetable*}

\begin{deluxetable*}{cc|cccc|cccc}
\tablecaption{Best-Fit Model Parameters by Bin}
\tablewidth{\linewidth}
\tablehead{
\colhead{Range of $\sigma$ (km/s)} &
\colhead{$\langle\log(\sigma)\rangle$} \vline &
\colhead{$\chi^2$} &
\colhead{$\langle Age\tablenotemark{a}\rangle$ (Gyr)} &
\colhead{$\langle[Z/H]\rangle$} &
\colhead{$\langle[\alpha/Fe]\rangle$} \vline &
\colhead{$\chi^2$} &
\colhead{$S(Age\tablenotemark{b})$} &
\colhead{$S([Z/H])\sbr{res}$} &
\colhead{$S([\alpha/Fe])\sbr{res}$} }

\startdata

	\multicolumn{10}{c}{Single Stellar Population Model (SSP); Age = time since single burst}\\ \hline
	$35-59$	 &	1.706	&	3.14	&	5.91	&	-0.002 &	0.189	&	25.6	&	0.186	&	0.001	&	0.105	\\
	$59-84$	 &	1.844	&	6.64	&	5.53	&	0.089	 &	0.186	&	7.32	&	0.212	&	0.119	&	0.001	\\
	$84-124$	&	2.022	&	6.49	&	7.08	&	0.185	 &	0.214	&	19.7	&	0.142	&	0.079	&	0.046	\\
	$124-171$	&	2.160	&	4.64	&	9.10  &	0.190  &	0.260	&	21.6	&	0.151	&	0.075	&	0.048	\\
	$171-314$	&	2.312	&	0.00  &	10.8	&	0.240	 &	0.280	&	33.4	&	0.141	&	0.070	&	0.041	\\ \hline
	
	\multicolumn{10}{c}{Abruptly Quenched Model (AQ); Age = time since quenching}\\ \hline	
	$35-59$		&	1.706	&	3.84	&	3.17	&	-0.009	&	0.183	&	39.9	&	0.228	&	0.001	&	0.115	\\	
	$59-84$		&	1.844	&	7.91	&	2.66	&	0.081	&	0.180	&	8.61	&	0.331	&	0.121	&	0.001	\\	
	$84-124$	&	2.022	&	7.55	&	3.99	&	0.180	&	0.211	&	22.1	&	0.238	&	0.085	&	0.039	\\	
	$124-171$	&	2.160	&	5.51	&	6.15	&	0.188	&	0.257	&	24.3	&	0.271	&	0.080	&	0.047	\\	
	$171-314$	&	2.312	&	0.00	&	9.06	&	0.240	&	0.280	&	44.9	&	0.136	&	0.090	&	0.037	\\	\hline

	\multicolumn{10}{c}{Strangulation Model (STR); Age = time since quenching}\\ \hline	
	$35-59$		&	1.706	&	4.31	&	5.44	&	-0.018	&	0.175	&	41.2	&	0.123	&	0.001	&	0.116	\\	
	$59-84$		&	1.844	&	9.67	&	5.05	&	0.056	&	0.169	&	4.95	&	0.218	&	0.109	&	0.001	\\	
	$84-124$	&	2.022	&	8.70	&	6.14	&	0.169	&	0.206	&	25.5	&	0.087	&	0.084	&	0.001	\\	
	$124-171$	&	2.160	&	5.80	&	7.90	&	0.184	&	0.255	&	22.2	&	0.127	&	0.074	&	0.001	\\	
	$171-314$	&	2.312	&	0.02	&	10.57	&	0.238	&	0.280	&	43.6	&	0.117	&	0.063	&	0.036 \\ \hline
	
	\multicolumn{10}{c}{Exponential Star Formation Rate Model (EXP); Age = e-folding time}\\ \hline	
	$35-59$	&	1.706		&	7.44	&	2.28	&	-0.040	&	0.161	&	27.2	&	0.711	&	0.001	&	0.095	\\
	$59-84$	&	1.844		&	16.5	&	2.39	&	0.005	&	0.145	&	18.6	&	0.829	&	0.109	&	0.001	\\
	$84-124$	&	2.022	&	15.0	&	2.10	&	0.133	&	0.190	&	30.0	&	0.366	&	0.098	&	0.039	\\
	$124-171$	&	2.160	&	8.28	&	1.81	&	0.160	&	0.246	&	27.5	&	0.613	&	0.085	&	0.015	\\
	$171-314$	&	2.312	&	0.14	&	1.44	&	0.233	&	0.277	&	15.2	&	0.903	&	0.082	&	0.001	\\ \hline

	\multicolumn{10}{c}{Frosting Model (FR); Age = time since secondary burst}\\ \hline	
	$35-59$		&	1.706	&	7.88	&	1.37	&	-0.055	&	0.160	&	32.8	&	0.146	&	0.001	&	0.105	\\	
	$59-84$		&	1.844	&	15.1	&	1.23	&	0.000	&	0.132	&	13.9	&	0.159	&	0.122	&	0.001	\\	
	$84-124$	&	2.022	&	16.2	&	1.48	&	0.119	&	0.184	&	27.1	&	0.085	&	0.099	&	0.001	\\	
	$124-171$	&	2.160	&	10.9	&	1.78	&	0.153	&	0.241	&	26.0	&	0.097	&	0.099	&	0.001	\\	
	$171-314$	&	2.312	&	0.63	&	2.49	&	0.224	&	0.274	&	19.9	&	0.199	&	0.087	&	0.001	\\	\hline
	
	\multicolumn{10}{c}{Old Single Stellar Population Model (OSP); Age = time since single burst (13Gyr)}\\ \hline
	
	$35-59$   & 1.706 & 75.6 & 13.00 & -0.194  & 0.303 & 75.4 & 0.000 & 0.050 & 0.133\\    
	$59-84$   & 1.844 & 107  & 13.00 & -0.151  & 0.303 & 47.0 & 0.000 & 0.137 & 0.067\\
	$84-124$  & 2.022 & 136  & 13.00 & -0.059  & 0.272 & 35.6 & 0.000 & 0.109 & 0.05 \\
	$124-171$ & 2.160 & 89.9 & 13.00 & 0.044   & 0.338 & 38.4 & 0.000 & 0.105 & 0.057\\
	$171-314$ & 2.312 & 15.8 & 13.00 & 0.173   & 0.293 & 50.2 & 0.000 & 0.096 & 0.041

	\enddata
\tablenotetext{a}{Age Parameters: SSP: $t\sbr{SSP}$; EXP: $\tau$; AQ: $t\sbr{AQ}$; FR: $t\sbr{FR}$; STR: $t\sbr{STR}$}
\tablenotetext{b}{Age Parameters: SSP: $\log(t\sbr{SSP})$; EXP: $\tau$; AQ: $\log(t\sbr{AQ})$; FR: $\log(t\sbr{FR})$; STR: $\log(t\sbr{STR})$}
\label{tab:params}
\end{deluxetable*}

\subsubsection{Fits to Scatter}
\label{sec:scatfit}
With at least three lines, we can iteratively calculate the best fit intrinsic scatter $S_{1,2,3}$ in the three model parameters $P_{1,2,3}$, where 
where $P_1$ represents the age-related parameter ($\log(t\sbr{SSP})$ for the SSP case) and $P_{2,3}$ are metallicity and \ale\ respectively:
\begin{equation}
S^{2}\sbr{pred}=\sum^{3}_{i=1}\left(\frac{\partial I }{\partial P_{i}}\right)^2 S_{i}^2
\label{eq:scatfit}
\end{equation}
where the derivatives $\frac{\partial I }{\partial P_{i}}$ are calculated numerically from the model grids. 

The above formula assumes that, at a given $\sigma$, the intrinsic scatter in age, for example, is independent of that of metallicity. However, we know that these parameters are not in fact independent. \cite{Wor94}, and later \cite{TraFabWor00} showed that there is a correlation between age and metallicity at a fixed mass. \cite{SmiLucHud08} found $\partial [Z/H]/ \partial \log(t\sbr{SSP}) = -0.68$ and $\partial [\alpha/\rm{Fe}] /\partial \log(t\sbr{SSP}) = 0.34$, using the same Shapley data as in this paper. 

To allow for correlated scatter, we modify equation (\ref{eq:scatfit}) as follows. We allow for a scatter in the age parameter, and a correlated scatter in metallicity and \ale. The sense of the correlation is that if a galaxy is older than the median by 0.1 dex, it is more metal-poor by --0.068 dex. We also allow for ``extra'' or residual uncorrelated scatter in the latter two parameters.  We can now predict the scatter in each index as a function of the predicted scatter in each (SSP) model parameter as
\begin{eqnarray}
S^2\sbr{pred} & = & \bigg(\frac{\partial I}{\partial t}+\displaystyle\sum_{i=2}^{3}\left(\frac{\partial I}{\partial P_i}\right)\left(\frac{\partial P_i}{\partial t}\right)\bigg|_{log(\sigma)}\bigg)S^2_{t}\nonumber \\
	& & +\displaystyle\sum_{i=2}^3\left(\frac{\partial I}{\partial P_i}\right)^2 S^2_{i, \rm{res}}
\label{eq:corrscatfit}
\end{eqnarray}
where $t$ represents the age-related parameter (e.g.\ $t = \log(t\sbr{SSP})$ for the SSP case) and $P_{2,3}$ are metallicity and \ale\ respectively. 

By examining the response of the indices in the other star formation history models, in comparison with the SSP,  it is possible to calculate the slope of the correlation between, for example, metallicity  and the ``age'' parameter $t$ (e.g.\ $\tau$ in the case of EXP models). These correlations for both metallicity and \ale\ for all models are tabulated in Table \ref{tab:parcor}.

\begin{deluxetable}{lcc}
\tablecaption{Parameter Correlations}
\tablewidth{0pt}
\tablehead{
\colhead{Model} & 
\colhead{$\frac{\partial [Z/H]}{\partial t\tablenotemark{b}}$} & 
\colhead{$\frac{\partial [\alpha/Fe]}{\partial t\tablenotemark{b}}$}}
\startdata
SSP     & -0.68\tablenotemark{a} & 0.34\tablenotemark{a}\\ 
EXP     & 0.172  & -0.086\\ 
AQ       & -0.367 & 0.183\\ 
FR      &  -0.523 & 0.262\\
STR     & -0.610 & 0.305
\enddata
\tablenotetext{a}{\cite{SmiLucHud08}}
\tablenotetext{b}{t $\equiv$ model age parameter. See table \ref{tab:params} note b.}
\label{tab:parcor}
\end{deluxetable}

\subsubsection{Distribution of Stellar Population Parameters as a Function of Velocity Dispersion }

\begin{figure*}[htbp]
	\centering
	\mbox{\subfigure{\includegraphics[width=.33\linewidth]{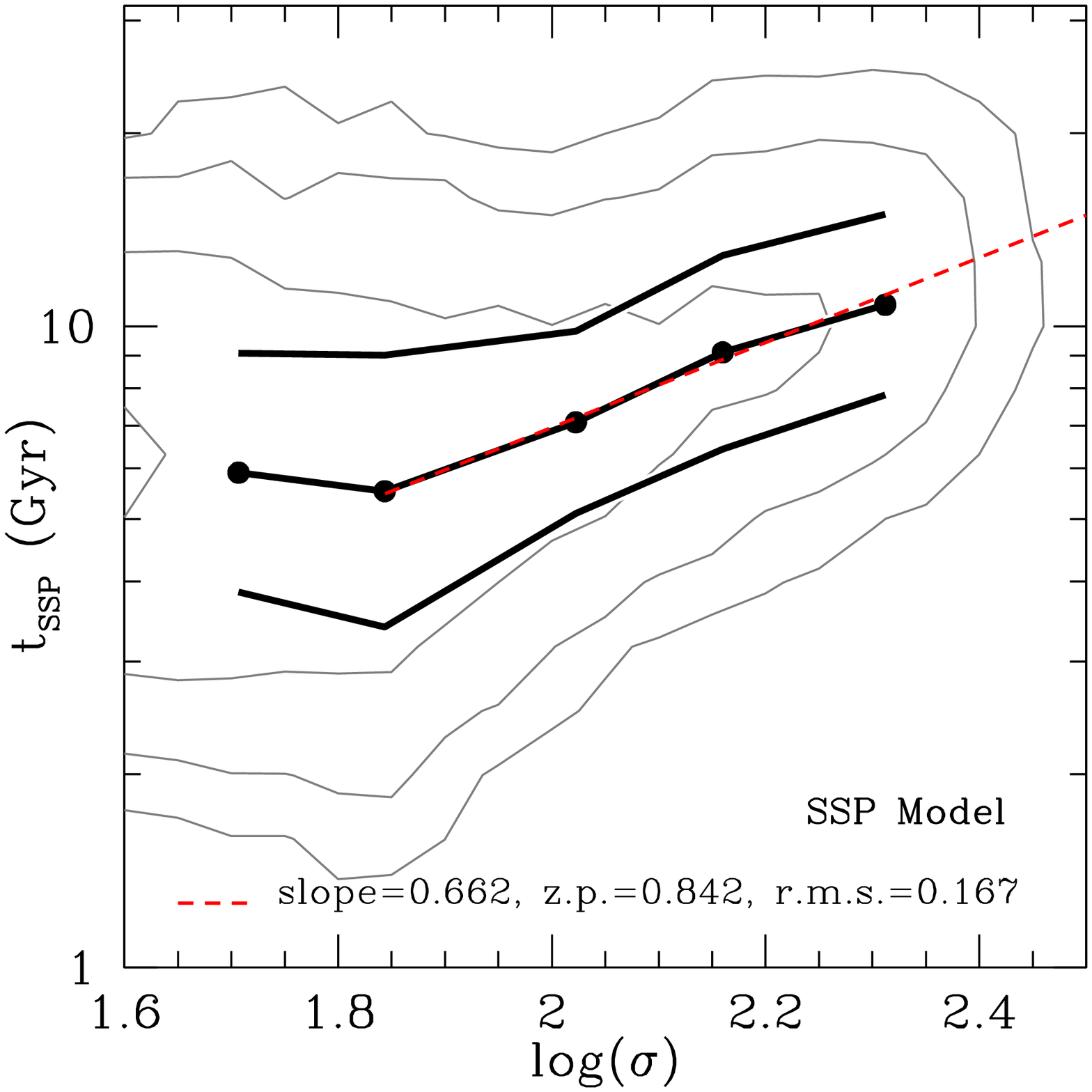}}
				\subfigure{\includegraphics[width=.33\linewidth]{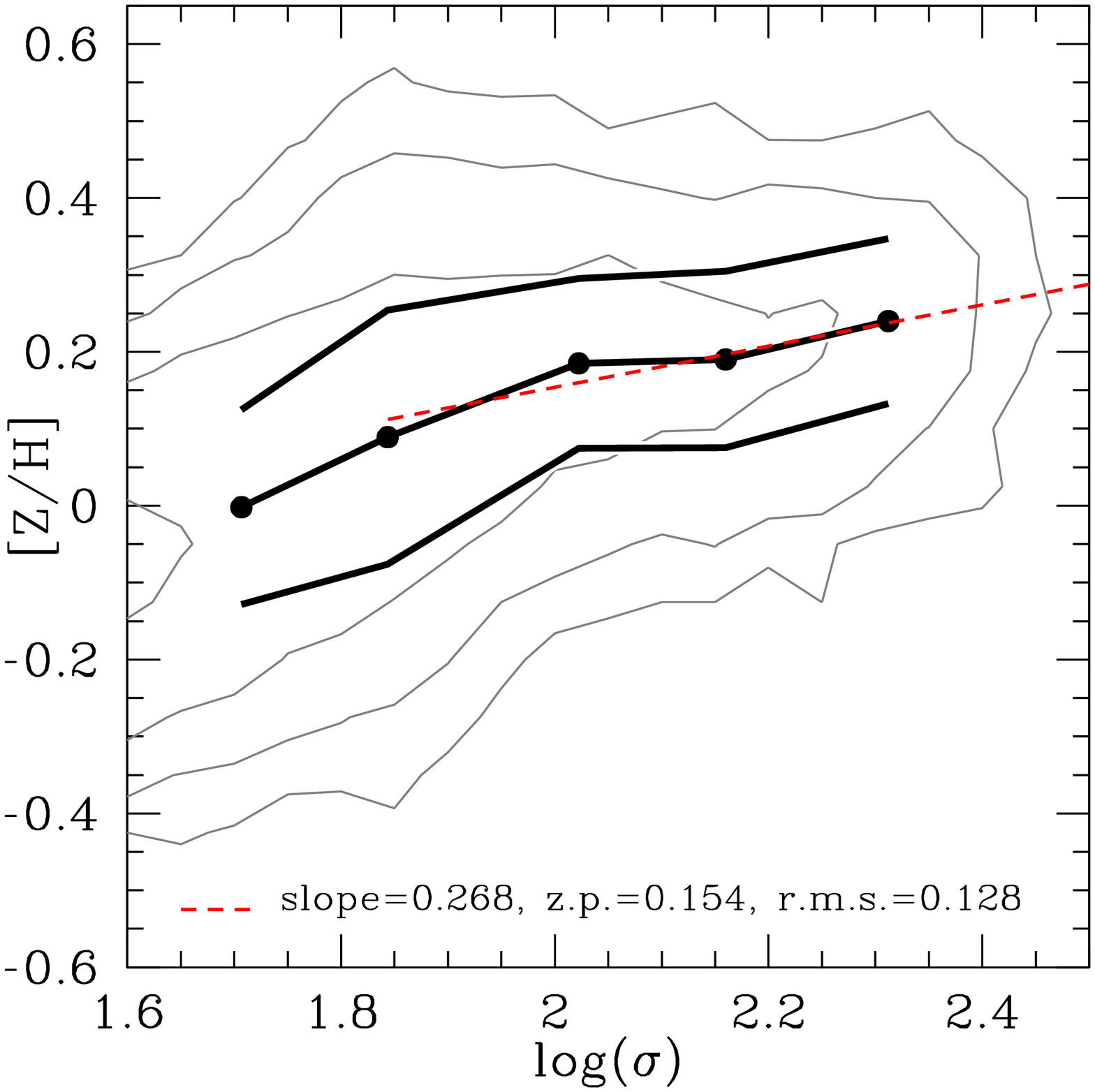}}
				\subfigure{\includegraphics[width=.33\linewidth]{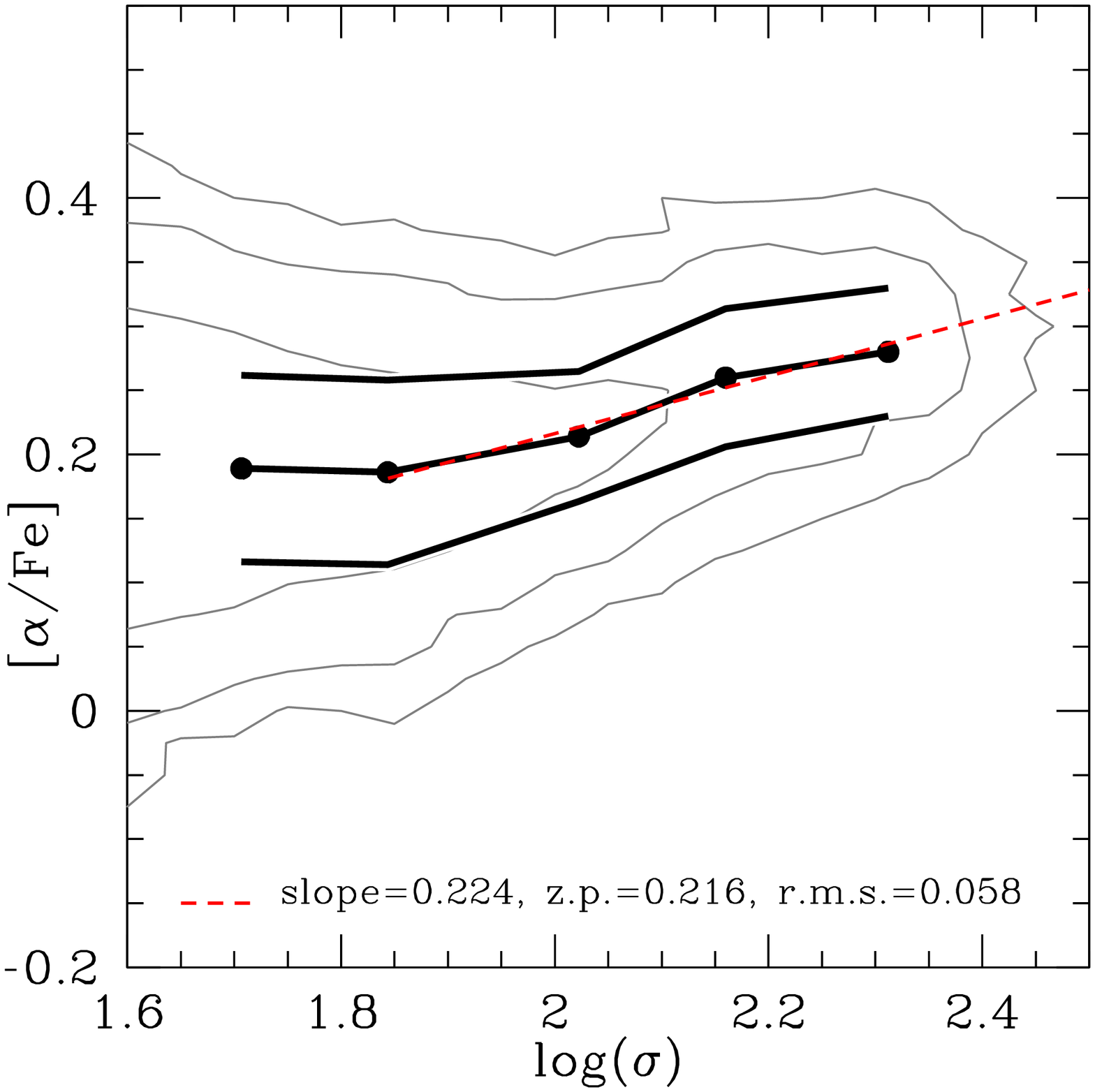}}}
	\mbox{\subfigure{\includegraphics[width=.33\linewidth]{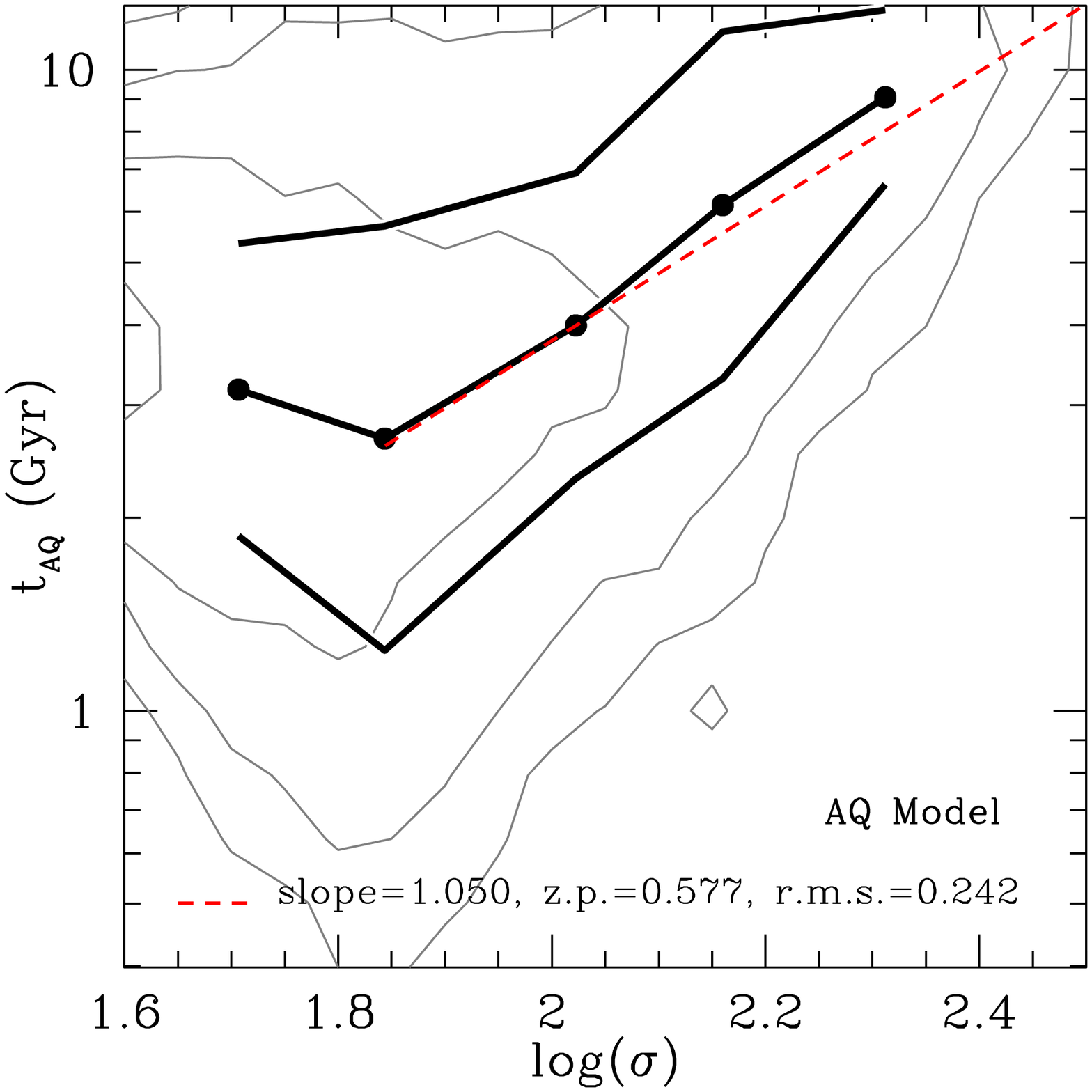}}
				\subfigure{\includegraphics[width=.33\linewidth]{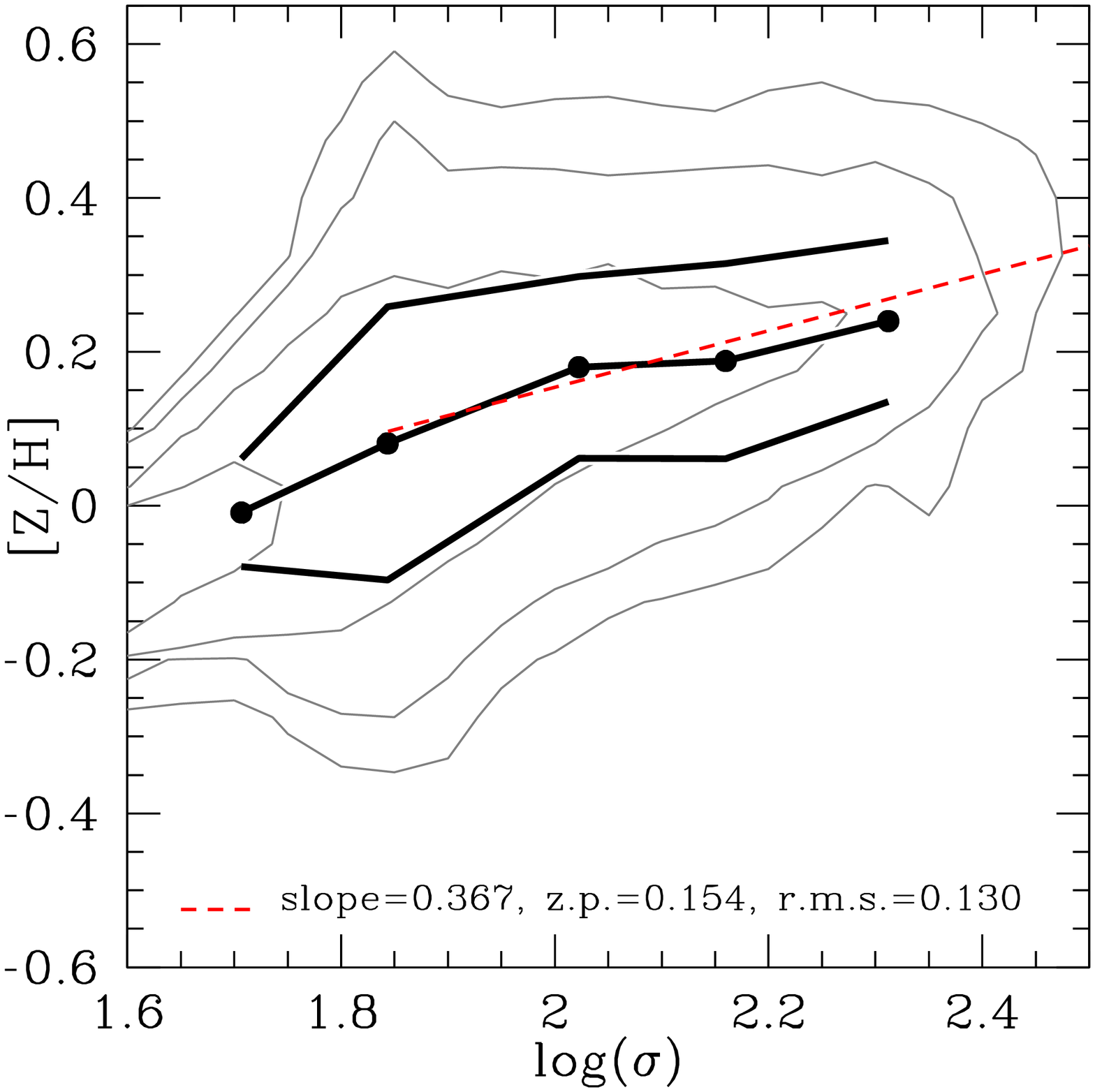}}
				\subfigure{\includegraphics[width=.33\linewidth]{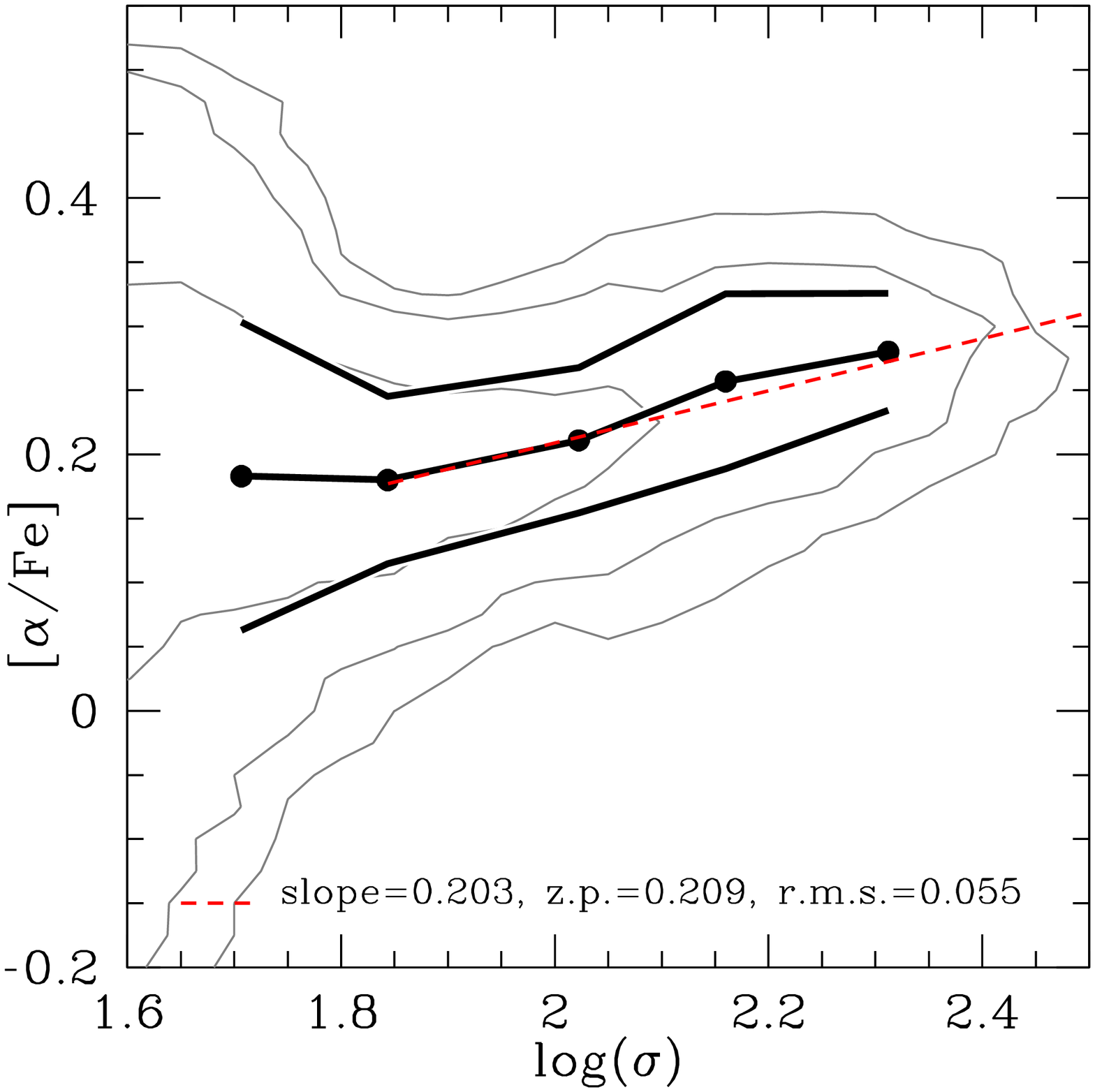}}}
	\mbox{\subfigure{\includegraphics[width=.33\linewidth]{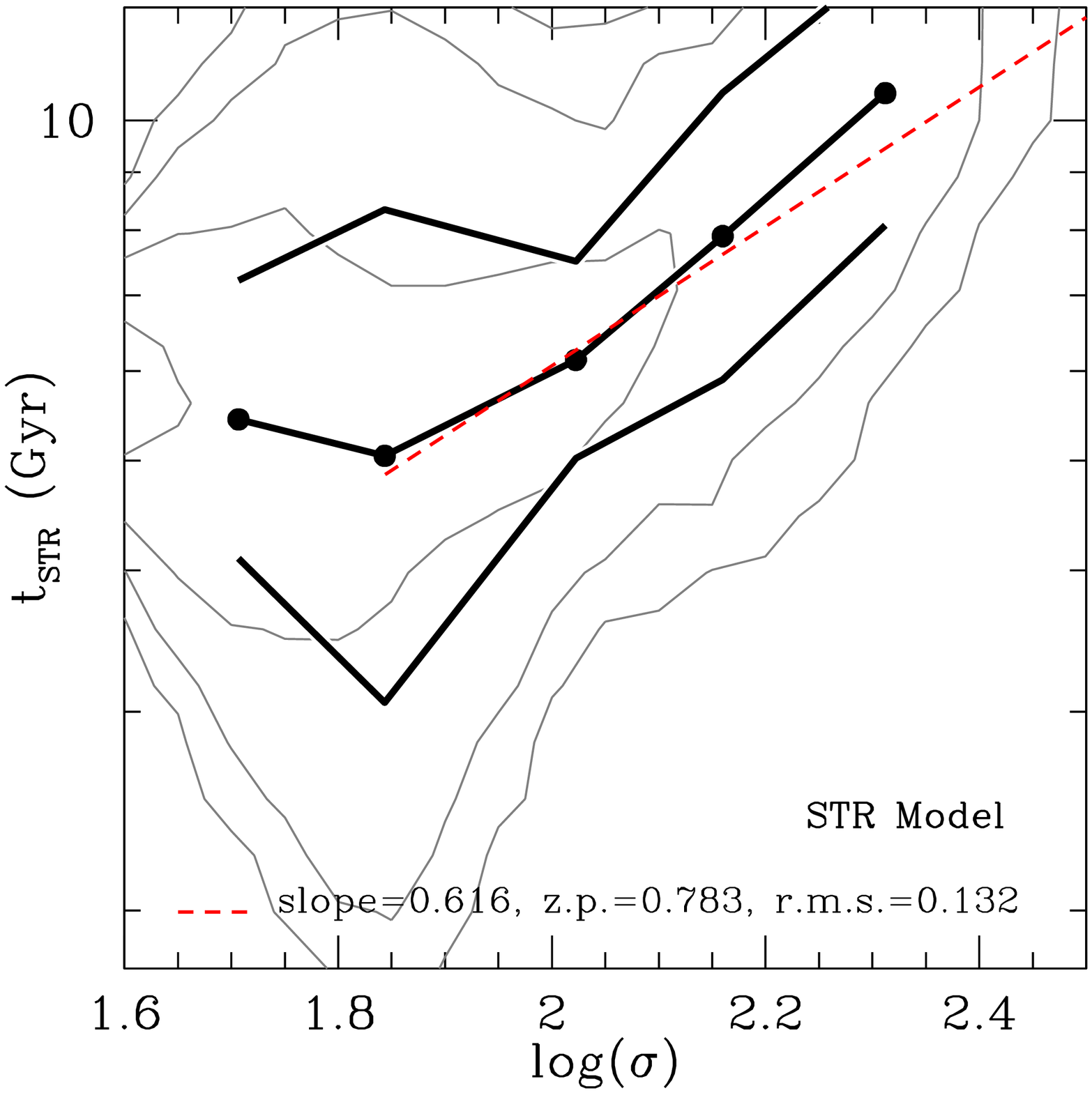}}
				\subfigure{\includegraphics[width=.33\linewidth]{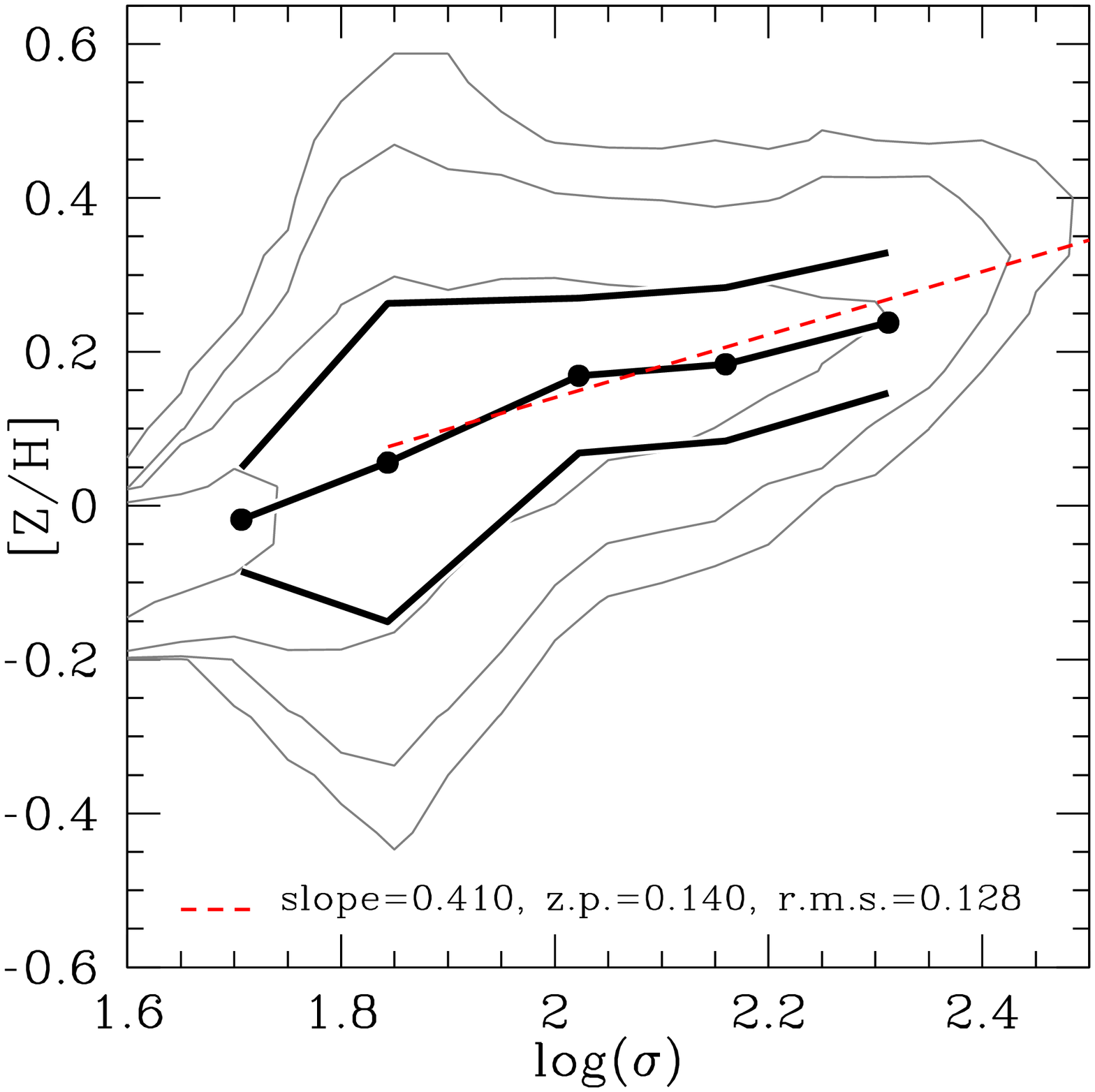}}
				\subfigure{\includegraphics[width=.33\linewidth]{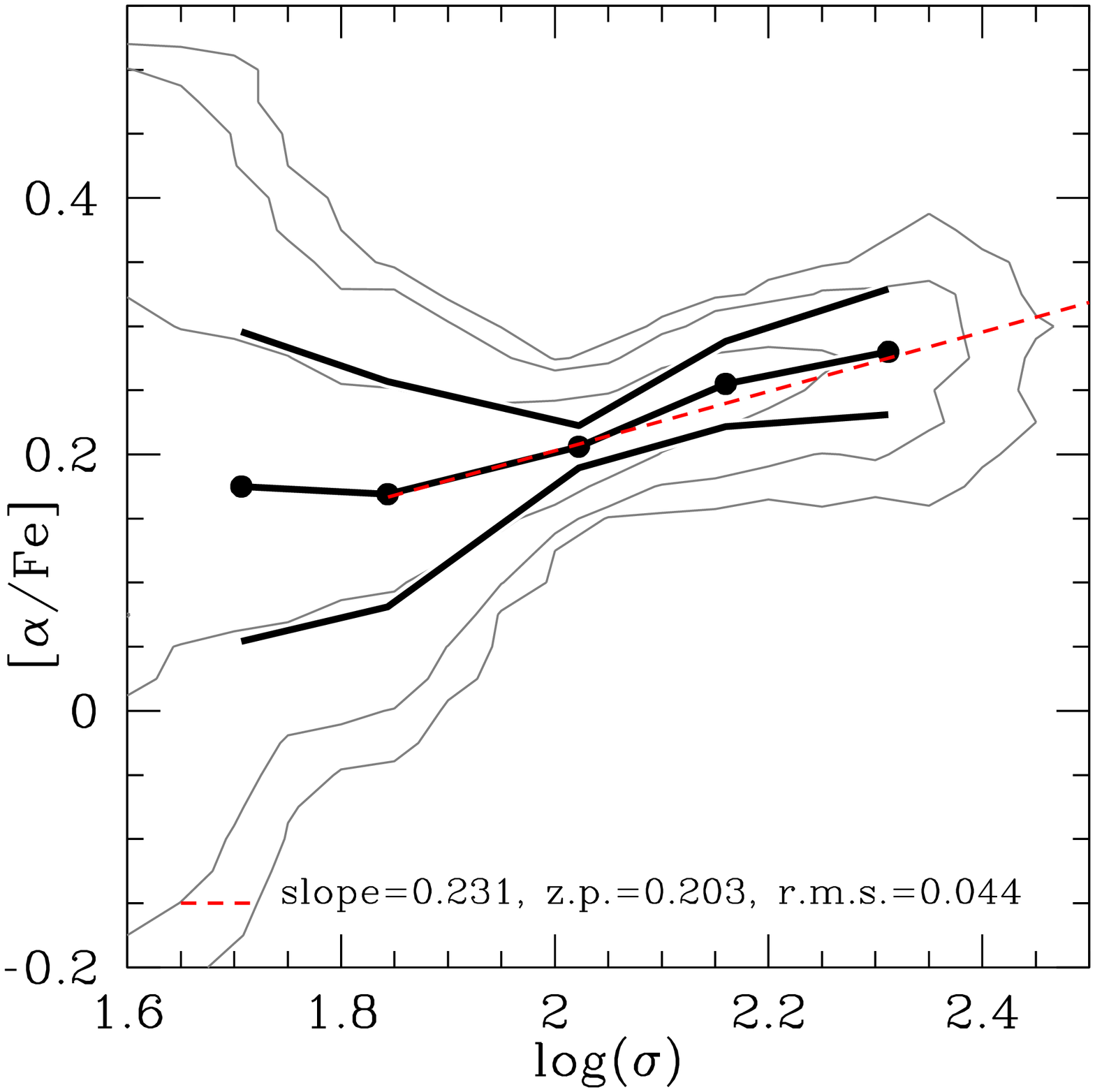}}}	

	\caption[Best Fit Model Parameters]{Best-fit model parameters as a function of log($\sigma$). The filled points show the median values of the parameters and the solid black lines show the $\pm1\sigma$ range. Fitted slopes (shown by red dotted lines) are to data with $\log(\sigma) > 1.844$ only. Rows from top to bottom correspond to SSP, AQ, and STR.  Age parameters are as in Table \ref{tab:params}. Note increasing metallicity, \ale\ and age with increasing velocity dispersion. Synthetic cluster density contours are shown in grey, at 0.7dex increments (see Section \ref{sec:synth}).}
	\label{fig:params1}
\end{figure*}

\begin{figure*}[htbp]
	\centering
	\mbox{\subfigure{\includegraphics[width=.33\linewidth]{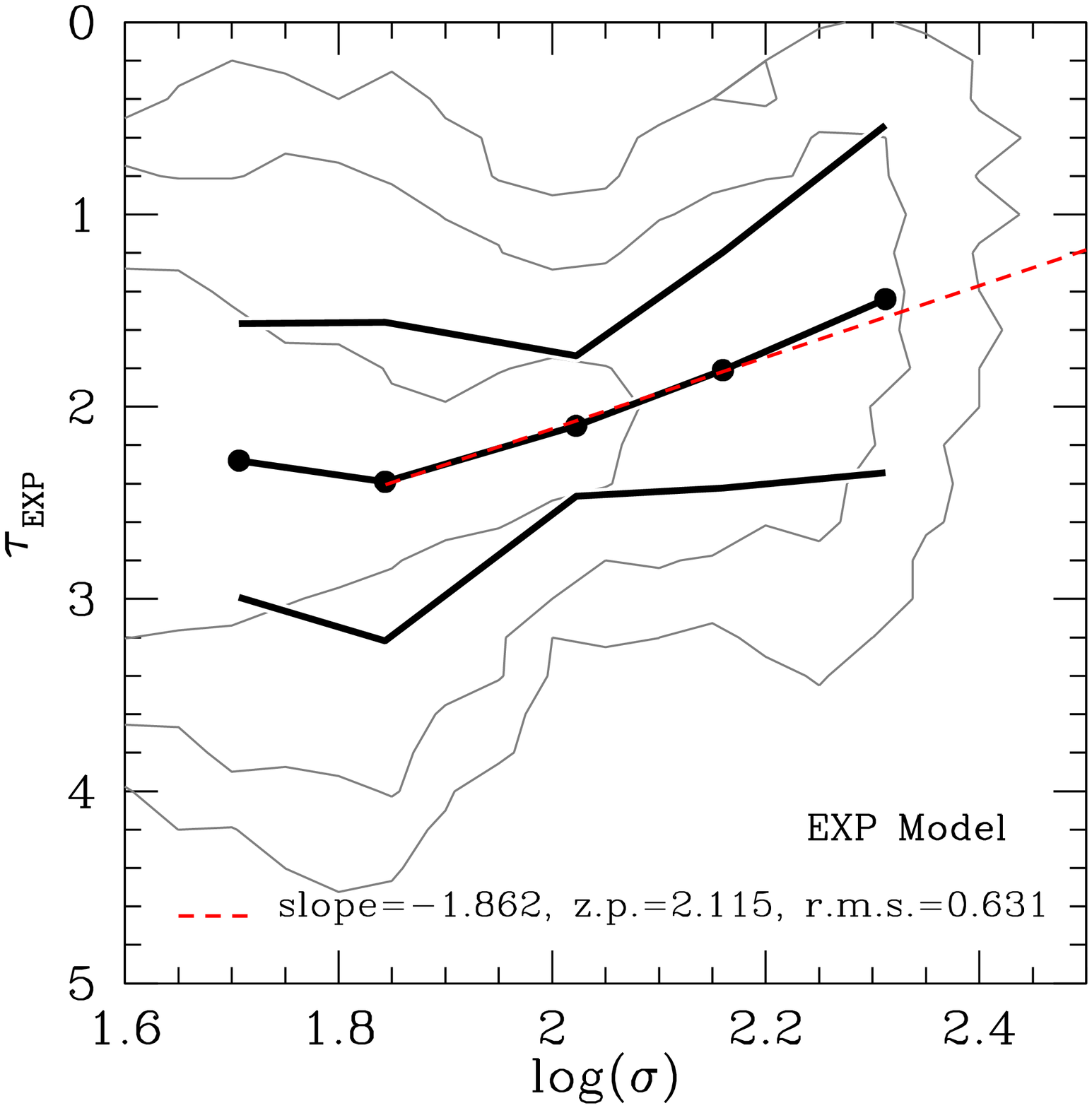}}
				\subfigure{\includegraphics[width=.33\linewidth]{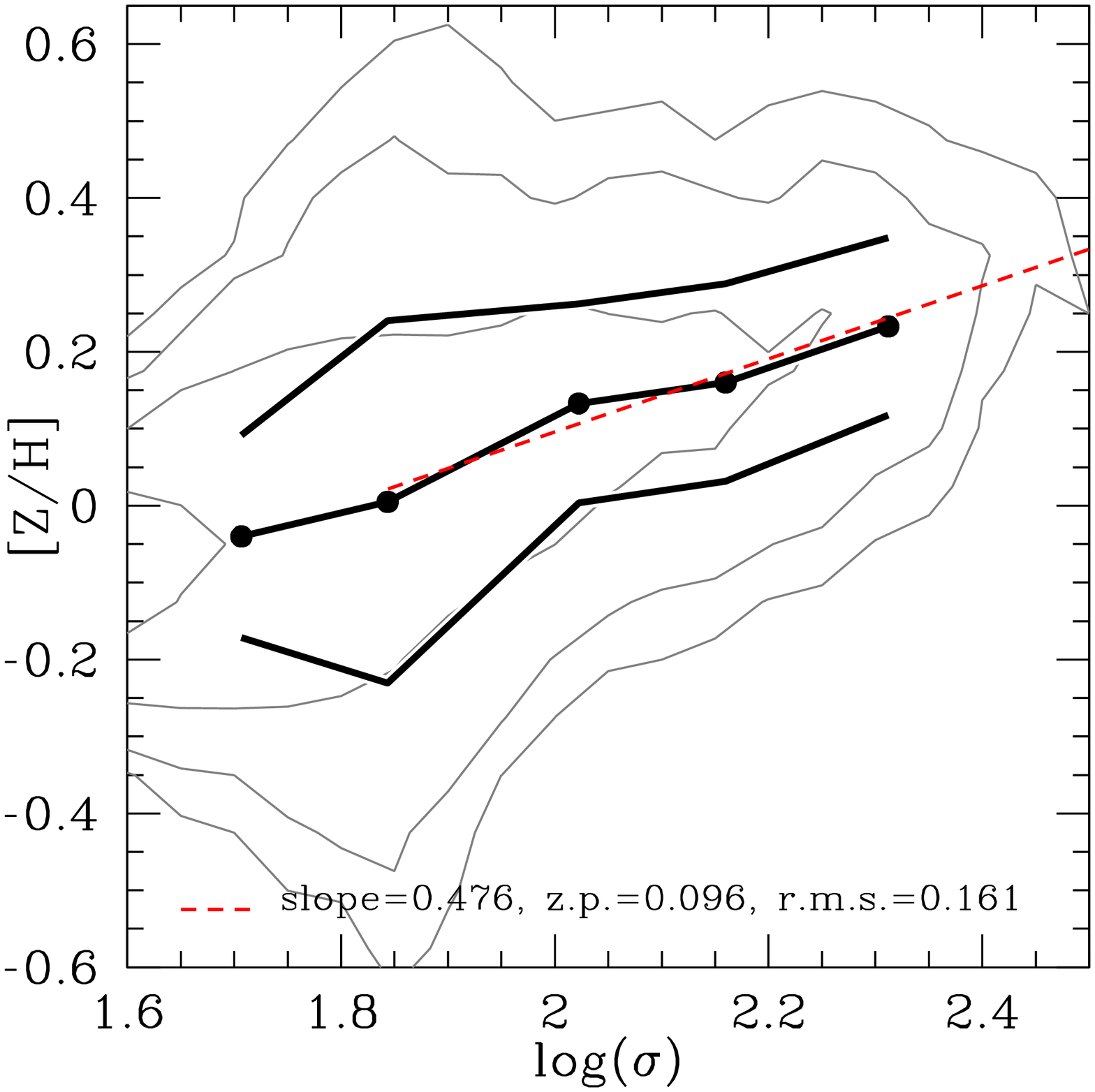}}
				\subfigure{\includegraphics[width=.33\linewidth]{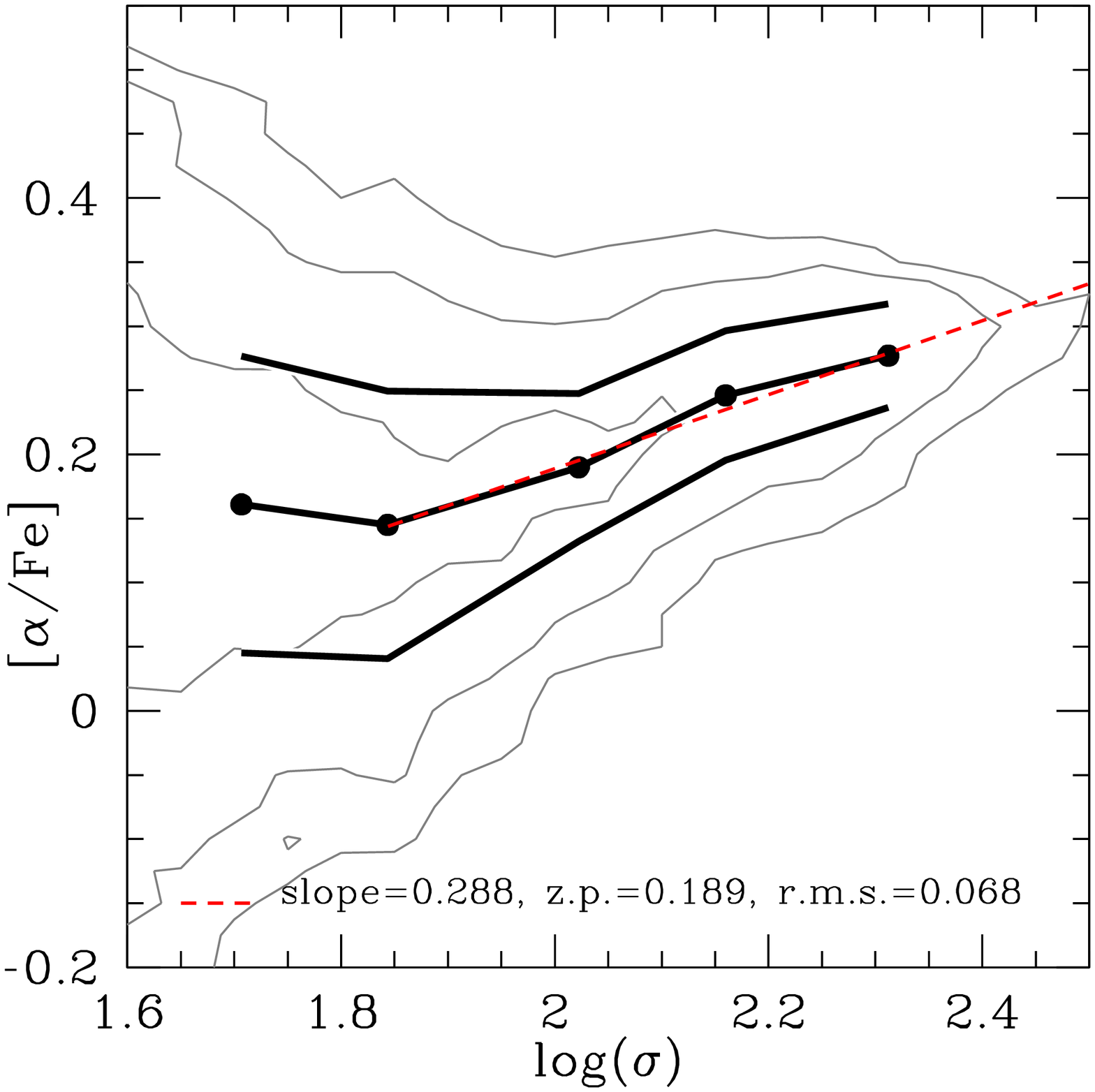}}}	
	\mbox{\subfigure{\includegraphics[width=.33\linewidth]{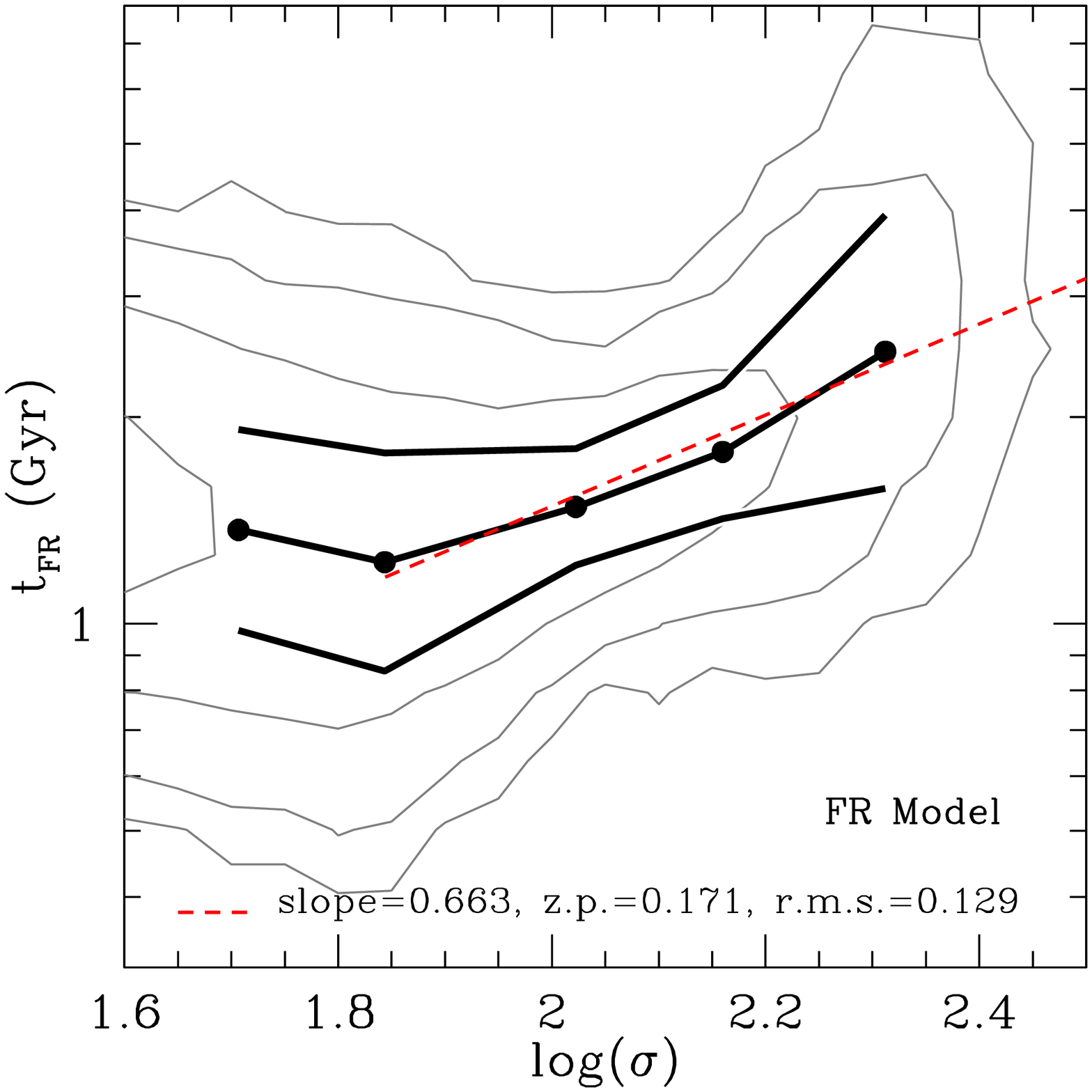}}
				\subfigure{\includegraphics[width=.33\linewidth]{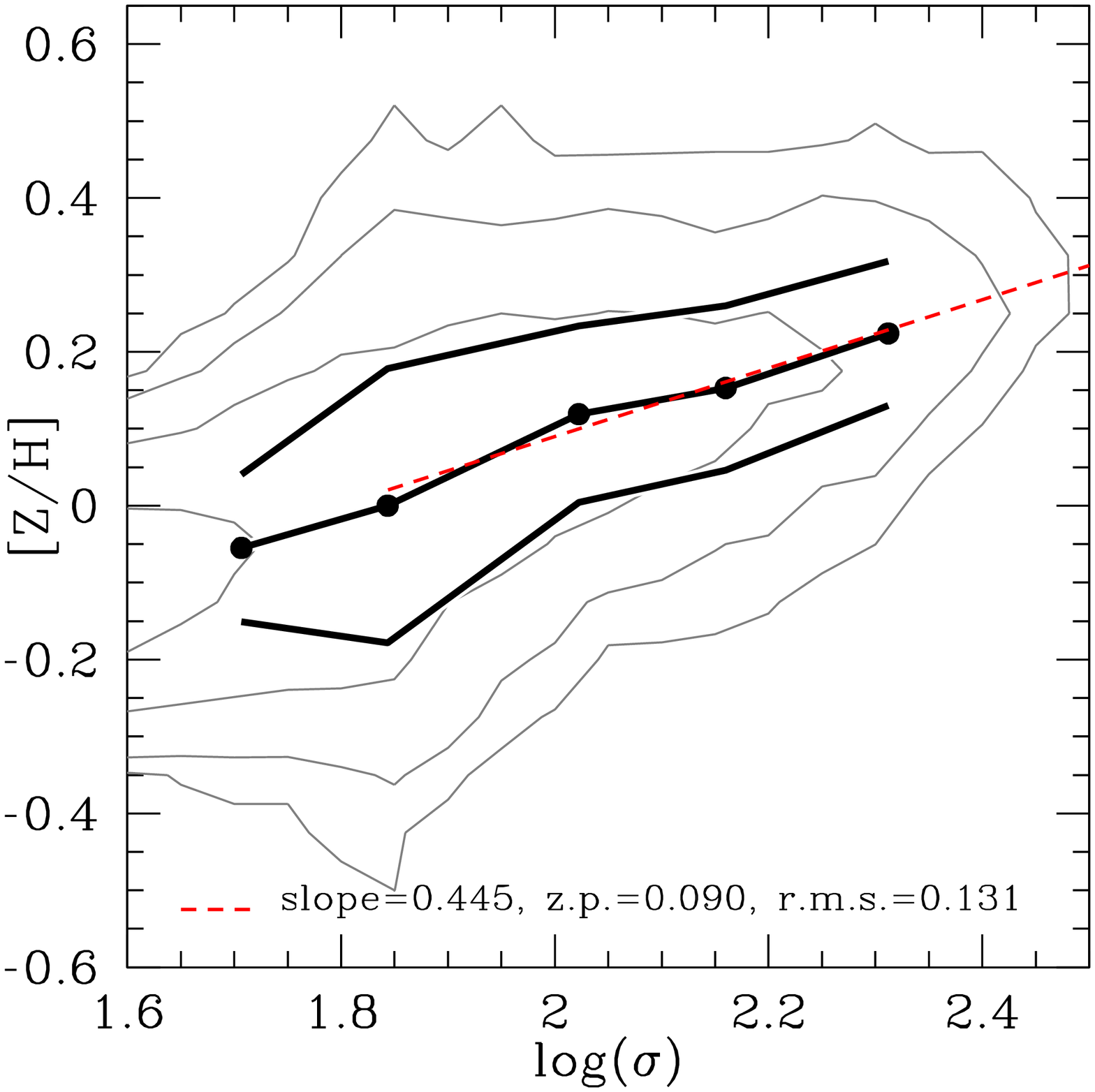}}
				\subfigure{\includegraphics[width=.33\linewidth]{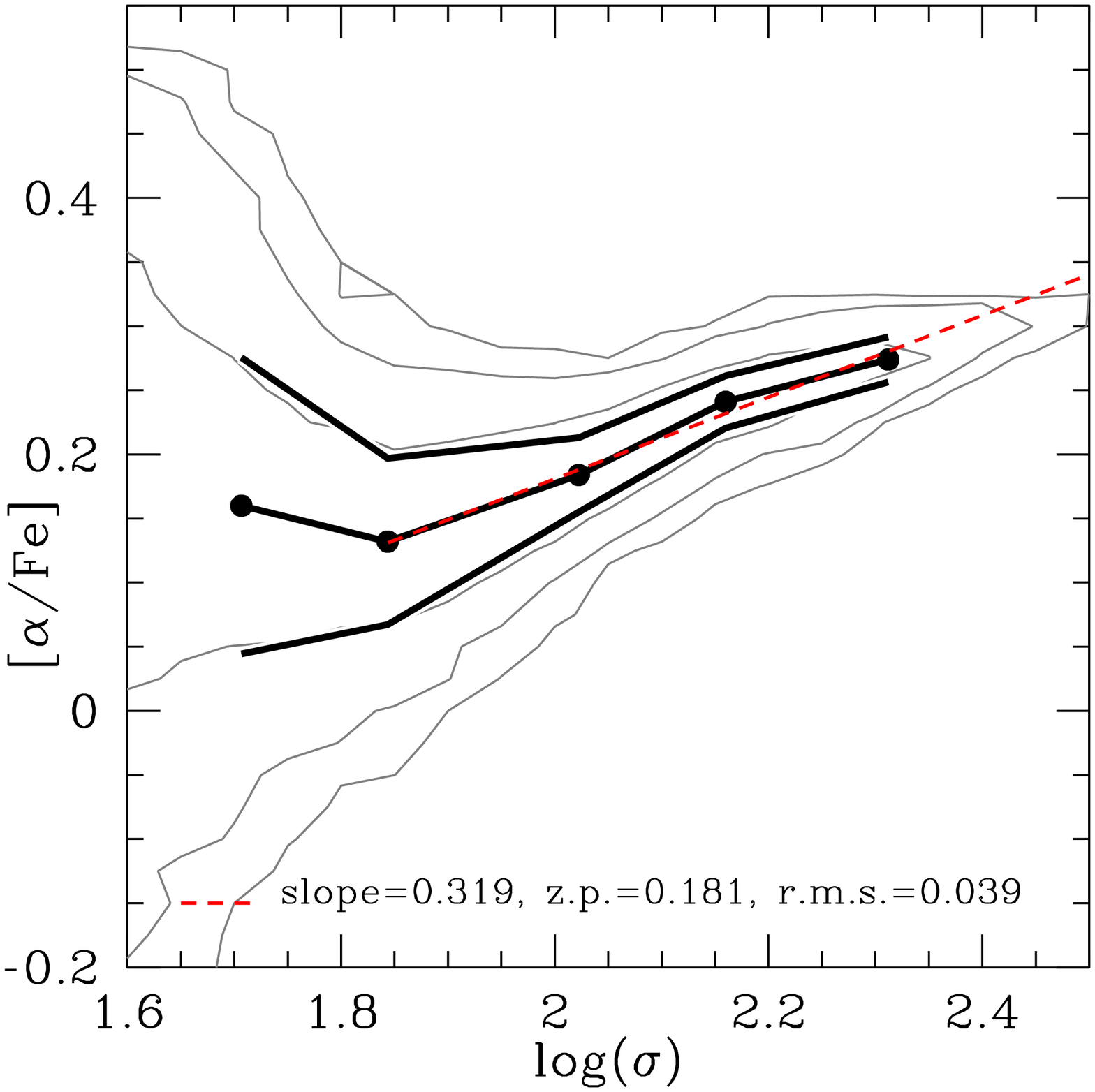}}}
			
	\mbox{\subfigure{\includegraphics[width=.33\linewidth]{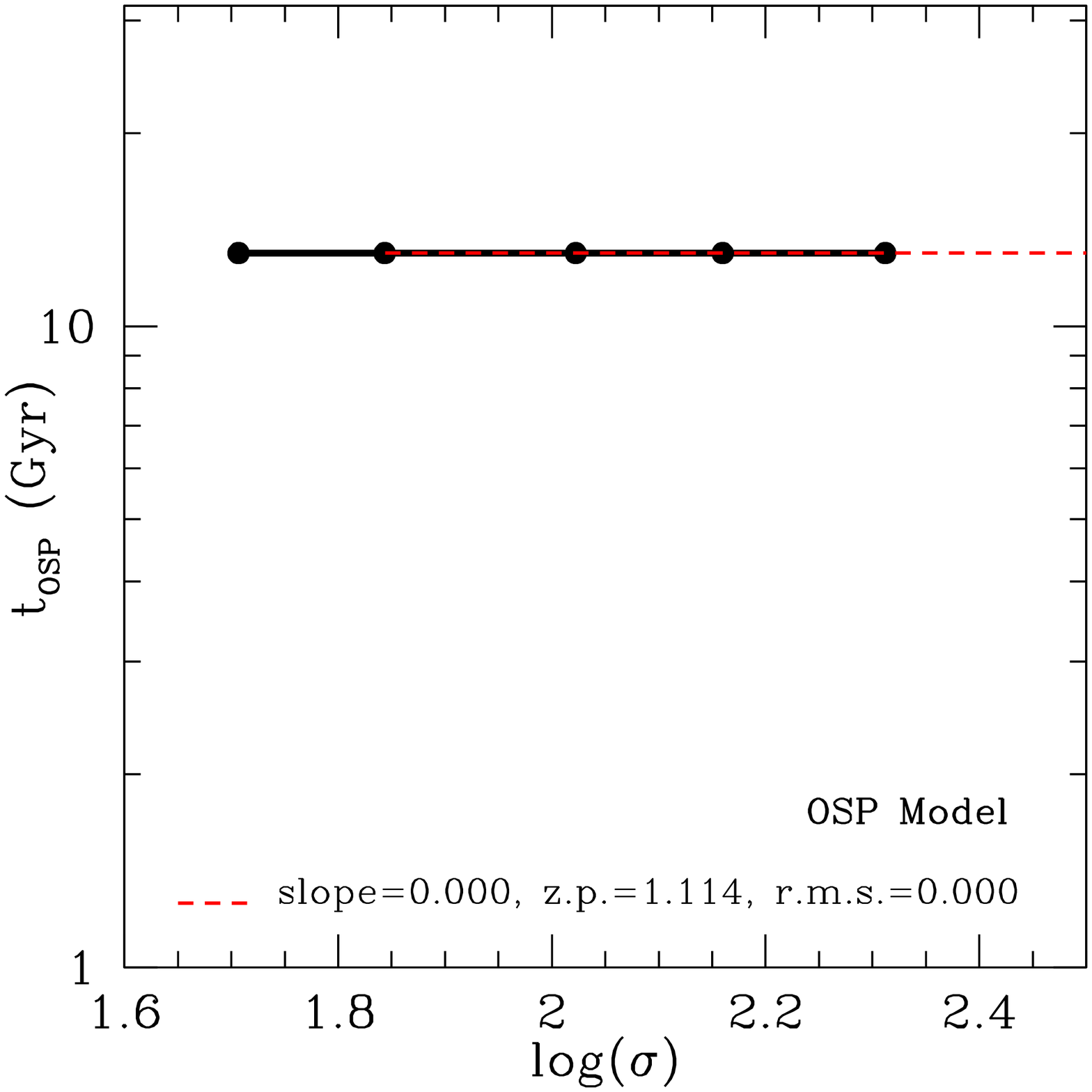}}
				\subfigure{\includegraphics[width=.33\linewidth]{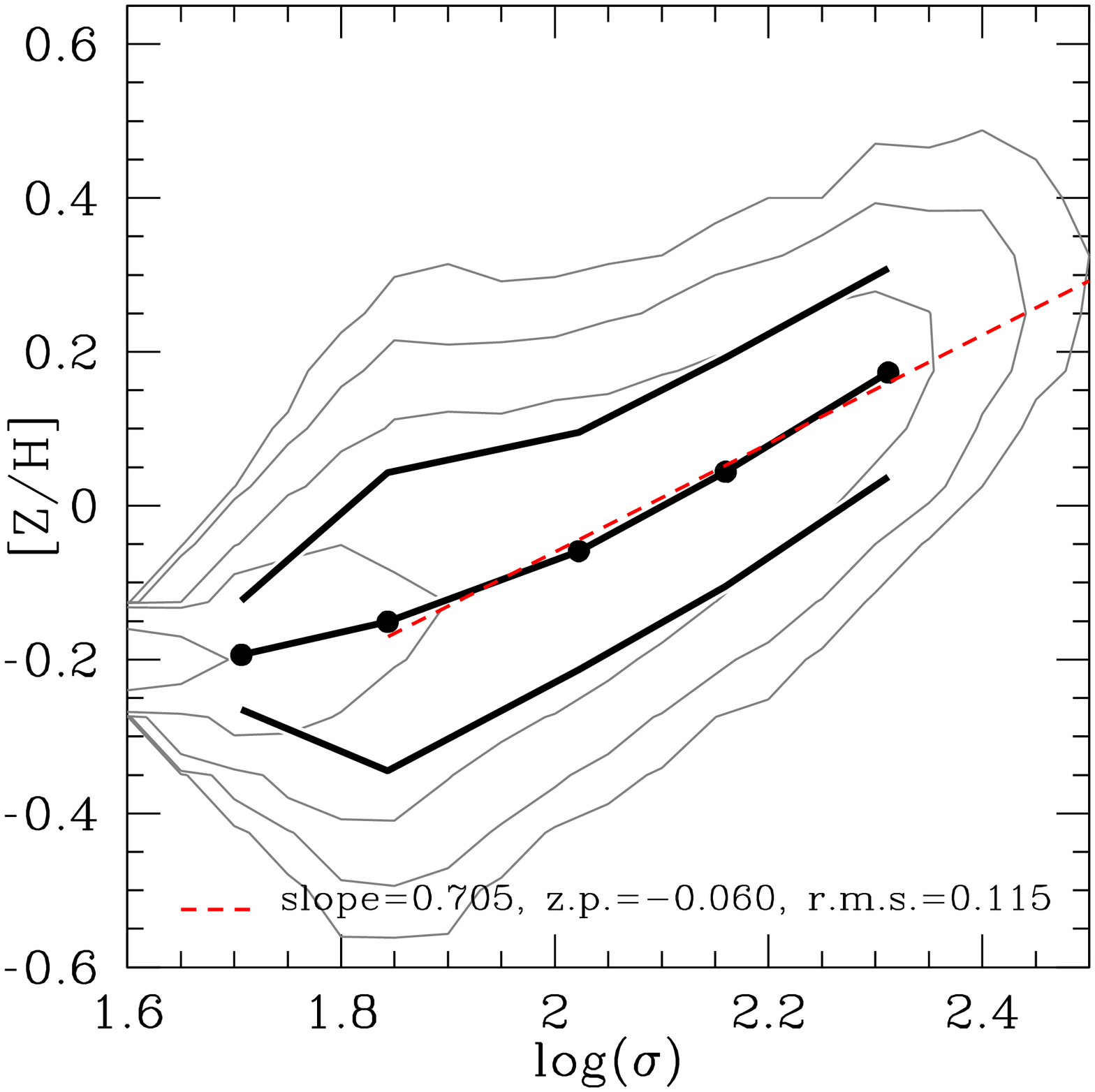}}
				\subfigure{\includegraphics[width=.33\linewidth]{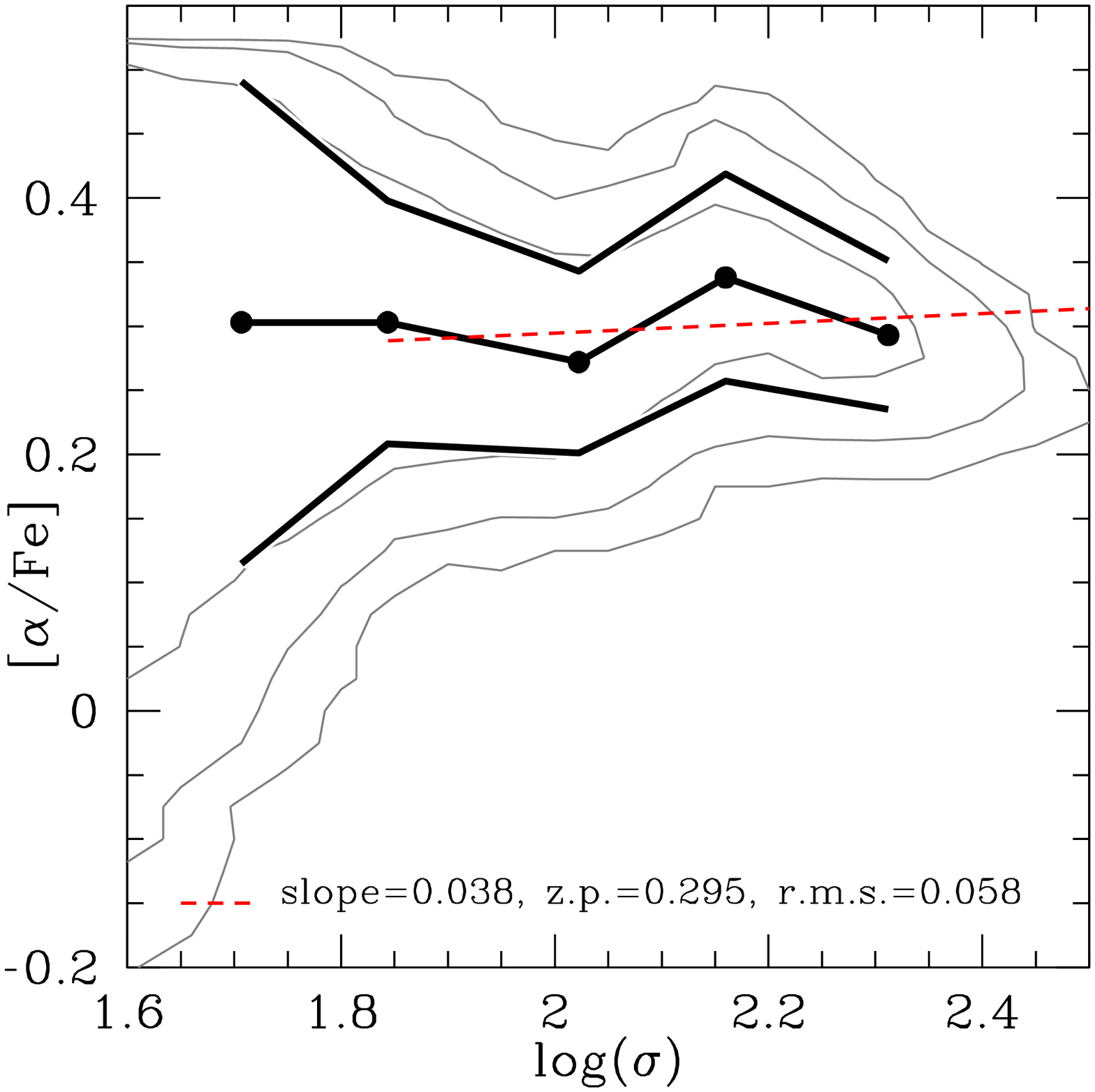}}}
				
	\caption[Best Fit Model Parameters]{Best-fit model parameter trends with log($\sigma$). Rows from top to bottom correspond to EXP, FR, and OSP. Lines, contours and symbols are as in Figure \ref{fig:params1}. Age parameters are as in Table \ref{tab:params}. Note increasing metallicity, \ale\ and age with increasing velocity dispersion.}
	\label{fig:params2}
\end{figure*}

Having outlined how the line index scatters are predicted given correlated scatter in the model parameters, we proceed to fit these scatters to the observed values. Table \ref{tab:params} presents the model parameters yielding the best fits to the line index data. Also tabulated are the fit scatters in each parameter for each bin and model. Figures \ref{fig:params1} and \ref{fig:params2} illustrate the parameter versus velocity dispersion relations, and their associated scatter.
We also show the linear fits to the parameters, ignoring the lowest $\sigma$ bin, i.e. using data with $\sigma > 70$ \kms\ ($\log(\sigma) > 1.844$); the quoted zero points are calculated at $\log(\sigma)=2.00$.

\subsection{Systematics}

\subsubsection{Aperture Effects}

Our data are observed through a 1'' diameter fiber, corresponding to 0.95 kpc radius at the distance of Shapley. Thus the stellar populations found above correspond to the stars within that aperture. Nevertheless, for some purposes, it is interesting to consider how this might be extrapolated to refer to all stars within the effective radius, $R_e$.  As our data are all at a common distance we cannot solve for population gradients internally. Internal radial gradients in early-type galaxies have been studied by a number of authors, and are summarized in Table 4 of \cite{RawSmiLuc08b}. They find that gradients in age and \ale\ are small, but the gradient in $Z$ is $\Delta Z/\Delta \log r\sbr{ap} = -0.18 \pm 0.02$. We can correct our data to $R_e$ using this result, together with eq. \ref{eq:Resig}. We find 
\begin{eqnarray}
[Z/H](R_e) - [Z/H](ap) = \ \ \ \ \ \ \ \ \ \ \ \ \ \ \ \ \ \ \ \ \ \ \ \ \ \nonumber \\ 
\begin{cases}
 - 0.043 & \text{$\log(\sigma) < 2.23$,}\\
 - 0.50 (\log \sigma / 2.23) - 0.043 & \text{$\log(\sigma) > 2.23$.}
\end{cases}
\end{eqnarray}
Thus the correction is small: $-0.08$ for the highest velocity dispersion bin, and $\sim -0.04$ for the other bins.  
 
\subsubsection{Line Indices: Choice of Index and Effect of Index Calibration Corrections}
\label{sec:linechoice}
\begin{figure}[htpb]
	\centering
	\includegraphics[width=\linewidth]{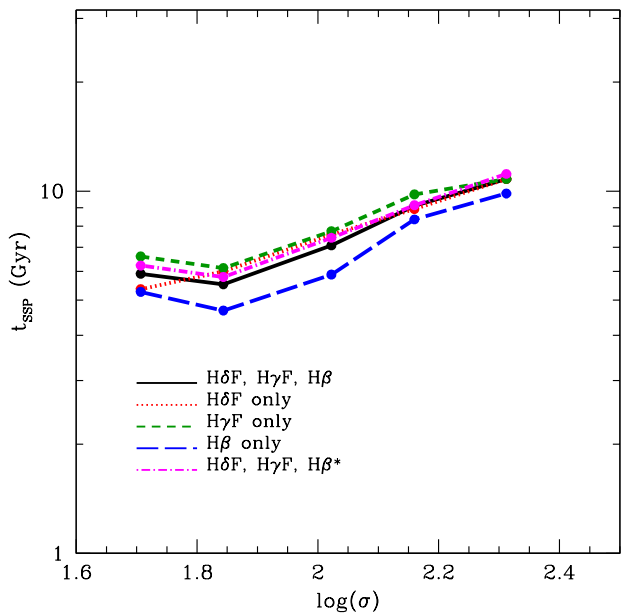}
	\caption{SSP age as a function of $\log(\sigma)$ for different choices of Balmer line used in the fits. (The Fe and Mg lines are used in all cases.) In the H$\beta$* case, we have fit H$\beta$ using a calibration fit to all velocity dispersion bins (see Section \ref{sec:linechoice} for details). In all cases, the choice of Balmer index makes little difference at the youngest ages.}
	\label{fig:balmer}
\end{figure}

It is interesting to see whether the conclusions, particularly those for age, change significantly depending on which lines are used in the analysis. Here we focus on the Balmer lines, reproducing the fits with only one Balmer line rather than all three simultaneously.  The results of these tests are shown in Fig. \ref{fig:balmer}.  We find that the median age in the second-lowest bin ($\log \sigma = 1.844$) changes by less than 1 Gyr (15\%) depending on which Balmer line is used. In H$\gamma$ and H$\beta$ we see a flattening or an upturn in age at the lowest velocity dispersion bin (as in Fig. \ref{fig:balmer}), but if we only use H$\delta$, the lowest bin is younger. 

One concern regarding the Balmer line indices is the possibility of weak infilling by AGN emission at high mass. \cite{SmiLucHud07} Figure 7 shows the fraction of RSGs with AGN-like emission in the Shapley sample. At $M_R<-20.5$, and thus high mass and velocity dispersion, there is still significant AGN contribution on the red-sequence. While galaxies with clear emission have been removed from our sample, it is possible that weak AGN emission may cause infilling in $H\beta$, particularly for the highest velocity dispersion galaxies. If so, it is possible we have overcorrected $H\beta$ by the procedure adopted in Section \ref{sec:fit}. An alternative calibration would have been to fit a zero-point iteratively to the average of all bins rather than to the highest velocity dispersion bin alone. In fact, if we fit the observed $H\beta$ indices to all velocity dispersion bins, then all models agree better with the predicted values of $H\beta$. The weak effect of this alternative correction on the derived ages is shown in Figure \ref{fig:balmer}. 

A related concern is the effect on our conclusions of the choice to calibrate all lines to \cite{NelSmiHud05}, which itself was not calibrated to the Lick system. To test the effect of alternative calibration schemes, we have also compared our Shapley linestrength-$\sigma$ relations to those from 12 Coma early-type galaxies observed by \cite{TraFabDre08}, which we expect to be well-calibrated to the Lick system. We then fit additional offsets to our Nelan-calibrated Shapley linestrength-$\sigma$ relations to match the \cite{TraFabDre08} linestrengths as a function of $\sigma$. In particular, the Balmer line offsets were estimated as {H$\delta$F,H$\gamma$F,H$\beta$} $\sim{}$ {0.12,0.2,0.1} (in addition to the correction to \cite{NelSmiHud05} tabulated in Table \ref{tab:corr}). We then refit our models  with this alternative calibration scheme, and find that it yields ages that are only slightly younger (by $<0.5$ Gyr  or 10\% for the youngest galaxies). This change in stellar population parameters is smaller than other systematics, such as the effect of using only $H\beta$ as discussed in the previous paragraph.

\subsubsection{Systematics Associated with Model Parameters}
\label{sec:sysparam}

In Section \ref{sec:csp}, in certain models, star formation was assumed to start 13 Gyr ago, which corresponds to a rather high redshift. As an alternative, we will consider a different starting time and investigate its effect on the fits. For this purpose, we choose a redshift of 2, corresponding to approximately 10.3 Gyr, as a more realistic choice, as this is close to the peak star formation in the universe \citep{HopBea06} . When we refit our star formation history models, we find that the CSFH age parameters shift towards older ages to compensate for the absence of very old stars. For example, for the EXP model, for the $\expec{\sigma} = 70$ \kms\ bin, $\tau$ is reduced, from its default value of 2.4 Gyr, to only 1.7 Gyr.  In the AQ model, the lookback time to quenching increased from 2.7 Gyr to 3.3 Gyr.
Other velocity dispersion bins are affected similarly. We also find that metallicity increases by approximately $7\%$ in the ``mostly old'' models, whereas the intermediate age models show no significant change in metallicity.  Finally $\alpha$-enhancements remain virtually unchanged from the 13 Gyr case.  We will discuss the effects of these changes on stellar mass-to-light ratios in Section \ref{sec:sysfx}.

\subsubsection{Effect of Other Population Synthesis Models}

\cite{SmiLucHud09} found a generally good agreement between our TMBK-based SSP models and the EZ-Ages model of \cite{Sch07} for age, metallicity and \ale. However, they noted that there was a small [Fe/H]-dependent bias in the derived ages in the sense that the EZ-Ages ages were younger by $\sim 0.25$ dex per dex of [Fe/H]. 
Given the small range in $\langle$[Fe/H]$\rangle$ covered by our sample (only $0.05$ dex for $\sigma > 70 \kms$), the effect of the choice of models on the Age-$\sigma$ relation and hence on colors and mass-to-light ratios is negligible.

\subsection{Comparison with previous work}

For all models, the general stellar population trends observed here for $\sigma > 70$ \kms, are that age, metallicity and \ale\ all increase with velocity dispersion, are in agreement with a number of previous results \citep{CalRosCon03,HeaPanJim04,NelSmiHud05,ThoMarBen05,GalChaBri06, SmiLucHud07}. 

\begin{deluxetable*}{l|c|ccc}
\tablecaption{SSP Scaling Relations}
\tablewidth{0pt}
\tablehead{
\colhead{Reference} &
\colhead{$\sigma$ range} &
\colhead{Age} & 
\colhead{Z/H} & 
\colhead{$\alpha$/Fe}%
}
\startdata
\cite{NelSmiHud05} & $50\kms<\sigma<400\kms$ & $0.59\pm0.13$ & $0.53\pm0.08$ & $0.31\pm0.06$\\ 
\cite{SmiLucHud07} & $30\kms<\sigma<300\kms$ & $0.52\pm0.06$ & $0.34\pm0.04$ & $0.23\pm0.04$\\ 
\cite{SmiLucHud07} & $100\kms<\sigma<300\kms$ & $0.64\pm0.12$ & $0.38\pm0.09$ & $0.36\pm0.07$\\ 
This work & $70\kms<\sigma<300\kms$ & 0.66 & 0.27 & 0.22
\enddata
\label{tab:ssprel}
\end{deluxetable*}

The scaling relations we find for $\sigma > 70$ \kms are tabulated in Table \ref{tab:ssprel}. Note that these are based on the fits to synthetic clusters (discussed below in Section \ref{sec:synth}) and are weighted to low velocity dispersions. We also show results from \cite{SmiLucHud07}, using the same Shapley data (and indices) for all galaxies and for $\sigma > 100$ \kms. In general our scaling relations are between the ``all $\sigma$'' and "$\sigma > 100$ \kms"\ relations found by \cite{SmiLucHud07}, although our scaling relation for $[Z/H]$ is somewhat flatter. In comparison with \cite{NelSmiHud05}, the scaling of age with velocity dispersion is similar, however this analysis yields shallower relationships with $[Z/H]$ and $[\alpha/{\rm Fe}]$. Note that our determination of \ale\ is based on Mgb and so primarily reflects the [Mg/Fe] abundance ratio, whereas the \ale\ of \cite{NelSmiHud05} was also based on CN. A full table of comparisons is given in \cite{SmiLucHud07}.

The lowest-velocity dispersion bin ($\sigma < 70 \kms$) shows a either a flattening or an increase in the mean age of galaxies, in comparison to $\sigma > 70 \kms$.  This is as a result of the downturn or flattening of the Balmer-line indices seen in Fig \ref{fig:6fitlines}. We note that this change in the ages of low velocity dispersion galaxies can be seen in the fits of \cite{SmiLucHud07}, in the sense that their scaling relations are flatter when all galaxies are included and become steeper when only high velocity dispersion galaxies ($\sigma > 100$) \kms\ are used in the fit. Furthermore, this trend is also seen in central Coma cluster galaxies at low velocity dispersions \citep{SmiLucHud09}. Consequently, we argue that this effect is real, and signals the end of ``downsizing'' on the cluster red-sequence. We discuss selection effects that may affect the derived stellar population parameters in Section \ref{sec:lowvd}, but conclude there that correction for such effects would strengthen our conclusion that this ``upsizing'' effect is real.

In Section \ref{sec:lowvd}, we also show  that the ``pinching'' of the scatter $S$ in the lowest velocity dispersion bin is a selection effect. In reality, the scatter in age increases at decreasing velocity dispersion. This increased scatter may also help to reconcile previously contradictory results in the literature. For example, \cite{SanGorCar06b} find no ``downsizing'' effect in the Coma cluster but do find downsizing in Virgo.  In both clusters they have a number of objects with high-$\sigma$ plus a small number (4-5) with $\log(\sigma) < 1.844$.   Given the large intrinsic scatter in age that we find at low-$\sigma$ ($>$0.2 dex or $\sim 3$ Gyr), we argue that a sample of 4-5 galaxies is too small to secure a robust detection of the mean age of the low-$\sigma$ population. Similarly, with regard to \cite{TraFabDre08}, who also find no downsizing with a 12 galaxy sample, we argue that a sample of this size is too small to detect the downsizing effect, given the large intrinsic scatter in the age-$\sigma$ relation. 

\section{Synthetic Galaxy Clusters}
\label{sec:synth}

In section \ref{sec:fit}, we have shown how we derive the medians and dispersion of age, metallicity and \ale\ as a function of velocity dispersion. For a given choice of stellar population parameters, $P$, we can generate stellar mass-to-light ratios, and hence colors and mass-to-light ratios, as outlined below in Section \ref{sec:MLcol}. To populate a synthetic cluster, the additional ingredients required are the velocity dispersion distribution, effective radii and stellar masses. 

\subsection{The Velocity Dispersion Distribution}
\label{sec:VDD}
The next step is to generate a synthetic cluster populated by galaxies with appropriate statistical distributions of velocity dispersion, effective radius, stellar mass, age and metallicity, as well as derived parameters (luminosity). Because velocity dispersion is the driving parameter, a prerequisite input is the velocity dispersion distribution (VDD). We adopt a lower limit of $\sigma = 20\kms$ for the velocity dispersion distribution for our synthetic clusters. While we will not create velocity-dispersion-limited samples with velocity dispersions this low, because of the large scatter in the Faber-Jackson relation, it is necessary to include low velocity-dispersion galaxies in order to ensure completeness at a given magnitude.

To model the VDD, we begin with the fits of \cite{SheBerSch03} to a
sample of de~Vaucouleurs-profile early-type galaxies in the SDSS
\begin{equation}
\phi(\sigma)d\sigma=\phi_{\ast}\left(\frac{\sigma}{\sigma_{\ast}}\right)^\alpha\frac{\exp\left[-\left(\sigma/\sigma_{\ast}\right)^\beta\right]}{\Gamma\left(\alpha/\beta\right)}\ \beta\ \frac{d\sigma}{\sigma}
\label{eq:VDdist}
\end{equation}
where the best fit parameters are $\phi_{\ast}=0.0020\pm0.0001$, $\alpha=6.5\pm1.0$, $\beta=1.93\pm0.22$ and $\sigma_{\ast}=88.8\pm17.7 \kms$.

\begin{figure}[htbp]
	\centering
	\includegraphics[width=\linewidth]{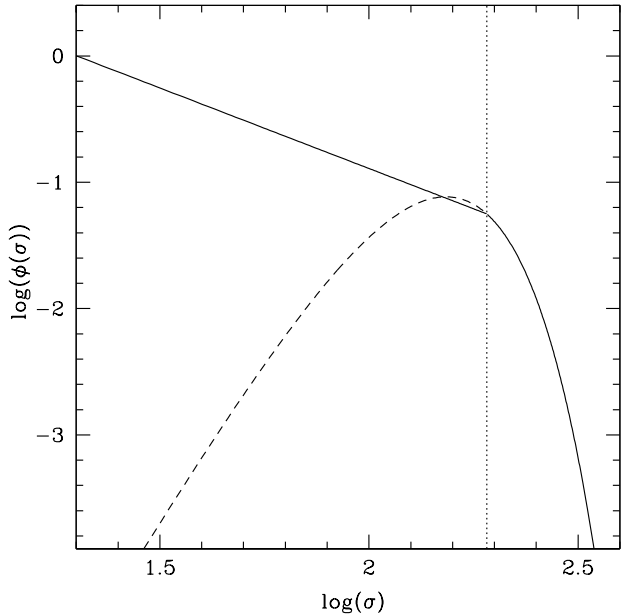}
\caption[VDD Function]{The normalized velocity dispersion distribution (VDD) of early-types from \cite{SheBerSch03}, with extrapolation to low velocity dispersions shown by the dashed line. Late-type galaxies dominate at $\sigma<200\kms$.  We approximate this with a power law (solid line) and fit the slope such that the resulting dwarf-to-giant ratio of all models reasonably matches the $z \sim 0$ observed dwarf-to-giant ratio of \cite{GilBal08}. }
\label{fig:VDD}
\end{figure}

According to \citet[Figure 6]{SheBerSch03}, one expects that late-type
galaxies dominate the VDD, at least for velocity dispersions lower
than 200 km/s. However, the exact level of the contribution of these
galaxies to the VDD is very uncertain. Nevertheless, to the velocity dispersion
limit to which we would like to synthesize galaxies ($\sim 20\kms$), we expect non-early-types to make a non-negligible contribution to the RSG cluster population. Thus we model the ``faint'' end of the VDD as a power law in $\sigma$:
\begin{equation}
\phi(\sigma) \propto 
\sigma^{\xi} 
\qquad (\sigma < 200 \kms) \,.
\end{equation}
We choose to fit the ``faint-end'' slope, $\xi$, by adjusting it to fit the observed $z \sim 0$ red-sequence dwarf-to-giant galaxy ratio (DGR) \citep{GilBal08}. 
We find this slope to be $\xi = -1.275$, yielding a DGR of 3-4 at $z=0$. 
The final adopted distribution of $\sigma$ is shown in Figure~\ref{fig:VDD}.
For each of our CSFH models, we create $50000$ galaxies from this distribution, 
with $\sigma > 20$ \kms.

Thus the procedure is as follows: generate a velocity dispersion from the VDD described above. For a given velocity dispersion, the mean and scatter in the stellar population parameters are known (the binned values are interpolated or extrapolated where necessary). For each of the stellar population parameters, we can thus generate a realization consistent with the statistical distributions, allowing for correlation between offset in ``age'' and the offset in metallicity, and likewise between ``age'' and \ale. The distribution of the simulated populations are shown by the contours in Figures \ref{fig:params1} and \ref{fig:params2}.

\subsection{Mass-To-Light Ratios and Colors}
\label{sec:MLcol}
Having generated the stellar population parameters for our synthetic,
clusters, we now turn to predicting observables, such as color. The
models of Maraston \citep{Mar98,Mar05} provide grids of stellar
mass-to-light ratios in both the Johnson-Cousins and SDSS filter sets,
which can be used to compute colors as a function of age and  metallicity.  Here we adopt the
\cite{Kro01} IMF with red/intermediate horizontal branch. Using the
Salpeter IMF would increase the $M_*/L$ ratios by a factor of
$\sim{}1.6$ \citep{ThoMarBen05}; we discuss the effects of the choice
of IMF in Section \ref{sec:imfchoice}.

We then fit scaling relations of mass-to-light ratio and color as a function of velocity dispersion. For the fits, only galaxies with velocity dispersion $\log(\sigma)>1.844$ are used, as these are not affected by the ``upturn'' in age at low velocity dispersion (Figures \ref{fig:params1}, \ref{fig:params2}).  In the Appendix, Tables \ref{tab:MLJC} and  \ref{tab:MLSDSS} summarize the slopes and intercepts as a function of velocity dispersion for Johnson-Cousins and SDSS passbands.

\subsection{Effective Radii, Stellar Masses and Luminosities}
\label{sec:effrad}
In order to predict luminosities and the Fundamental Plane, we need to assign a stellar mass and an effective radius to each synthetic galaxy. 

We assign an $R_e$ based on the $\sigma$, following the
broken-power-law model of our Coma data in Section \ref{sec:comadata},
allowing for scatter in this relation by generating a random Gaussian deviate with
the same root-mean-square scatter as given in Figure \ref{fig:ComaRe}.  Total mass is then assigned according to the virial relation 
\begin{equation}
M = c_2 \frac{\sigma_e^2 R_e}{G} 
\label{eq:Msol}
\end{equation}
where here we assume $\sigma_e \sim \sigma$ (see Section \ref{sec:ML}).  The factor $c_2$ depends on the luminosity distribution, velocity dispersion anisotropy and rotation. Perhaps surprisingly, \cite{CapBacBur06} found that a constant $c_2 = 5$ was an excellent approximation to detailed dynamical mass reconstructions. This indicates that there is little evidence for non-homology, i.e. a dependence of $c_2$ on mass or velocity dispersion. The same result, that early-types are homologous, has also been found using gravitational lensing as a mass estimator \citep{BolTreKoo08}. Note that \cite{CapBacBur06} measured $R_e$ in the $I$-band, which is close to the $r$ band used here for measuring $R_e$, so their value of $c_2$ may be transferred to this analysis. Denoting the fraction of the total mass in stars as $f_{\ast}$, we then have $M_{\ast} = f_{\ast} M$. Clearly for our analysis, $c_2$ and $f_*$ are degenerate. We adopt $c_2 = 5$ and $f_* = 1$ (no dark matter), and return to discuss these assumptions in Section \ref{sec:FPDM}.

Finally, the magnitudes are then given by, for example in the $R$-band,
\begin{equation}
M_{R} = M_{R_{\odot}} -2.5\log\left(\frac{M_{\ast}}{M_*/L_R}\right) 
\end{equation}
where the stellar $M_*/L_R$ is given by the Maraston models for the known stellar population parameters, constructed as described in Section \ref{sec:csp}.

\subsection{Selection Effects at Faint Magnitudes }
\label{sec:lowvd}

As noted above, the model fits show an increase in all age-related parameters at low velocity dispersion. \cite{KelIllFra06} pointed out that a magnitude-limited sample will lead to biases in stellar population parameters at fixed mass because, for example, older galaxies will be fainter and hence will be excluded by the magnitude limit. This bias was also studied by \cite{GraFabSch07}, who calibrated the bias empirically, and found a larger correction to metallicity than to age.

\begin{figure*}[htbp]
	\centering
 	\mbox{\subfigure{\includegraphics[width=.33\linewidth]{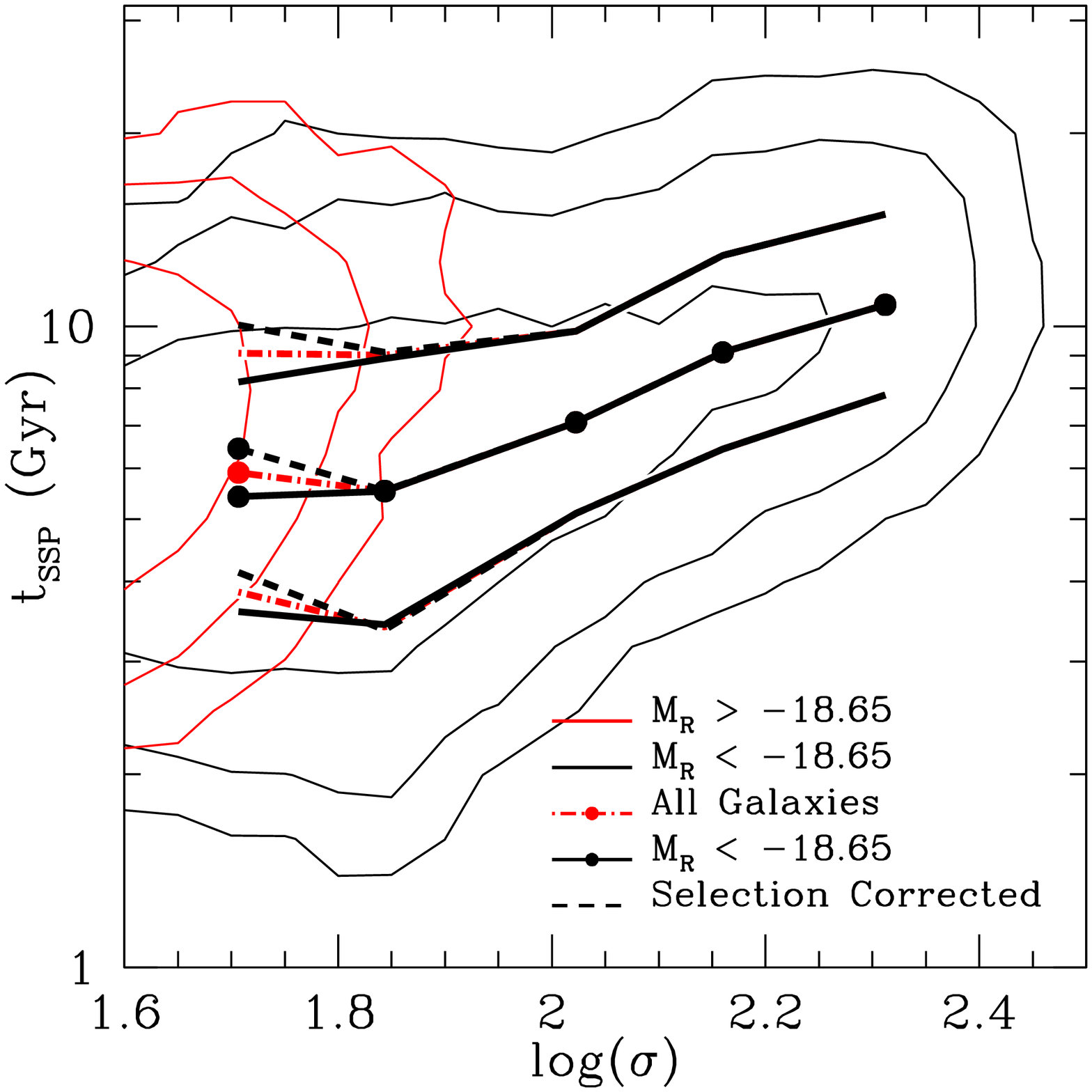}}
 				\subfigure{\includegraphics[width=.33\linewidth]{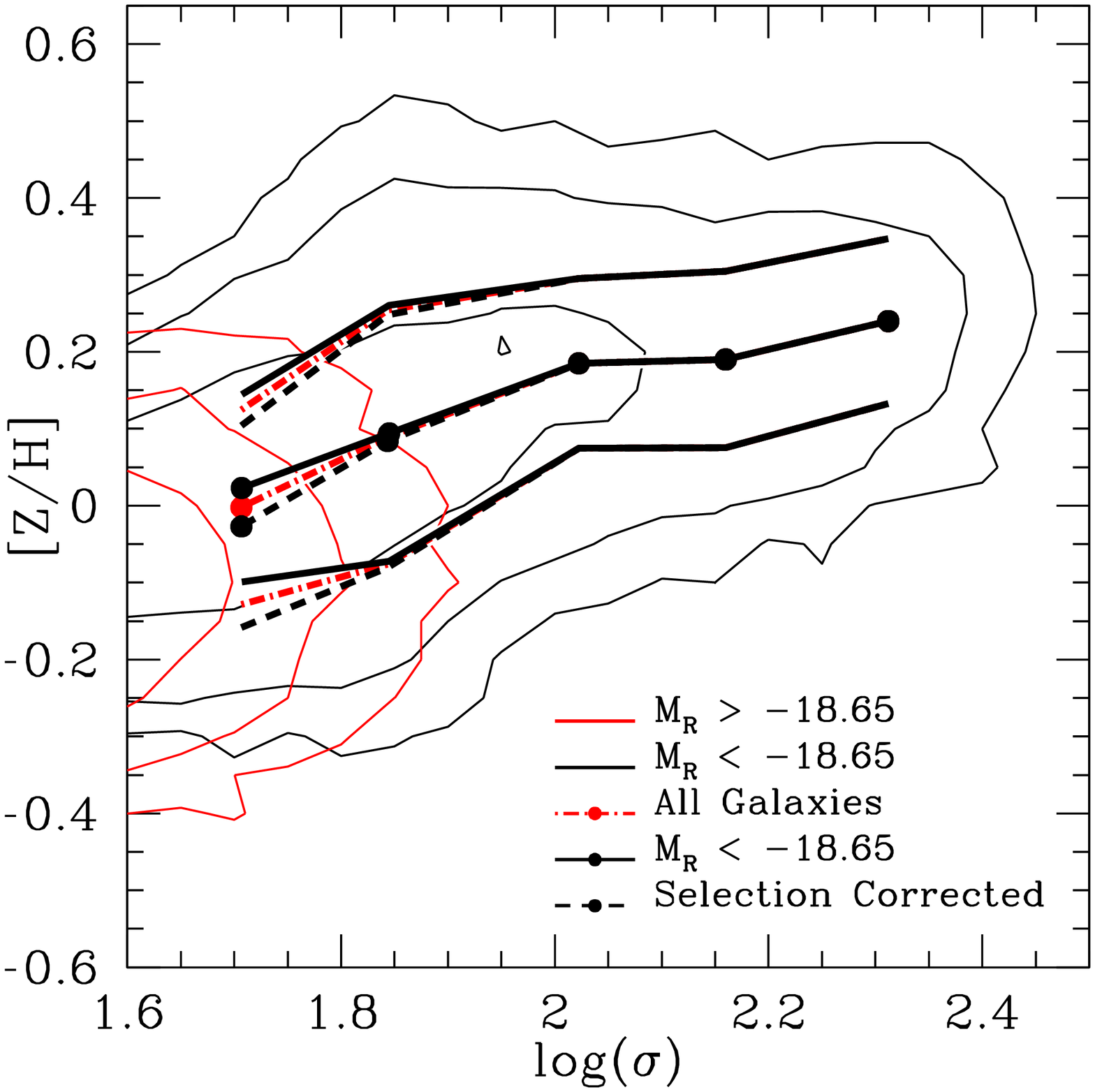}}
 				\subfigure{\includegraphics[width=.33\linewidth]{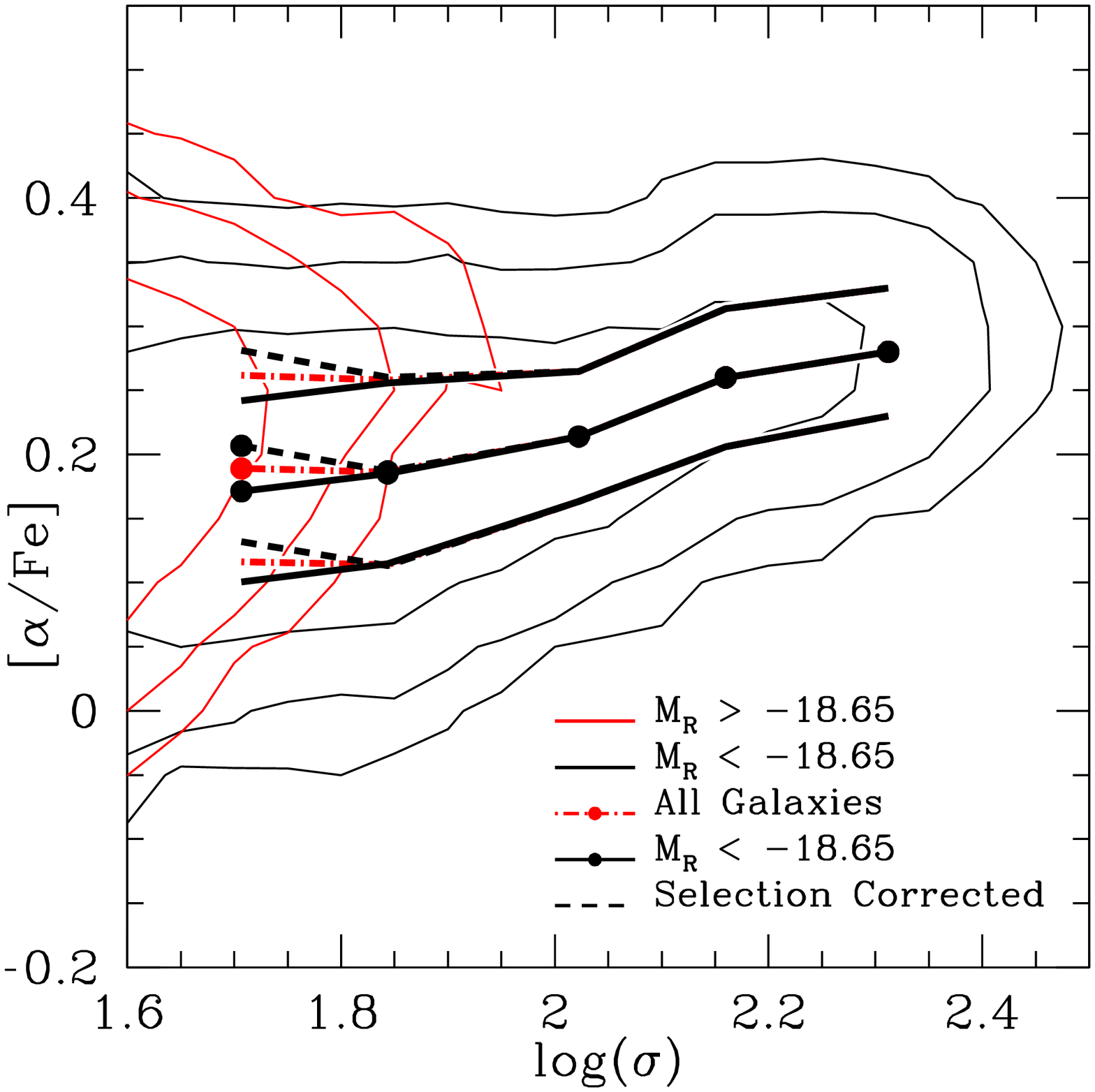}}   }
	\caption[Selection Effects]{Effect of a magnitude limit on the modeled SSP stellar population parameters. The Shapley data, on which the stellar populations of our synthetic clusters are based, were limited to $M_R < -18.65$. Thus the derived stellar populations, and hence the mock clusters, may be affected by biases due to this magnitude limit. However, having fixed the stellar populations, there is no explicit magnitude limit in the synthetic clusters, i.e. some galaxies have magnitudes fainter than $-18.65$. This allows us to use the synthetic clusters to test the biases due to the magnitude limit.  Black contours represent galaxies with $M_R < -18.65$, whereas red contours represent galaxies with $M_R > -18.65$ which would have been excluded had our synthetic clusters been magnitude limited. The dot - dashed red lines are fits to the stellar populations of the synthetic clusters with the faint galaxies included in the entire population, whereas the solid black lines are the fits when the faint galaxies are removed. When faint galaxies are removed from the sample, we find that, at the low velocity dispersions, the population is generally younger, more metal and less $\alpha$-enhanced. The scatter in each stellar population parameter also decreases. Therefore, in order to correct for this selection effect in the real data, we would have to \emph{reverse} the offset from this test, resulting in an underlying population that is older, more metal-poor and more $\alpha$-enhanced with greater scatter in all parameters at dispersions $\log(\sigma) < 1.85$. The dashed black lines show our best estimate of the underlying population parameters after correction for this effect.}
\label{fig:seleffects}
\end{figure*}

Here, we use our synthetic cluster populations to determine the effect of this bias on our modeled stellar population parameters.  Our Shapley sample was limited to $m_R<18$, or $M_R<-18.65$, but our synthetic clusters are limited by velocity dispersion, not by magnitude. By removing the synthetic galaxies fainter than $-18.65$, we can quantify the biases that may result from this selection by refitting the model parameters to the magnitude-limited sample. Figure \ref{fig:seleffects} illustrates the effect, in the case of the SSP model. We find that, when faint galaxies are excluded, the remaining low-$\sigma$ galaxies are younger, more metal-rich and less-$\alpha$ enhanced. One might naively expect the bias in metallicity to go in the opposite sense: i.e. metal-poor galaxies are brighter and hence they should be preferentially included in a magnitude-limited sample. However, in our SSP model, the scatter in age and metallicity at fixed $\sigma$ is assumed to be anti-correlated (Section \ref{sec:scatfit}). Since the effect of age on mass-to-light is greater than that of metallicity, the bias towards younger brighter galaxies implies that we are also selecting relatively metal-rich galaxies. The correlated scatter between age and \ale\ goes in the opposite sense, and hence so does the correction to \ale. The effects of this selection bias are small, however, and affect only the lowest velocity dispersion bin. Now because the original Shapley data were indeed magnitude-limited, then to account properly for this selection effect, a correction should be applied (in the opposite sense to that found when faint galaxies are removed from the synthetic cluster sample). The corrected stellar population parameters are shown by the dashed line in Figure \ref{fig:seleffects}. Thus the correction, if applied, would make the low velocity dispersion galaxies older, less metal-rich and more $\alpha$-enhanced.  The same bias also affects the scatter in stellar population parameters: because galaxies are removed from the magnitude-limited sample, the scatter in, for example, age is reduced. Thus a correction for selection would increase the scatter at low velocity dispersion. We do not correct for these effects in this paper, because Figure \ref{fig:seleffects} shows that the effects are small. Nevertheless, the sense of the correction strengthens our conclusion that the ``upsizing'' at low velocity dispersion is real. Note, however, that the selection does not affect the robust ``downsizing'' trend seen at higher velocity dispersion ($\sigma > 70 \kms$).

\section{Results}
\label{sec:results}

For each model, we now have stellar population parameters and derived parameters, such as magnitudes and colors,  for a synthetic cluster galaxy population.  We can now compare these to independent observations in order to constrain the model. Later in this section, we will compare the predictions to scaling relations such as the FP. We begin however with a comparison sensitive only to the stellar populations: the color-$\sigma$ relation.

\subsection{Comparison of color--$\sigma$ relations}

\begin{figure*}[htbp]
	\centering
 	\mbox{\subfigure{\includegraphics[width=.33\linewidth]{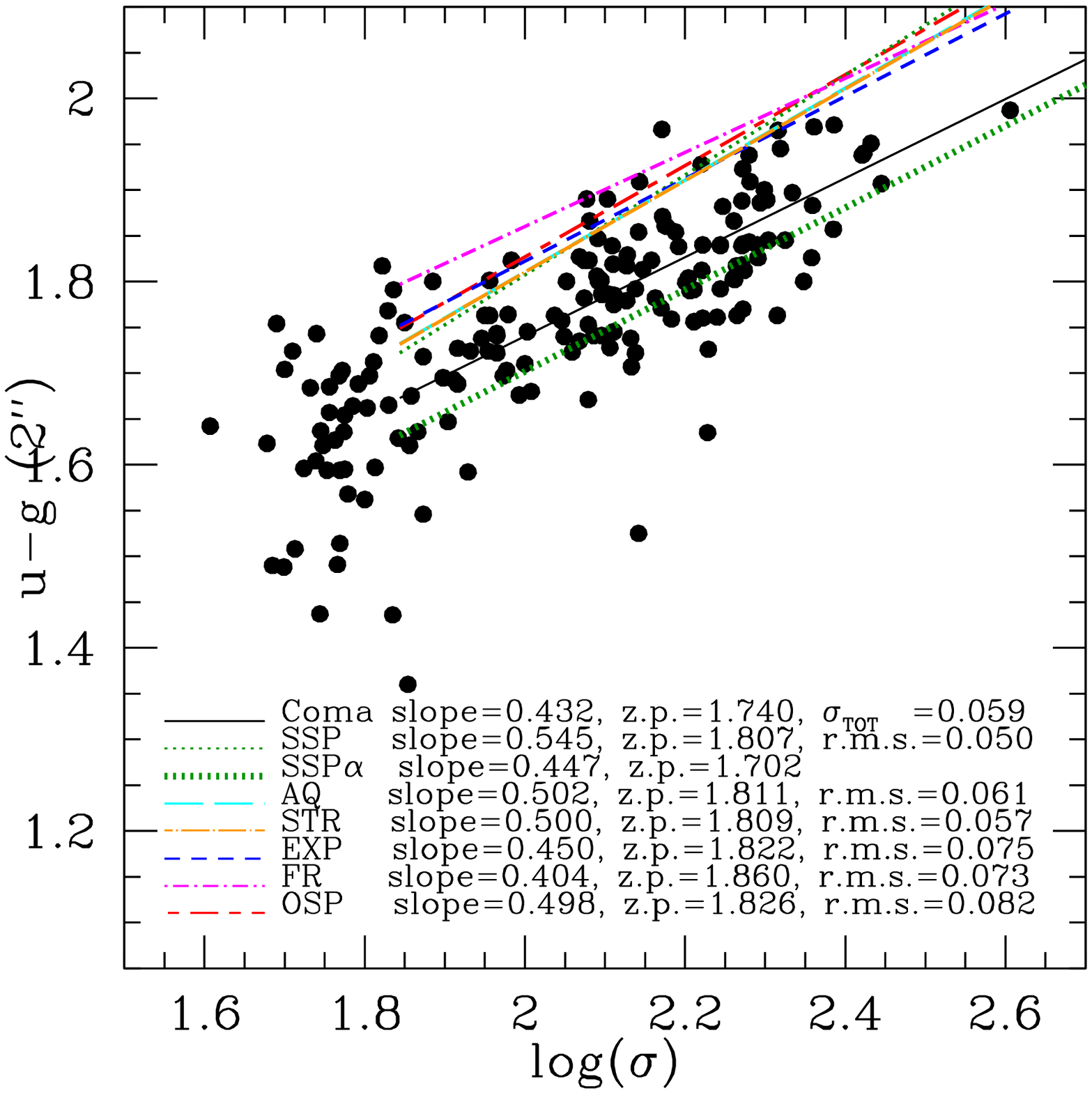}}
 				\subfigure{\includegraphics[width=.33\linewidth]{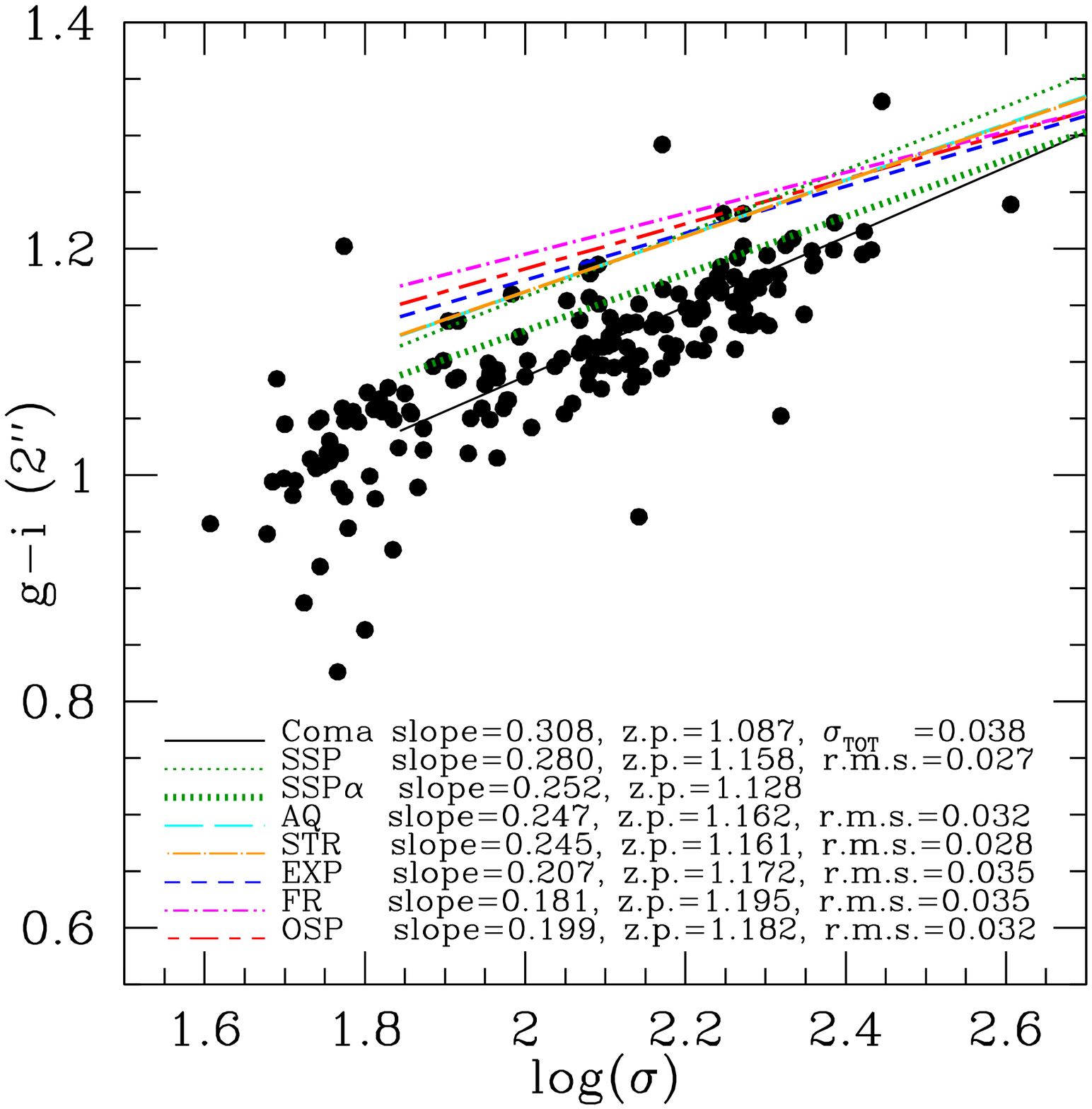}}
 				\subfigure{\includegraphics[width=.33\linewidth]{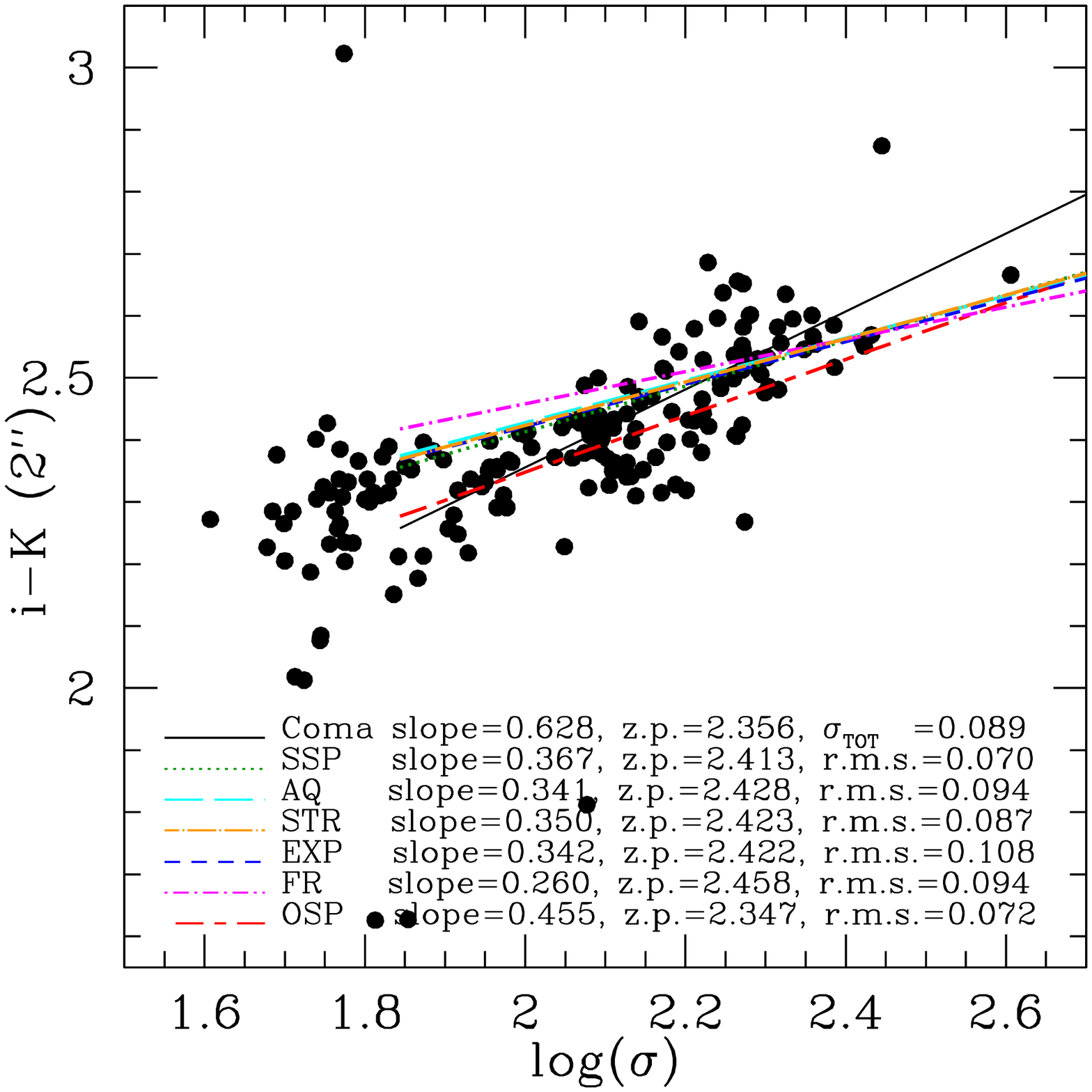}}   }
	\caption[Coma colors]{Color - $\sigma$ relations compared to model predictions. Colors are measured in a 2'' aperture on Coma to match the 1'' aperture of the Shapley data, at twice the  distance. Fits to the Coma data are shown by the solid line, and robust fits to the medians, hence insensitive to outliers. Only galaxies within $r_{200}/2$ (filled circles) and $\sigma > 70 \kms$ are used in the fit. Line styles and colors as in Fig. \ref{fig:6fitlines}, except for the heavy green dotted line is the SSP model with a correction for \ale.  The \ale-correction reduces the color offsets in $u-g$ and $g-i$, as discussed in the text.}
	\label{fig:Comacol}
\end{figure*}

Figure \ref{fig:Comacol} compares predicted $u-g$, $g-i$ and $i-K$ colors in our models to the Coma cluster data measured in an aperture of the same physical size ($\sim{}0.95$ kpc, 2") as the Shapley line index data.  The quantitative comparisons are restricted to galaxies with cluster-centric radii $r_{200}/2$ to match our models derived from the central regions of Shapley clusters. 

Although variable \ale\ is used to fit the line indices using the models of \cite{ThoMarBen03,ThoMarKor04}, the stellar $M_*/L$ predictions of \cite{Mar05} do not allow for the effect of \ale. We can estimate the effects of \ale\ using the models of \cite{CoeBruCha07}, who compute tracks and colors for an $\alpha$-enhanced model as well as models with solar abundance ratios. Their colors are tabulated as a function of [Fe/H] and [$\alpha$/Fe]. Fitting the colors as a function of age, \ale, and metallicity (where [Z/H] = [Fe/H] + 0.75[$\alpha$/Fe] for the enhancement pattern adopted by \cite{CoeBruCha07}), we can find the differential effect of [$\alpha$/Fe] (at fixed [Z/H]) on the colors. Since \ale\ is a weak function of velocity dispersion, in principle both the slopes and the zero-points of the color--$\sigma$ relations could change.  Using their self-consistent SSP model in a differential sense, ie. to derive an \ale-dependent color correction, we find that the optical colors shift blueward with increasing \ale, at fixed Z/H.  We illustrate this for the SSP case for $u-g$ and $g-i$ in Figure \ref{fig:Comacol}, where the lower SSP model is with \ale.

Note that while our models do not allow for dust, in reality, of course dust may be present, particularly for low-$\sigma$ galaxies, that may be of later morphological type. Correcting the data for this possible effect would shift the low-$\sigma$ data blueward, and would make slope steeper.

In $u-g$, without allowing for \ale, all models are too red. The best fit slope is the FR model and all the other models overpredict the observed slope, by factors as large as 33\% (SSP).  However, if we allow for \ale, the SSP prediction for $u-g$ shifts blueward by $-0.090$ at $\log \sigma = 1.84$ and $-0.136$ at $\log \sigma = 2.31$. This makes the SSP model very slightly bluer (by a few hundredths of a magnitude) than the Coma data, and flattens the slope to 0.45, in good agreement with the observed slope ($0.43\pm0.04$). We have not calculated the effect of \ale\ on the other models, but one might expect the sense of this correction to be the same as for the SSP case, i.e.\ to flatten the model slope and shift the model to bluer colors on average. Note that $u-g$ is, however, sensitive to dust, and so the small reddening  of the data may be due to dust extinction.

In the $g-i$ color, all of the models are again too red and, with the exception of the SSP model, are also too shallow compared to the data.  The effect of \ale\ is weaker in $g-i$ than in $u-g$: the models shift blueward by $-0.026$ at $\log \sigma = 1.84$ and $-0.039$ at $\log \sigma = 2.31$ for the SSP case. Thus allowing for \ale, all models remain slightly too red (by $\sim 0.04$ mag in the SSP case), and of course, a disagreement in this direction cannot arise from dust.  The SSP $(g-i)-\log \sigma$ slope flattens slightly to 0.25, but is still a reasonable fit to the slope of the data ($0.31\pm0.03$). \cite{MarStrTho08} have noted that, by using empirical libraries from \cite{Pic98}, the colors of solar-metallicity models change, particularly in $g$ and $r$.   However, the sense of the change would be to make the models even redder in $g$-$i$. In contrast to the good agreement with the \ale-corrected SSP case, the ``mostly-old'' models (EXP, FR, OSP) have slopes only two-thirds of that observed before considering \ale, which one expects would flatten them further. Furthermore, correcting for dust, if preferentially present in low-$\sigma$ galaxies, would steepen the observed slope, exacerbating the conflict with the ``mostly-old'' models.  

In $i-K$, again the model slopes are all too shallow. The OSP model has the strongest metallicity trend, and since $i-K$ is more metallicity sensitive than age sensitive, it yields the steepest $i-K$ slope. However, the SSP model is the next steepest model after the OSP case.  While the models of \cite{CoeBruCha07} do not predict the effect of \ale\ on $i-K$, those of \cite{PerSalCas09} do predict its effect on $I-K$:  $\sim0.04$ for the \ale\ relevant at $\sigma \gtrsim 70 $ \kms. Thus qualitatively we would expect \ale\ to shift colors to the blue and very slightly flatten the predicted color-$\sigma$ relation, as with the optical colors.

We conclude that, overall, the SSP model and, to a lesser extent, the AQ model are good fits to the optical color-$\sigma$ slopes, once the effects of $\alpha$-enhancement are accounted for.  The FR and EXP models do not fit the observed trends well in $g-i$. None of the models fit the observed $i-K$ slope well, even the OSP model which has a strong dependence of metallicity on $\sigma$.  The comparison of color-$\sigma$ zero-points is less certain, particularly because of the effects of \ale\ and of dust. No model fits the zero-point of, for example, $g-i$, although the SSP model zero-point is closest to the Coma data. We conclude that colors are not a strong discriminant of the CSFH models.

\subsection{The Faber-Jackson relation}

As described in Section \ref{sec:effrad}, stellar masses are generated based on the observed $R_e-\sigma$ relation, and from these and the mass-to-light ratios one can calculate magnitudes and surface brightnesses.

\begin{figure}[htbp]
	\centering
	\includegraphics[width=\linewidth]{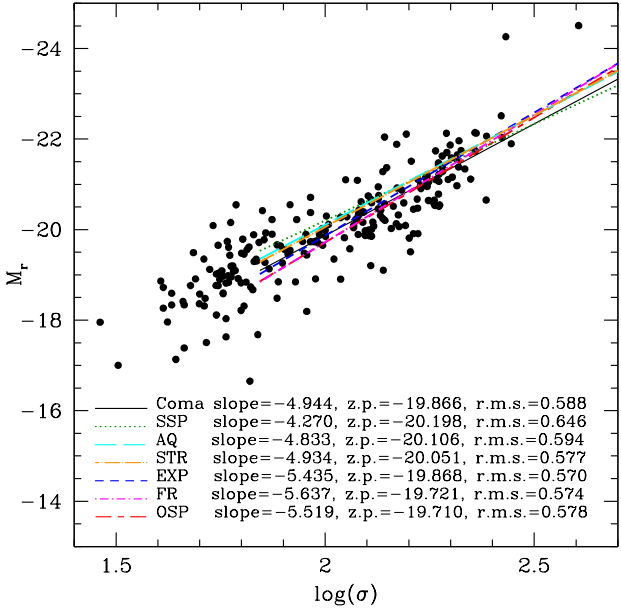}
	\caption[Coma Faber Jackson]{Coma Faber-Jackson relation compared to model predictions. Line styles and colors are as in Fig. \ref{fig:6fitlines}. 
	
	The fit to Coma data uses galaxies within a cluster-centric $r_{200}/2$ only and with $\sigma > 70$ \kms. Note the transition in the data from $L \propto \sigma^2$ to $L \propto \sigma^4$ at $\log \sigma \sim{} 2.2$, resulting from break in the $R_e - \sigma$ relation (Figure \ref{fig:ComaRe}). The same transition is seen in FJ of the synthetic clusters, but here we simply plot the best linear fit to their FJ relation.}
	\label{fig:ComaFJ}
\end{figure}

Figure \ref{fig:ComaFJ} illustrates the resulting Faber-Jackson (FJ) relations both for the Coma data and the models. Note that over the range of velocity dispersions covered here, the observed slope of the FJ is close to $-5$, i.e. $L \propto \sigma^2$. This is in agreement with previous work \citep{MatGuz05}. Only at large velocity dispersions ($\log(\sigma)  > 2.23)$ does FJ approach the ``classical'' relation $L \propto \sigma^4$. This change in slope is a direct result of the change in slope of the $\log(R_e)-\log(\sigma)$ relation discussed above (Section \ref{sec:effrad}).  The FJ relation is not a strong model discriminator because its slope and scatter are driven by the $\log(R_e)-\log(\sigma)$ relation, and not by stellar populations and their associated mass-to-light ratios.

\subsection{The Fundamental Plane}
\label{sec:FP}

\begin{figure}[htbp]
	\centering
	\includegraphics[width=\linewidth]{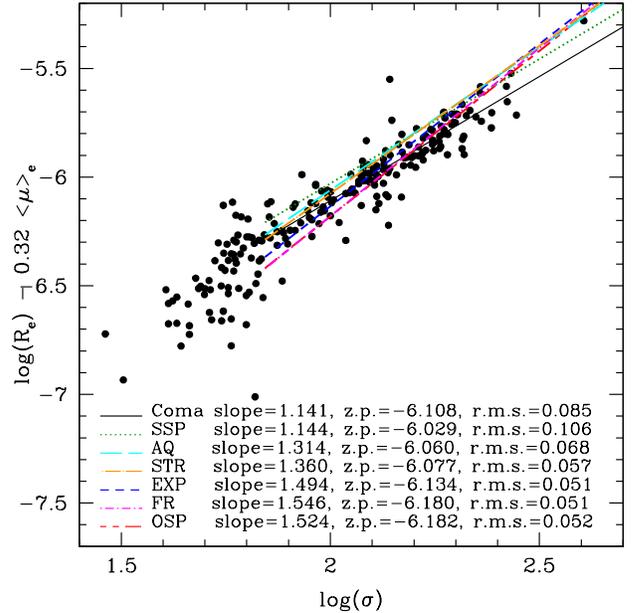}
\caption[Fundamental Plane]{The Fundamental Plane. The $r$-band data from Coma are reproduced in each panel by the open circles. The scatter in $[\log(R_e) - 0.32 \langle \mu_e \rangle]$ for Coma is 0.085. Line styles and colors are as in Fig. \ref{fig:6fitlines}.   In order of increasing extra ``tilt'' relative to the data, these are SSP, AQ, STR, EXP, FR and OSP.  We note however that the SSP slightly overproduces the scatter in the FP.  However all models, with the exception of SSP and AQ predict, at fixed $R_e$, too faint a surface brightness at low $\sigma$.}
\label{fig:FP}
\end{figure}

The FP is considerably tighter than the FJ relation, and hence offers a more powerful test of dynamical mass-to-light ratios. Figure \ref{fig:FP} shows the $r$-band  FP data for Coma where we have chosen to regress the combination $[\log(R_e) - 0.32 \langle \mu_e \rangle]$ on $\log \sigma$.

If the tilt of the FP with respect to the virial relation were due only to variation in average stellar populations as a function of $\sigma$, then we would expect the slope of the model FP to agree with that of the Coma data.  Thus the observed slope can be used to determine the additional tilt that is due to DM and/or non-homology. We find that the SSP and AQ models require the least additional tilt from DM \& non-homology.  We discuss this in more detail in Section \ref{sec:FPDM} below. Furthermore, the ``mostly-old'' (EXP, FR, OSP) models predict a surface brightness too faint compared to the data, at low velocity dispersions and at a fixed effective radius. The addition of dark matter would exacerbate this situation as it would increase the predicted velocity dispersion and so would move the models further to the right.

\begin{figure}[htbp]
	\centering
	\includegraphics[width=\linewidth]{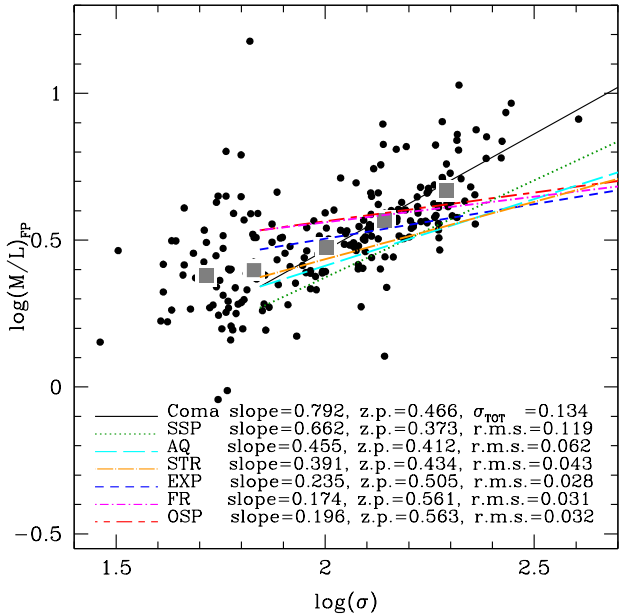}
\caption[MLFP]{The $r$-band mass-to-light ratio data from Coma.  Observed data from the Coma Cluster are shown as black points, and the fit to these is indicated by the solid black line. Binned medians are plotted as grey squares. These bins represent the same $\log(\sigma)$ ranges used when fitting to Shapley (see Table \ref{tab:indices}). Model fits to the \emph{stellar} $M_*/L$ are shown by the colored broken lines (line styles and colors as in Fig. \ref{fig:6fitlines}). Again, we only fit these lines to $\log(\sigma)>1.844$, see Section \ref{sec:lowvd}. We find the models consisting of intermediate mass stars (SSP, AQ, and STR) fit best, whereas the models consisting of old stars (OSP, FR, EXP) fit the poorest. Note here that the SSP and AQ models underpredict $M\sbr{FP}/L$ at all $\sigma$'s.}
\label{fig:MLFP}
\end{figure}

An alternative way to address this issue is by constructing the \emph{dynamical} mass-to-light ratio directly from the observed FP parameters \citep{DreLynBur87,BenBurFab92,JorFraKja96}. These can then be compared to \emph{stellar} mass-to-light ratio predictions, as shown in Fig. \ref{fig:MLFP}. The most striking aspect of this is how closely the Coma dynamical $M/L$, as a function of $\sigma$,  follows the distribution of stellar \emph{ages} as a function of $\sigma$ (see Fig.\  \ref{fig:params1}). Specifically we see the same steep slope at high $\sigma$ and the same upturn or flattening, and increased scatter at low $\sigma$. This strongly suggests that the FP tilt and scatter are driven primarily by stellar age effects.

As with Fig. \ref{fig:FP}, the additional tilt of the SSP model (i.e. the tilt that is attributable to dark matter and/or non-homology, in addition to the stellar populations), is small.  However, as also noted above, at $\log \sigma = 1.844$, the OSP, FR and EXP models overpredict the FP-based mass-to-light ratios at the  low-$\sigma$ end by factors 1.45 - 1.7. The numerical values are given in Table \ref{tab:MLshift}. In contrast, the SSP leaves some room ($\sim 20$\%) for dark matter.

\subsection{Comparison to dynamical mass-to-light ratios}
\label{sec:ML}

\begin{figure*}[htbp]
	\centering
 	\mbox{\subfigure{\includegraphics[width=0.5\linewidth]{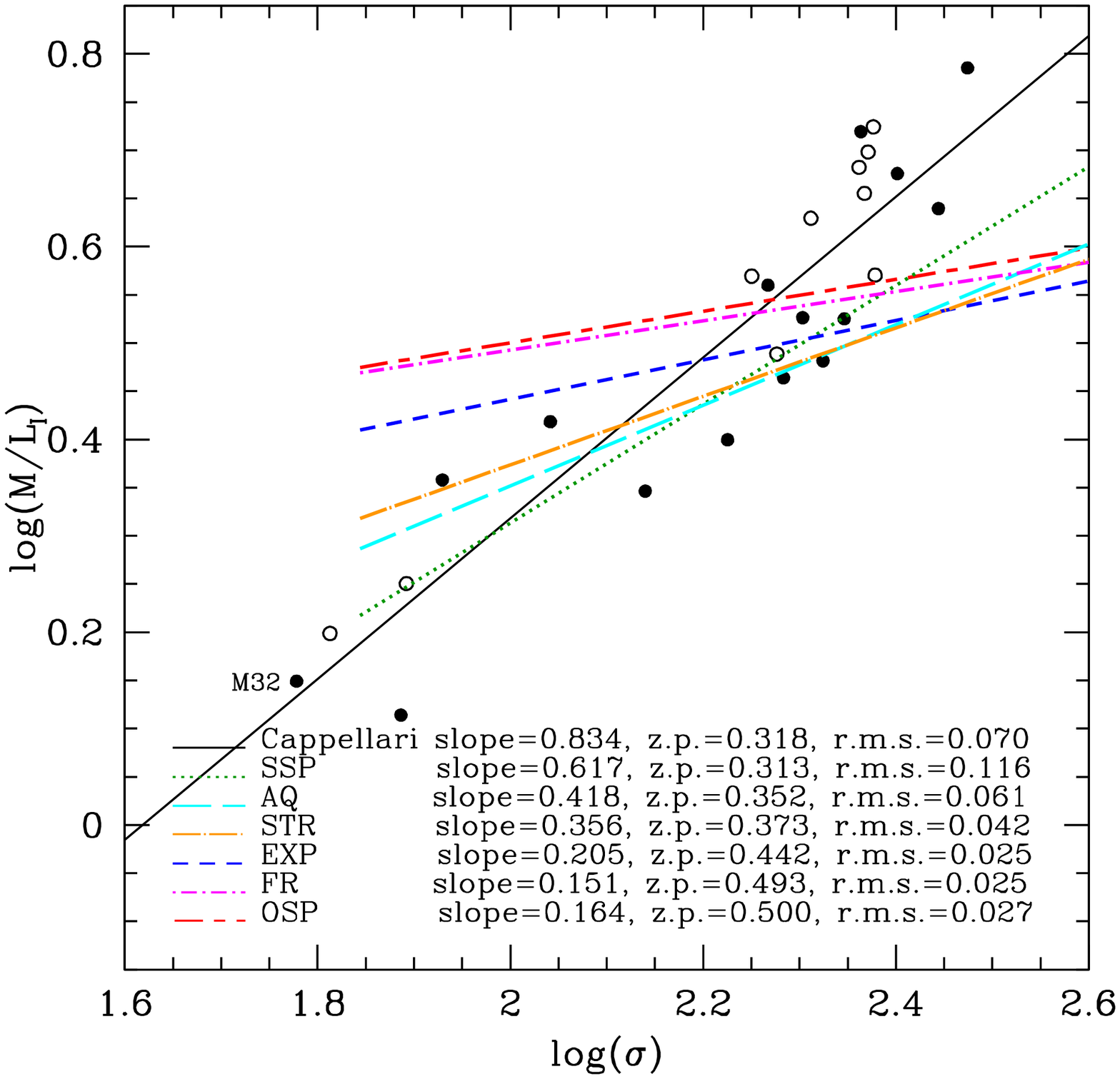}}
 				\subfigure{\includegraphics[width=0.5\linewidth]{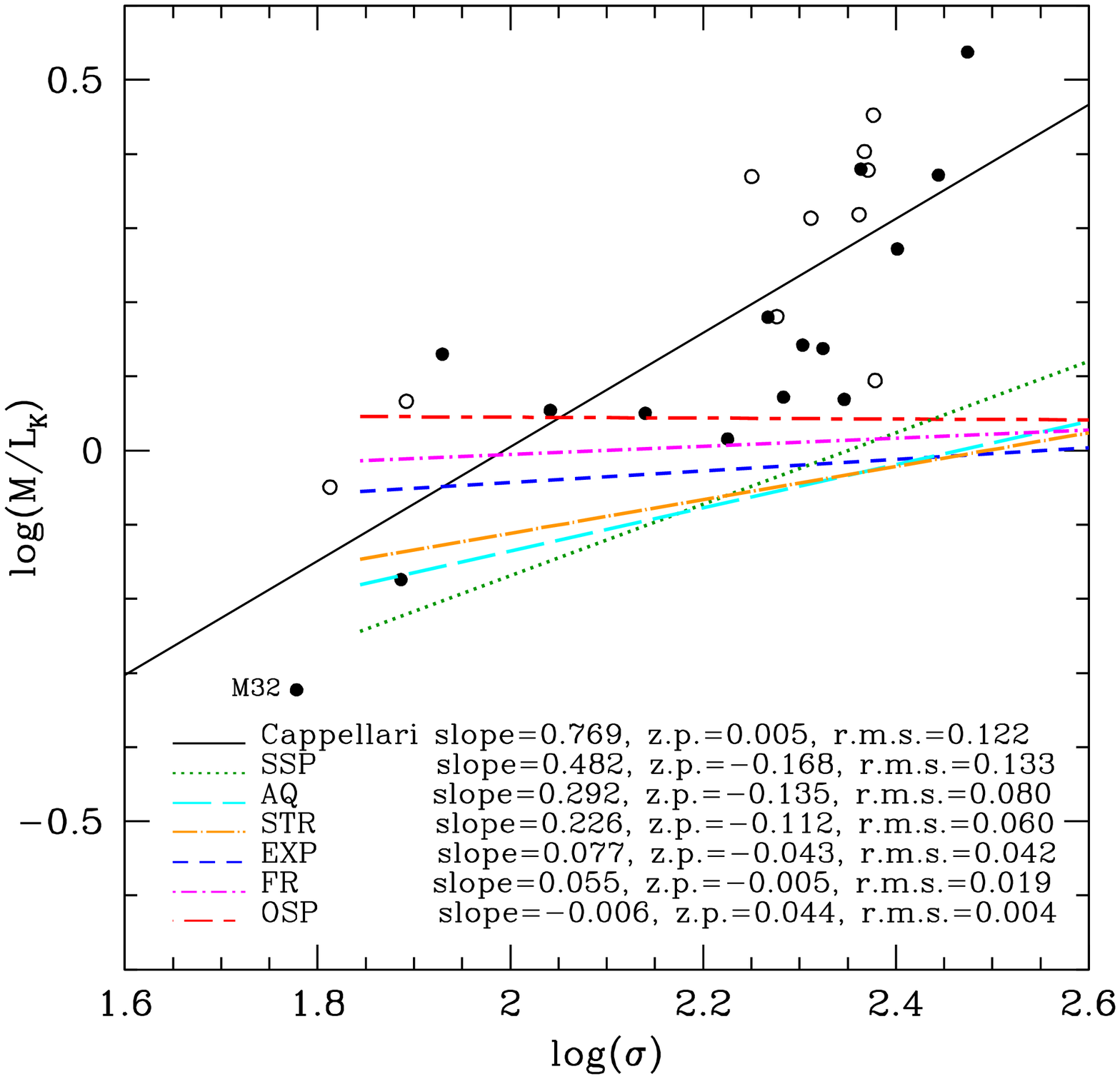}}}
			\caption[Cappellari et al. M/L$_I$ and M/L$_K$ Model Comparisons]{Mass-to-light ratios as a function of velocity dispersion. The left hand panel shows results in the $I$-band; $K$-band results are shown on the right. The \emph{dynamical} mass-to-light ratios from SAURON \citep{CapBacBur06} are shown by the black points (open circles are field galaxies, filled circles are cluster galaxies), and the fit to these is indicated by the solid black line. Model fits to the \emph{stellar} $M_*/L$ are shown by the colored broken lines (line styles and colors as in Fig. \ref{fig:6fitlines}). Again, we only fit these lines to $\log(\sigma)>1.844$, see Section \ref{sec:lowvd}. Note that all models except for SSP overpredict the $M/L$ at low $\sigma$. The models consisting mostly of old stars (OSP, FR, EXP) are the poorest fits at low $\sigma$.}
	\label{fig:MLICap}
\end{figure*}

A more direct comparison is with dynamical mass-to-light ratios in the $I$ and $K$-band from the Schwarzchild model of the SAURON group \citep{CapBacBur06}. The SAURON results are based on full orbit reconstruction using the observed luminosity profile and the observed line-of-sight velocity distribution. Their method thus accounts for spatial and kinematic non-homology and so a comparison between a given model and the data allow us to determine the fraction of dark matter as a function of velocity dispersion. 
Fig. \ref{fig:MLICap} compares the stellar mass-to-light ratios of our synthetic galaxies with the data from \cite{CapBacBur06}, as a function of velocity dispersion%
\footnote{Note that the \cite{CapBacBur06} data tabulate $\sigma_e$, the velocity dispersion at the effective radius, whereas our velocity dispersions are measured at a fixed metric aperture of 0.95 kpc radius. For most of the velocity dispersion range, $R_e$ is constant (Fig. \ref{fig:ComaRe}). Since the velocity dispersion aperture correction is $\sigma/\sigma_e = (R\sbr{ap}/R_e)^{-0.06}$, this yields $\log \sigma_e = \log \sigma - 0.015$, which is a negligible correction.}.
We find that the slopes of the OSP, EXP, and FR models are much flatter than the SAURON mass-to-light ratios. The AQ and STR models are less flat and the SSP slope is a close match. As noted above the masses determined by \cite{CapBacBur06} include both dark and stellar matter, whereas our TMBK models predict only the stellar mass-to-light ratio. Note that aperture effects are small: correcting to $R_e$ (where the \cite{CapBacBur06} mass-to-light ratios are measured) yields changes in $\log(M_*/L) < 0.01$. Thus a serious problem with the OSP, FR and EXP models is that the predict \emph{stellar} $M_*/L_I$ \emph{in excess of the total} $M\sbr{dyn}/L_{I}$, by significant factors (1.93, 1.91 and 1.67, respectively) at $\log \sigma = 1.844$.  On this basis, these models are ruled out. However, note that even the AQ model has a stellar $M_*/L_I$ 26\% larger than the data. Table \ref{tab:MLshift} outlines the offsets between data and model at $\log(\sigma)=1.844$.

Consistent dynamical mass-to-light scaling relations are found in other dynamical studies. For example, \cite{VanVan07b}, using a compilation of data sources including \cite{CapBacBur06}, find a steep relation in the $B$-band, with $M\sbr{dyn}/L_B \propto \sigma^{0.992\pm0.054}$.  This is significantly steeper than all models except for SSP, for which the slope is 0.773 in the $B$-band. Their intercept at $\sigma = 100$ km/s is $\log(M/L) = 0.597$, and is consistent with the SSP model, but, as was the case with the $I$-band, is larger than that of all other models at low velocity dispersion.

From Figure \ref{fig:MLICap}, we see that for the SSP case, stellar populations account for 74\% of the total observed $M\sbr{dyn}/L_I$ ``tilt". Even in the $K$-band, 63\% of the tilt is due to stellar population effects. These reduce to 50\% and 38\% for the AQ model.   For consistency between our models and the SAURON data, we require the baryonic-mass to total-mass ratio to scale as $f_* \propto \sigma^{-0.22}$ for SSP, and $\propto \sigma^{-0.42}$ for the AQ case.

The SAURON data are a mixture of cluster and field early-types, and it may be argued that it is not well-matched to our predictions, which were derived from cluster galaxies alone. 
It is clear from Fig. \ref{fig:MLICap}, however, that there is no significant difference between cluster and field in $M\sbr{dyn}/L$.  At low velocity dispersions ($\sigma < 100$ \kms) there are 5 galaxies: \cite{CapBacBur06} assign three of these to clusters (although one of these is M32) and the other two to the field.  Their H$\beta$ values range from 2-3 according to their Fig. 16, the upper end of which is larger than our median value of 2.05 at the same velocity dispersions. Interestingly, one of the high H$\beta$, low $M/L$ galaxies is a cluster member.  We have investigated how the slopes of the SAURON $M/L$ relations are affected by dropping between 2 and 4 of the SAURON galaxies with the highest $H \beta$ (and hence lowest $M\sbr{dyn}/L$). We find that this does not significantly affect our conclusions: the ``mostly-old'' class of model, when tuned to reproduce the observed spectral indices, yields stellar $M/L$ ratios that are too large compared to the dynamical values.

\begin{deluxetable}{l|rrr}
\tablecaption{M/L Offsets at Low Velocity Dispersion}
\tablewidth{\columnwidth}
\tablehead{
\colhead{Model} &
\colhead{$\Delta \log(M/L_I)$\tablenotemark{a}} &
\colhead{$\Delta \log(M/L_K)$\tablenotemark{a}} &
\colhead{$\Delta \log(M/L_r)$\tablenotemark{b}}
}
\startdata
SSP & +0.029 & -0.128 &  -0.092 \\
AQ  & +0.099 & -0.066 &  +0.00 \\
STR & +0.129 & -0.032 &  +0.027 \\
EXP & +0.222 & +0.060 & +0.157\\ 
FR  & +0.281 & +0.101 &  +0.239 \\
OSP & +0.286 & +0.160 &  +0.235 \\
\enddata
\tablenotetext{a}{Offset between data and model at $\log(\sigma)=1.844$, see Fig. \ref{fig:MLICap}}
\tablenotetext{b}{Offset between data and model at $\log(\sigma)=1.844$, see Fig. \ref{fig:FP}}
\label{tab:MLshift}
\end{deluxetable}

\section{Extending the Models}
\label{sec:mod}

Our models as described above are necessarily simplistic. In this section we explore the effects on our results of (i) systematics, (ii) altering the IMF, and (iii) adding dark matter.

\subsection{Systematic Effects on Mass-to-Light Ratios and Colors}
\label{sec:sysfx}

As mentioned in Section \ref{sec:sysparam}, the arbitrary choice of restricting the star formation of CSFH models to begin at a look-back time of $t\sbr{start} = 13$ Gyr requires investigation. Again, we choose a look-back time of 10.3 Gyr for comparison, which corresponds to $z\sim{}2$. First, we note the effect on the colors and mass-to-light ratios of the choice of $t\sbr{start}$  is small. In all models, we find that, if the star formation starts at 10.3 Gyr then, \emph{at the fixed observed values of the line indices}, the predicted stellar $M_{*}/L$ values are slightly higher than in the case with $t\sbr{start} = 13$ Gyr, with the effect being largest for the bluest bands. This arises because, for example in the case of the EXP model, when $t\sbr{start}$ is decreased, then in order to fit the observed line indices, $\tau$ is also reduced compared to the $t\sbr{start}= 13$ Gyr case, with the consequence that $M_{*}/L$ increases. Other models are affected in a similar way. In the $r$ and $I$ bands in the mid-$\sigma$ range, we find a $12\%$ increase in $M_{*}/L$ for the EXP model and $7\%$ increase for the AQ model. For all models, the slope of the $M_{*}/L$ -$\sigma$ relations are negligibly affected by the choice of $t\sbr{start}$. Colors are also largely unaffected by this choice, with the largest offset being for the bluest bands. This offset in color towards the blue is $\lesssim 0.05$ in magnitude for all colors given in Tables \ref{tab:MLJC} and \ref{tab:MLSDSS},
for all velocity dispersion bins, in all models, compared to the 13 Gyr case. This effect can be explained as follows: although $t\sbr{start}$ is now more recent, the constraint on fitting the line indices forces the fitted `age' towards older populations, as discussed in detail in Section \ref{sec:sysparam}. Because the mass-to-light ratios are even higher when star formation starts later, this test strengthens the conclusion that the `mostly-old' models (EXP, FR, OSP) are a poor fit. 

Next we consider the effect of Balmer line choice, as discussed in Section \ref{sec:linechoice}. In that Section, we found that, for example, fitting only the H$\beta$ Balmer line had the largest effect on the ages\footnote{We note that the other scenarios, such as calibrating the H$\beta$ line using all velocity dispersion bins, had a much smaller effect on the ages (see Figure \ref{fig:balmer}), and hence presumably on the mass-to-light ratios, and so we do not discuss those here.}. As with the choice of $t\sbr{start}$, the effect of using only H$\beta$ on the predicted colors is negligible.  The effect on stellar mass-to-light ratios is somewhat greater: for the SSP model, this choice leads to a decrease in mass-to-light ratios on the order of $10\%$ at $\log \sigma \sim 1.85$ in the $r$, $I$ and $K$ bands. For the EXP and FR model, however, only a $3\%$ and $1$\% decreases in $M_{*}/L$ are found. Thus, this effect is too small to avoid our rejection of the FR and EXP models as discussed in Section \ref{sec:ML}, as these models would still overpredict the stellar mass-to-light ratios at low velocity dispersion, albeit by a slightly lesser amount.

\subsection{Choice of IMF}
\label{sec:imfchoice}
In calculating stellar mass-to-light ratios, we have adopted the \cite{Kro01} IMF. Other choices often considered in the literature are the \cite{Sal55} and \cite{Cha03}  IMFs. These models have similar slopes at high masses, but differ in the low mass ($M < 1 M\sbr{\sun}$)  regime: while the Salpeter IMF is a pure power-law, the Kroupa and Chabrier IMFs have knees at masses of $0.5 M_{\sun}$ and $1 M_{\sun}$ respectively. We expect the choice of IMF to have little effect on the derived stellar ages and metallicities, since most of the light is from main-sequence turn-off and red-giant branch stars. For the ages considered here (2 Gyr to 13 Gyr), the turn-off masses are well above the IMF ``knee''.

The choice of IMF, however, has greater impact on the stellar mass-to-light ratios. Using the Salpeter IMF would increase the $M_*/L$ ratios by a factor of $\sim{}1.6$ \citep{ThoMarBen05} and so would \emph{increase} the conflict for all models at low $\sigma$  between the stellar and dynamic mass-to-light ratios from  the SAURON and FP comparisons. The Chabrier IMF produces marginally lower stellar mass-to-light ratios compared to a Kroupa IMF, but this would be insufficient to reconcile the OSP, EXP and FR models with the dynamical mass-to-light ratios.

More radical IMF modifications have been suggested by \cite{Van08}, who proposes an IMF in which the critical mass at the knee varies as a function of age, with older populations having higher critical masses.  It is difficult to compare the results of this model in detail, as full population synthesis models have yet to be constructed. Referring to \cite{Van08} Figure 15, the effects on $M_*/L$ may be modest, amounting to at most a 20\% reduction compared to the Chabrier IMF for a 3 Gyr population.  It seems very difficult to reduce the high stellar $M/L$ values of the FR and OSP models, as most of their population is old, and for older populations with a larger critical mass ($M_{c} \gtrsim 0.3 M_{\sun}$), $M_{*}/L$ actually increases due to remnants. 

We conclude that the choice of IMF cannot lower the stellar mass-to-light ratios sufficiently to ease the conflict with the OSP, FR and EXP models.

\subsection{Effects of Dark Matter and Non-homology}
\label{sec:FPDM}

For simplicity, our models have been assumed to be 100\% stellar, i.e. $f_* = 1$.  Various studies have suggested that, like spiral galaxies, early-type galaxies also have non-negligible mass fractions of dark matter (DM) in their central regions. For example, from lensing studies, \cite{BolTreKoo08} find $f_* = 0.62\pm0.07$ within the effective radius, for galaxies with $175 \kms < \sigma < 400 \kms$. It is clear that adding DM will increase the total mass-to-light ratios by a factor $1/f_*$.  As noted above, the OSP, EXP and FR models already over-predict the dynamical $M/L$  for low $\sigma$ galaxies, so including DM would only increase the conflict. 

The combined influence of DM  and non-homology can also be deduced from the tilt of the FP. Following \cite{BoyMaQua06}, let us assume a scaling $M\sbr{dyn} \propto M_{\ast}^{1+\mu}R_{e}^{\nu}$. We define the scaling of the stellar mass-to-light as a function of $\sigma$ as $(M_{\ast}/L) \propto \sigma^{m}$, where the value of $m$ is given as a function of model and passband in Table \ref{tab:MLSDSS}.  The theoretical FP
\begin{equation}
R_e = (M\sbr{dyn}/L)^{-1} \frac{c_2}{2\pi} \sigma^{2}\langle I_e\rangle ^{-1}
\end{equation}
becomes 
\begin{eqnarray}
\label{eq:FPDM}
	\log(R_e)\  & \propto &\  \left(\frac{2-m(1+\mu)}{1+\nu+2\mu}\right)\log(\sigma) \\
						& & -\  \left(\frac{1+\mu}{1+\nu+2\mu}\right)\log \langle I_e \rangle \nonumber
\end{eqnarray}
after allowing for stellar populations and dark matter/non-homology.
If we assume that the dynamical mass-to-light ratio scales with mass $M$ as
$M\sbr{dyn}/L \propto M^{T}$, then the ``tilt" $T$ of the FP can be expressed as 
\begin{equation}
T = \frac{2-\alpha}{2+\alpha}
\end{equation}
where $\alpha$ is the coefficient of $\log \sigma$ in equation \ref{eq:FPDM}. We can also derive tilts for the cases with only stellar populations and no DM ($\mu = 0$, $\nu = 0$): 
\begin{equation}
T_{*} = \frac{2-\alpha_{*}}{2+\alpha_{*}} = \frac{m}{4-m}
\end{equation} 
as well as with DM (either $\mu$ or $\nu$ non-zero).

Table \ref{tab:DMFP} restates the value of $m$ for each model from Table \ref{tab:MLSDSS} in the $r$-band. We find that all of the models with the exception of the SSP are too steep (too high $\alpha$, close to the virial relation) in comparison to the observed FP.  We can then adjust the DM/non-homology scaling (the value of $\mu$ or $\nu$) so as to match the observed FP slope in Figure \ref{fig:FP}. This yields the values in the second and third columns of Table \ref{tab:DMFP}.  For the SSP model, the DM/non-homology scaling is essentially zero. For the other models, stronger scaling is required: for AQ $\mu =  0.06$, or  $\nu = 0.150$.  As a result, we find that in the SSP case, essentially all of the $r$-band FP tilt is due to stellar populations. For the AQ model, 2/3 of the tilt is due to stellar populations and 1/3 is due to DM and/or non-homology. Note that because of the age-dependence of the $K$-band stellar $M/L$, for the SSP case we find that most of the predicted $K$-band tilt is due to stellar population effects. In the case of the AQ model,  we expect that half of the $K$-band tilt is due to stellar population effects.

If we repeat this exercise using the dynamical $M_{FP}/L$ values calculated from the FP and shown in Fig. \ref{fig:MLFP}, we can determine the tilt assuming that dark matter and non-homology depend directly on $\sigma$. For the SSP case, we find that $f_* \propto \sigma^{-0.13}$ and for the AQ case the exponent is $-0.34$.   To extend these scaling relations to the case of non-homology, replace $f_*$ with the combination ${f_*}(c_2/5)$. These scalings are slightly flatter than the corresponding values derived from the SAURON comparison: $f_* \propto \sigma^{-0.22}$ and $\propto \sigma^{-0.42}$, respectively.  Since the SAURON results are free of the effects of non-homology, the comparison between the FP and SAURON scalings then yields $c_{2} \propto \sigma^{0.09}$ for both SSP and AQ cases.  This non-homology is weak, and is not a major driver of the tilt of the FP.

\begin{deluxetable}{l|c|cc|ccc}
\tablecaption{Effect of Dark Matter on $r$-band FP Tilt}
\tablewidth{0pt}
\tablehead{
\colhead{Model} &
\colhead{$m$\tablenotemark{a}} &
\colhead{$\mu$\tablenotemark{b}} &
\colhead{$\nu$\tablenotemark{c}} &
\colhead{$T_*$\tablenotemark{d}} &
\colhead{$T$\tablenotemark{e}} &
\colhead{$T_{*}/T$\tablenotemark{f}}%
}
\tablewidth{\linewidth}
\startdata
SSP    &    0.662    &    0.001    &    0.003    &    0.198    &    0.200    &    0.993    \\
AQ    &    0.455    &    0.065    &    0.150    &    0.128    &    0.197    &    0.652    \\
STR    &    0.391    &    0.085    &    0.190    &    0.108    &    0.194    &    0.558    \\
EXP    &    0.235    &    0.140    &    0.310    &    0.062    &    0.194    &    0.322    \\
FR    &    0.174    &    0.165    &    0.355    &    0.045    &    0.194    &    0.234    \\
OSP    &    0.196    &    0.155    &    0.340    &    0.052    &    0.194    &    0.270    \\

\enddata
\tablenotetext{a}{($M_{\ast}/L)_r \propto \sigma^{m}$. From table \ref{tab:MLSDSS}}
\tablenotetext{b}{$\mu$ required in eq. \ref{eq:FPDM} (with $\nu = 0$) to reproduce the $\log(\sigma)$ coefficient in the observed Coma FP.}
\tablenotetext{c}{same as (b), but with $\mu=0$}
\tablenotetext{d}{$M_{\ast}/L \propto M^{T_*}$: tilt due to stellar populations}
\tablenotetext{e}{$M/L \propto M^{T}$: tilt of the FP}
\tablenotetext{f}{$T_*/T$ is the fraction of the tilt due to stellar populations}
\label{tab:DMFP}
\end{deluxetable}

\section{Discussion}
\label{sec:discuss}

\subsection{Summary of Model Fits}

We consider each model separately, and summarize the comparison with the observables. The OSP case is already ruled out by the Balmer linestrengths, but we note that it is also a poor fit to the colors, to the FP and dynamical $M/L$.

\begin{enumerate}

\item[FR:] The 2\% frosting  model consists mostly of old stars. While this model is a good fit to the slope of the $u-g$ color-$\sigma$ relation, it is a poor fit to the slopes of the $g-i$ and particularly the $i-K$ color-$\sigma$ relation. It overpredicts the FP-based $r$-band mass-to-light ratios, and those from SAURON  by significant factors (1.7 - 1.9).  We reject this model.

\item[EXP:] This model also consists mostly of old stars, with the young ages arising from the exponential tail of young stars.  This model, like FR,  is a good fit to the slope of the $u-g$ diagram, it is too flat fit in $g-i$. Like the FR model, the stellar $M/L$ also overpredicts the FP-based dynamical $r$-band mass-to-light ratios, and those from SAURON by factors 1.45 \& 1.65, respectively.   Given that we expect there to be some dark matter present in RSGs, we rank this model as ``disfavored''.

\item[AQ:] The AQ model consists of old through intermediate age stars.  In the color diagrams, its fit is better than EXP but poorer than SSP. In terms of mass-to-light ratio, it marginally overpredicts the FP and dynamical mass-to-light ratios, by factors 1.09 and 1.25.  Given the uncertainties, we rank this model as ``acceptable'', although we note that the implication  of this model is that there is little DM at low velocity dispersions.  

\item[STR] The behavior of this model is similar to the AQ model in most respects. 

\item[SSP:] This model is the best fit to the linestrengths. It also fits the slopes of the color diagrams the best overall: it is somewhat too steep in $u$-$g$, but the expectation that there may be small amounts of dust affecting the observed $u$-$g$ colors at low $\sigma$ goes in the right sense to account for this difference.  It is the best fit in $g-i$, and aside from the rejected OSP model has the closest fit in $i-K$. Its predicted stellar mass-to-light ratios are always comfortably below the FP-based by factors $\sim{} 0.8$ and appear to be similar to the dynamical mass-to-light ratios measured by the SAURON team.  We rank this model as ``good''.

\end{enumerate}

In summary,  the models that fit best to the faint end of the FP, and to the SAURON mass-to-light ratios, are the SSP and AQ models. The models which are ``mostly-old'' (EXP, FR, OSP) fail to fit the faint end of the FP or the low-$\sigma$ end of the SAURON $M/L$ ratios by factors of $\sim 1.45-1.7$. 

\subsection{Comparison of the predicted and observed scatter}

For all of the models, with the exception of SSP, the predicted scatter from stellar population effects is well within the scatters observed in the color-$\sigma$, M/L-$\sigma$, and FP relations. The SSP model, however, overpredicts the scatter in the $M/L$ and, to a lesser extent, in the FP.   This is a direct result of assuming that age is strongly anti-correlated with metallicity at a given velocity dispersion (Section \ref{sec:scatfit}). Recall that the scatter in stellar population parameters is constrained by the scatter in linestrengths. When age and metallicity are anti-correlated as we have assumed, their effects on the predicted linestrength tend to partially cancel. Consequently, the observed scatter in a given linestrength index can be fit with a larger scatter in age and a larger anti-correlated scatter in metallicity than would be the case if these two parameters were assumed uncorrelated. The same ``conspiracy'' keeps the scatter in colors low. Mass-to-light ratios are more sensitive to age, however, and so as a result, the scatter in $M_*/L$ is larger than it would be if we had assumed that age and metallicity were uncorrelated when fitting the linestrength scatters. As a result, this does not invalidate the SSP model itself, since it may be possible to fit the line strengths adequately with a weaker age--metallicity anti-correlation\footnote{See \cite{SmiLucHud09c} for an alternative approach to constraining the anti-correlation between age and metallicity.}.
For the other CSFH models, the stellar $M/L$ ratios are largely dominated by older stars and so have considerably less scatter. 

\subsection{Timescales of Star Formation Histories of Red-Sequence Galaxies}

The results here argue that star formation histories of RSGs were either a short SSP-like burst or of somewhat more extended duration as in the AQ model. It is difficult to distinguish between these two cases based on this low-redshift data alone. 

One method, although it is somewhat uncertain, is via \ale\ ratios. \cite{ThoMarBen05} have argued that, in order to obtain high \ale\ ratios, the duration of star formation must be less than $\sim 1$ Gyr. Specifically, they propose a relation $[\alpha/\rm{Fe}] = 1/5 - 1/6 \log(\Delta t)$, where $\Delta t$ is the full-width-at-half-maximum of a Gaussian star formation history. For RSGs of all masses, $[\alpha/\rm{Fe}] > 0.2$ (Fig. \ref{fig:params1} and \ref{fig:params2} ) so this implies $\Delta t < 1$ Gyr. Thus based on \ale\ ratios and the \cite{ThoMarBen05} models, the SSP would be preferred over the AQ model, for which the duration of star formation can extend over a range of up to $\sim 10$ Gyr at $\sigma \sim 70$ \kms.
However, it must be emphasized that the translation between $[\alpha/\rm{Fe}]$ and star-formation timescale depends on many assumptions about the chemical enrichment histories of early-type galaxies, that remain uncertain at present.

Of course, more complex star formation histories than the ones described here are possible. For example, a plausible model is one in which the bulge component formed in a short SSP-like burst, whereas the disk component was quenched.  Clearly the results from this model would lie somewhere between the SSP and AQ cases.

\subsection{Implications for the tilt of the FP}

We have shown that the best-fitting model is the SSP, and that, as a result of the scaling of SSP age with velocity dispersion, there is a strong trend of stellar mass-to-light ratio with velocity dispersion. 
Consequently, we find that 75\%-100\% of the $r/I$-band FP tilt is due to stellar population variations as a function of $\sigma$. 

A strong SSP-age-driven scaling has generally not been assumed in previous analyses of the FP, leading to the conclusion that the ``extra'' tilt, e.g. due to DM variations, is much stronger than we find here.  
For example, \cite{TruBurBel04}, assumed an exponential star formation history model, and concluded that only 1/4 the tilt was due to stellar populations.  In fact, we would have come to a similar value (30\%), had we adopted the EXP model (see Table \ref{tab:DMFP}). The EXP model is, however, inconsistent with the observed dynamical mass-to-light ratios, as we have shown (Section \ref{sec:ML}). 

\cite{PadSelStr04} based their scaling analyses on the star formation  models of \cite{KauHecWhi03}, which are combinations of the EXP and FR models, and found a strong scaling of dynamical mass with stellar mass. A similar conclusion was reached by \cite{GalChaBri06}, who found $M\sbr{dyn}/M_* \propto M_*^{0.28}$,  and \cite{HydBer08}, who found $M\sbr{dyn}/M_* \propto M\sbr{dyn}^{0.17}$, both based on models similar to those of \cite{KauHecWhi03}.  As a result the assumed star formation history, these studies will overestimate the stellar mass at low masses and hence will overestimate the strength of the DM scaling. For the SSP model, there is essentially no FP tilt due to DM. For the AQ model, the scaling is weak: $M\sbr{dyn}/M_* \propto M_*^{0.06}$.

These results have important consequences for estimating the contributions of mergers and dissipation in forming the FP.  For example, dry mergers predict a scaling $M\sbr{dyn}/M_* \propto M_*^{0.12-0.25}$  \citep{BoyMaQua06}. \cite{RobCoxHer06} studied dissipational mergers  argued that the tilt of the FP is due to a varying amount of dissipation in mergers along the sequence as a function of mass. If this interpretation is correct, a consequence of our study is that there is little, if any, variation in dissipation as a function of mass.

One might assume that the tilt of the FP in the near-infrared would be less sensitive to the effects of stellar populations, and hence that the $K$-band FP tilt should reflect primarily the effects of dark matter and non-homology.  However, this is approximately valid only in mostly-old SFH models, which cannot simultaneously match line indices and the absolute mass-to-light ratio. For SSP or AQ models, the K-band tilt can be strongly affected by stellar population variations.
 
\subsection{Implications for stellar masses and the stellar mass density of the Universe in red galaxies}

\begin{figure}[htbp]
	\centering
	\mbox{\subfigure{\includegraphics[width=\linewidth]{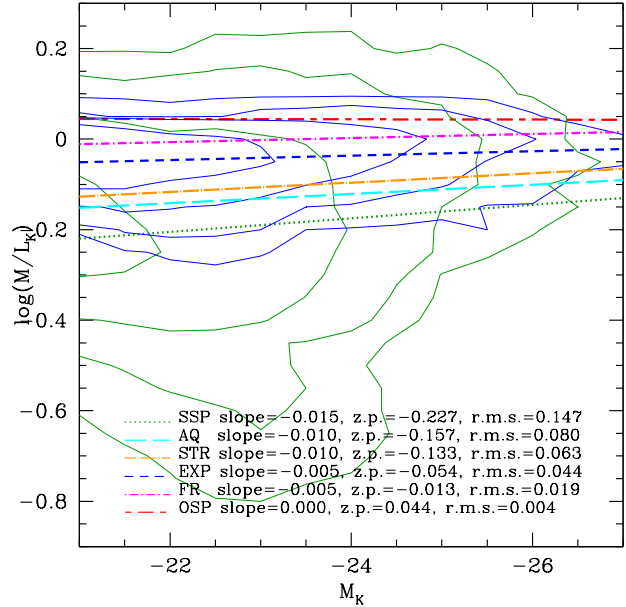}}}
        \caption{$M/L_K$ vs $M_K$. Best fit lines are calculated for
          those galaxies below $M_K<-21$, where the data are complete
          is $\log(\sigma)$. We have overlaid the contours for the SSP
          and EXP case, to illustrate that the scatter in
          $\log(\sigma)$ is diagonal, and also to show the very
          different scatters in $\log(M/L_K)$ between models. The quoted zero-points (z.p.) are
          measured at $M_K=-24$ ($L_{\ast}$). Line styles and colors as in
          Fig. \ref{fig:6fitlines}. Green and blue contours are for the SSP and EXP populations, and levels are as in Fig. \ref{fig:params1}.
          Notice the larger scatter in the SSP case, which decreases the mean mass-to-light ratio.}
	\label{fig:MLKMK}
\end{figure}

We have found that models that are ``mostly-old'' are ruled out: for small mass systems, their stellar $M/L$ ratios exceed the total dynamical $M/L$ ratios by factors of $\sim 1.45-1.7$. This class of models, which includes exponential star formation histories and ``frosting'' scenarios, have often been used to model the star formation histories of galaxies \citep{ColNorBau01, KauHecWhi03,BelMcIKat03,GalChaBri06}. While this may be a reasonable model for late-type or blue galaxies, we have found that it is not a good fit for RSGs. Consequently, the stellar masses of red galaxies have been overestimated by this assumption. 

If we assume a different star formation history model, we will arrive at different conclusions, even at bright magnitudes. This is because the dependence of $M_*/L_K$ on age is quite strong for e.g. the SSP model, and hence the scatter in $M_*/L_K$ is also quite large. Thus many galaxies which are bright in $K$ are actually galaxies with relatively small stellar mass that are quite young, boosting their $K$-band luminosity and lowering their mass-to-light ratio.  Figure \ref{fig:MLKMK} shows the $K$-band mass-to-light ratio for the different star formation models. The ratios of the mean $K$-band mass-to-light ratios at $M_K = -24$ (the characteristic $K$-band luminosity) are 0.73:0.82:1 for SSP:AQ:EXP. Thus for the SSP case, the luminosity density in the early-type galaxies or RSGs as calculated from the $K$-band should be reduced by a factor of $\sim 0.73$.  This would reduce the stellar mass  density calculated by \cite{BelMcIKat03} from $3.2$ in units of $10^{-8} h M_{\sun}$ Mpc$^{-3}$ to $2.3$ in the same units, and would reduce the fraction of stellar mass in early types (assuming that late-types are unaffected) from 60\% to 50\%.  Repeating this calculation in the $g$-band, for example, the reduction in stellar mass in RSGs is similar to that found for morphologically-selected early-types in the $K$-band: a factor $\sim 0.7$. Thus the total fraction of all stellar mass which resides on the red sequence, as derived by \cite{BelMcIKat03}, would be reduced from 70\% to 60\%.

\section{Conclusions}
\label{sec:conc}

We have constructed 6 parametric star formation models, each representing a distinct star formation history. The modeled Lick indices were fit to observed linestrength data in the Shapley supercluster, and sets of best-fit model parameters were constructed for each velocity dispersion bin for each model.  Synthetic cluster populations were then constructed based on the statistical distributions of each stellar population parameter, and mass-to-light ratios were derived from the stellar populations.

Our main results are as follows:
\begin{enumerate}

\item From the fits to linestrength indices, we find that in all models there exists  a ``downsizing'' trend in luminosity weighted-age age for galaxies with $\sigma > 70 \kms$. Lower-$\sigma$ galaxies also tend to be less metal rich and $\alpha$-enhanced.  We find that, based on the fits to the Lick indices alone, the OSP model is strongly rejected. This agrees with previous studies \citep{NelSmiHud05}. The EXP and FR models are poorer fits to the linestrength data than the SSP model, but cannot rejected from this test alone.

\item The linestrength data suggest that the ``downsizing'' effect stops at $\sigma \sim  70$ \kms\ below which luminosity-weighted stellar ages either stop decreasing or begin to increase. This is not a selection effect, which would tend to bias the data in the opposite sense.

\item  The derived stellar mass-to-light ratios from the models were compared to dynamical mass-to-light ratios from the FP and from  \cite{CapBacBur06}.  The ``mostly-old'' class of star formation models (OSP, EXP and FR) have Kroupa-IMF \emph{stellar} mass-to-light ratios that are larger than the observed dynamical mass-to-light ratios for low-$\sigma$ galaxies, by factors $1.45$--$1.7$. For these models, the addition of dark matter would make the agreement worse, thus these ``mostly-old'' models are ruled out. We conclude that the $\sim 6$ Gyr luminosity-weighted ages of low-mass RSGs do indeed reflect the ``intermediate'' ages of the bulk of the stellar population, as parametrized by the SSP or AQ mdoels.  A hybrid of the SSP and AQ, in which, for example, the bulge of the system formed in a short intense burst, while the disk was quenched at a later time, would likely fit all of the data well.

\item We have shown that, for the ``intermediate-age'' scenarios, the scaling of stellar mass-to-light ratio with mass is strong, both in the optical bands and in the $K$-band. Consequently, stellar populations explain most of the FP tilt in the optical bands, and, for the SSP case, at least 50\% of the tilt in the $K$-band. This leaves much less room for tilt due to variations in the dark-to-stellar mass along the red sequence.

\item For the SSP model, the stellar masses are considerably reduced, particularly at low velocity dispersions, compared to the star formation histories assumed in previous studies. As a consequence, the stellar mass density in red galaxies may be as much as 30\%  lower than previously assumed. 

\end{enumerate}

A further test of the star formation histories of RSGs is via the redshift evolution of their red and blue fractions, of their dwarf-to-giant ratios and of their color-magnitude relation slopes and zero-points. In particular, the AQ and SSP models will show strong evolution from the blue cloud to the red-sequence, whereas the ``mostly-old'' models will have a well-populated red-sequence at intermediate lookback times ($z \sim 0.5$). 
We will explore these predictions in future work.

\acknowledgments

We thank Claudia Maraston for providing tables of mass-to-light ratios based on empirical stellar spectra in advance of publication.
MJH acknowledges support from NSERC. 
RJS is supported by STFC rolling grant PP/C501568/1 `Extragalactic Astronomy and Cosmology at Durham 2005--2010'.

\appendix
\section{Tables of Scaling Relations}

Tables \ref{tab:MLJC} and  \ref{tab:MLSDSS} summarize the slopes and intercepts as a function of velocity dispersion for Johnson-Cousins, infrared and SDSS passbands for each star formation history model. Zero points are measured at $\log(\sigma)=2.00$.

\begin{deluxetable*}{c|ccc||ccc||c|ccc}  
\tablecaption{Mass-To-Light Ratio, Magnitude, color - Velocity Dispersion Relations By Model (Johnson-Cousins [Vega])}
\tablewidth{0pt}
\tablehead{
\colhead{Band (X)} &
\colhead{$m=\frac{d\log(M/L_X)}{d\log(\sigma)}$\tablenotemark{a}} &
\colhead{z.p.\tablenotemark{b}} &
\colhead{r.m.s.} &
\colhead{$\frac{d M_X}{d\log(\sigma)}$\tablenotemark{a}} &
\colhead{z.p.\tablenotemark{b}} &
\colhead{r.m.s.} &
\colhead{color} &
\colhead{$\frac{d (X-X')}{d\log(\sigma)}$\tablenotemark{a}} &
\colhead{z.p.\tablenotemark{b}} &
\colhead{r.m.s.} }

\startdata

	\multicolumn{11}{c}{Single Stellar Population Model (SSP)}\\ \hline
U	&	0.956	&	0.757	&	0.135	&	-3.666	&	-18.283	&	0.661	&	U-B	&	0.457	&	0.571	&	0.041	\\
B	&	0.773	&	0.581	&	0.125	&	-4.123	&	-18.855	&	0.649	&	B-V	&	0.199	&	0.980	&	0.019	\\
V	&	0.693	&	0.449	&	0.120	&	-4.322	&	-19.835	&	0.643	&	V-R	&	0.099	&	0.475	&	0.009	\\
R	&	0.654	&	0.423	&	0.118	&	-4.421	&	-20.310	&	0.640	&	R-I	&	0.093	&	0.613	&	0.010	\\
I	&	0.617	&	0.313	&	0.116	&	-4.513	&	-20.923	&	0.638	&	I-J	&	0.231	&	1.107	&	0.046	\\
J	&	0.524	&	0.047	&	0.128	&	-4.744	&	-22.030	&	0.653	&	J-H	&	0.054	&	0.674	&	0.008	\\
H	&	0.503	&	-0.095	&	0.128	&	-4.798	&	-22.704	&	0.653	&	H-K	&	0.052	&	0.224	&	0.017	\\
K	&	0.482	&	-0.168	&	0.133	&	-4.851	&	-22.928	&	0.660	&	-	&	-	&	-	&	-	\\ \hline
	
	\multicolumn{11}{c}{Abruptly Quenched Model (AQ)}\\ \hline	
U	&	0.726	&	0.799	&	0.082	&	-4.231	&	-18.191	&	0.606	&	U-B	&	0.422	&	0.574	&	0.052	\\
B	&	0.557	&	0.621	&	0.069	&	-4.653	&	-18.764	&	0.597	&	B-V	&	0.186	&	0.981	&	0.022	\\
V	&	0.483	&	0.489	&	0.064	&	-4.839	&	-19.745	&	0.594	&	V-R	&	0.087	&	0.477	&	0.012	\\
R	&	0.448	&	0.462	&	0.062	&	-4.926	&	-20.222	&	0.593	&	R-I	&	0.076	&	0.615	&	0.012	\\
I	&	0.418	&	0.352	&	0.061	&	-5.002	&	-20.837	&	0.592	&	I-J	&	0.227	&	1.114	&	0.060	\\
J	&	0.327	&	0.082	&	0.072	&	-5.230	&	-21.952	&	0.601	&	J-H	&	0.041	&	0.677	&	0.011	\\
H	&	0.311	&	-0.061	&	0.074	&	-5.270	&	-22.628	&	0.602	&	H-K	&	0.048	&	0.227	&	0.021	\\
K	&	0.292	&	-0.135	&	0.080	&	-5.318	&	-22.856	&	0.607	&	-	&	-	&	-	&	-	\\ \hline

	\multicolumn{11}{c}{Strangulation Model (STR)}\\ \hline	
U	&	0.664	&	0.819	&	0.062	&	-4.353	&	-18.132	&	0.596	&	U-B	&	0.421	&	0.573	&	0.048	\\
B	&	0.496	&	0.642	&	0.049	&	-4.775	&	-18.705	&	0.590	&	B-V	&	0.191	&	0.979	&	0.020	\\
V	&	0.419	&	0.510	&	0.044	&	-4.966	&	-19.684	&	0.587	&	V-R	&	0.086	&	0.477	&	0.010	\\
R	&	0.385	&	0.483	&	0.042	&	-5.052	&	-20.161	&	0.587	&	R-I	&	0.073	&	0.616	&	0.011	\\
I	&	0.356	&	0.373	&	0.042	&	-5.125	&	-20.777	&	0.587	&	I-J	&	0.235	&	1.111	&	0.057	\\
J	&	0.262	&	0.105	&	0.052	&	-5.360	&	-21.888	&	0.594	&	J-H	&	0.039	&	0.677	&	0.009	\\
H	&	0.246	&	-0.038	&	0.054	&	-5.399	&	-22.564	&	0.595	&	H-K	&	0.051	&	0.226	&	0.018	\\
K	&	0.226	&	-0.112	&	0.060	&	-5.450	&	-22.790	&	0.599	&	-	&	-	&	-	&	-	\\ \hline

	\multicolumn{11}{c}{Exponential Star Formation Rate Model (EXP)}\\ \hline
U	&	0.478	&	0.896	&	0.063	&	-4.694	&	-17.931	&	0.597	&	U-B	&	0.380	&	0.583	&	0.063	\\
B	&	0.326	&	0.715	&	0.040	&	-5.073	&	-18.513	&	0.582	&	B-V	&	0.168	&	0.982	&	0.024	\\
V	&	0.259	&	0.582	&	0.031	&	-5.241	&	-19.496	&	0.578	&	V-R	&	0.072	&	0.481	&	0.013	\\
R	&	0.230	&	0.554	&	0.027	&	-5.313	&	-19.977	&	0.577	&	R-I	&	0.063	&	0.621	&	0.013	\\
I	&	0.205	&	0.442	&	0.025	&	-5.376	&	-20.597	&	0.575	&	I-J	&	0.230	&	1.107	&	0.071	\\
J	&	0.113	&	0.175	&	0.032	&	-5.607	&	-21.705	&	0.577	&	J-H	&	0.040	&	0.679	&	0.013	\\
H	&	0.096	&	0.031	&	0.036	&	-5.647	&	-22.384	&	0.578	&	H-K	&	0.048	&	0.225	&	0.020	\\
K	&	0.077	&	-0.043	&	0.042	&	-5.694	&	-22.609	&	0.581	&	-	&	-	&	-	&	-	\\ \hline

	\multicolumn{11}{c}{Frosting Model (FR)}\\ \hline	
U	&	0.395	&	0.972	&	0.069	&	-5.123	&	-17.743	&	0.596	&	U-B	&	0.341	&	0.613	&	0.062	\\
B	&	0.259	&	0.779	&	0.044	&	-5.464	&	-18.357	&	0.581	&	B-V	&	0.159	&	0.996	&	0.025	\\
V	&	0.195	&	0.640	&	0.034	&	-5.623	&	-19.353	&	0.577	&	V-R	&	0.062	&	0.489	&	0.012	\\
R	&	0.171	&	0.609	&	0.030	&	-5.684	&	-19.842	&	0.576	&	R-I	&	0.048	&	0.630	&	0.014	\\
I	&	0.151	&	0.493	&	0.025	&	-5.733	&	-20.472	&	0.574	&	I-J	&	0.201	&	1.123	&	0.063	\\
J	&	0.071	&	0.220	&	0.011	&	-5.934	&	-21.594	&	0.572	&	J-H	&	0.014	&	0.689	&	0.010	\\
H	&	0.066	&	0.072	&	0.013	&	-5.947	&	-22.283	&	0.573	&	H-K	&	0.026	&	0.233	&	0.017	\\
K	&	0.055	&	-0.005	&	0.019	&	-5.973	&	-22.517	&	0.574	&	-	&	-	&	-	&	-	\\ \hline

	\multicolumn{11}{c}{Old Single Stellar Population Model (OSP)}\\ \hline
U	&	0.451	&	0.960	&	0.073	&	-4.903	&	-17.774	&	0.603	&	U-B	&	0.423	&	0.584	&	0.069	\\
B	&	0.282	&	0.779	&	0.046	&	-5.326	&	-18.358	&	0.586	&	B-V	&	0.159	&	0.993	&	0.026	\\
V	&	0.218	&	0.641	&	0.035	&	-5.486	&	-19.352	&	0.582	&	V-R	&	0.068	&	0.484	&	0.011	\\
R	&	0.191	&	0.612	&	0.031	&	-5.553	&	-19.835	&	0.580	&	R-I	&	0.069	&	0.620	&	0.011	\\
I	&	0.164	&	0.500	&	0.027	&	-5.622	&	-20.456	&	0.579	&	I-J	&	0.303	&	1.057	&	0.048	\\
J	&	0.043	&	0.253	&	0.008	&	-5.926	&	-21.512	&	0.577	&	J-H	&	0.059	&	0.672	&	0.011	\\
H	&	0.019	&	0.112	&	0.004	&	-5.985	&	-22.185	&	0.576	&	H-K	&	0.062	&	0.210	&	0.010	\\
K	&	-0.006	&	0.044	&	0.004	&	-6.047	&	-22.394	&	0.577	&	-	&	-	&	-	&	-	
\enddata
\tablenotetext{a}{Measured with galaxies with $\log(\sigma)>1.844$}
\tablenotetext{b}{Measured at $\log(\sigma)=2.00$}
\label{tab:MLJC}
\end{deluxetable*}

\begin{deluxetable*}{c|ccc||ccc||c|ccc}  
\tablecaption{Mass-To-Light Ratio, Magnitude, color - Velocity Dispersion Relations By Model (SDSS [AB])}
\tablewidth{0pt}
\tablehead{
\colhead{Band (X)} &
\colhead{$m=\frac{d\log(M/L_X)}{d\log(\sigma)}$\tablenotemark{a}} &
\colhead{z.p.\tablenotemark{b}} &
\colhead{r.m.s.} &
\colhead{$\frac{d M_X}{d\log(\sigma)}$\tablenotemark{a}} &
\colhead{z.p.\tablenotemark{b}} &
\colhead{r.m.s.} &
\colhead{color} &
\colhead{$\frac{d (X-X')}{d\log(\sigma)}$\tablenotemark{a}} &
\colhead{z.p.\tablenotemark{b}} &
\colhead{r.m.s.} }
\startdata
	\multicolumn{11}{c}{Single Stellar Population Model (SSP)}\\ \hline
u	&	0.958	&	0.735	&	0.136	&	-3.529	&	-17.553	&	0.667	&	u-g	&	0.545	&	1.807	&	0.050	\\
g	&	0.740	&	0.528	&	0.123	&	-4.075	&	-19.360	&	0.651	&	g-r	&	0.196	&	0.838	&	0.018	\\
r	&	0.662	&	0.373	&	0.119	&	-4.270	&	-20.198	&	0.646	&	r-i	&	0.084	&	0.320	&	0.009	\\
i	&	0.628	&	0.289	&	0.117	&	-4.354	&	-20.518	&	0.643	&	i-z	&	0.109	&	0.296	&	0.015	\\
z	&	0.585	&	0.191	&	0.118	&	-4.463	&	-20.814	&	0.645	&	z-J	&	0.151	&	1.220	&	0.034	\\ \hline

	\multicolumn{11}{c}{Abruptly Quenched Model (AQ)}\\ \hline	
u	&	0.727	&	0.777	&	0.082	&	-4.151	&	-17.456	&	0.608	&	u-g	&	0.502	&	1.811	&	0.061	\\
g	&	0.527	&	0.568	&	0.067	&	-4.653	&	-19.266	&	0.597	&	g-r	&	0.179	&	0.840	&	0.022	\\
r	&	0.455	&	0.412	&	0.062	&	-4.833	&	-20.106	&	0.594	&	r-i	&	0.068	&	0.322	&	0.010	\\
i	&	0.428	&	0.328	&	0.061	&	-4.900	&	-20.428	&	0.594	&	i-z	&	0.100	&	0.298	&	0.021	\\
z	&	0.388	&	0.228	&	0.063	&	-5.000	&	-20.726	&	0.595	&	z-J	&	0.153	&	1.226	&	0.044	\\ \hline

	\multicolumn{11}{c}{Strangulation Model (STR)}\\ \hline	
u	&	0.664	&	0.797	&	0.061	&	-4.253	&	-17.403	&	0.586	&	u-g	&	0.500	&	1.809	&	0.057	\\
g	&	0.464	&	0.589	&	0.047	& -4.753	&	-19.212	&	0.579	&	g-r	&	0.181	&	0.839	&	0.020	\\
r	&	0.391	&	0.434	&	0.043	&	-4.934	&	-20.051	&	0.577	&	r-i	&	0.065	&	0.323	&	0.009	\\
i	&	0.366	&	0.349	&	0.042	&	-4.999	&	-20.374	&	0.577	&	i-z	&	0.102	&	0.298	&	0.020	\\
z	&	0.325	&	0.250	&	0.043	& -5.101	&	-20.671	&	0.578	&	z-J	&	0.158	&	1.223	&	0.041 \\ \hline
	
	\multicolumn{11}{c}{Exponential Star Formation Rate Model (EXP)}\\ \hline
u	&	0.477	&	0.876	&	0.064	&	-4.830	&	-17.201	&	0.589	&	u-g	&	0.450	&	1.822	&	0.075	\\
g	&	0.297	&	0.663	&	0.036	&	-5.280	&	-19.023	&	0.573	&	g-r	&	0.155	&	0.844	&	0.024	\\
r	&	0.235	&	0.505	&	0.028	&	-5.435	&	-19.868	&	0.570	&	r-i	&	0.053	&	0.328	&	0.010	\\
i	&	0.214	&	0.418	&	0.025	&	-5.488	&	-20.196	&	0.570	&	i-z	&	0.100	&	0.298	&	0.026	\\
z	&	0.174	&	0.319	&	0.023	&	-5.588	&	-20.494	&	0.569	&	z-J	&	0.154	&	1.220	&	0.050	\\ \hline

	\multicolumn{11}{c}{Frosting Model (FR)}\\ \hline	
u	&	0.393	&	0.952	&	0.069	&	-5.090	&	-17.003	&	0.593	&	u-g	&	0.404	&	1.860	&	0.073	\\
g	&	0.231	&	0.724	&	0.040	&	-5.494	&	-18.863	&	0.578	&	g-r	&	0.143	&	0.859	&	0.024	\\
r	&	0.174	&	0.561	&	0.031	&	-5.637	&	-19.721	&	0.574	&	r-i	&	0.038	&	0.337	&	0.011	\\
i	&	0.159	&	0.470	&	0.027	&	-5.675	&	-20.058	&	0.573	&	i-z	&	0.086	&	0.307	&	0.025	\\
z	&	0.124	&	0.367	&	0.018	&	-5.761	&	-20.365	&	0.572	&	z-J	&	0.134	&	1.229	&	0.044	\\ \hline

	\multicolumn{11}{c}{Old Single Stellar Population Model (OSP)}\\ \hline
u	&	0.454	&	0.939	&	0.074	&	-4.873	&	-17.033	&	0.599	&	u-g	&	0.498	&	1.826	&	0.082	\\
g	&	0.255	&	0.724	&	0.041	&	-5.370	&	-18.859	&	0.581	&	g-r	&	0.149	&	0.851	&	0.024	\\
r	&	0.196	&	0.563	&	0.032	&	-5.519	&	-19.710	&	0.578	&	r-i	&	0.050	&	0.330	&	0.008	\\
i	&	0.176	&	0.475	&	0.029	&	-5.569	&	-20.041	&	0.577	&	i-z	&	0.129	&	0.284	&	0.021	\\
z	&	0.124	&	0.382	&	0.020	&	-5.699	&	-20.324	&	0.576	&	z-J	&	0.204	&	1.182	&	0.032
\enddata
\tablenotetext{a}{Measured with galaxies with $\log(\sigma)>1.844$}
\tablenotetext{b}{Measured at $\log(\sigma)=2.00$}
\label{tab:MLSDSS}
\end{deluxetable*}

\bibliographystyle{apj}
\bibliography{mjh}

\clearpage

\end{document}